\documentclass[a4paper,onecolumn,11pt,]{quantumarticle}
\pdfoutput=1
\usepackage[utf8]{inputenc}
\usepackage[english]{babel}
\usepackage[T1]{fontenc}
\usepackage{amsmath}
\usepackage{hyperref}
\usepackage{amsfonts}
\usepackage{braket}
\usepackage{multicol}
\usepackage{amsfonts}
\usepackage{braket}
\usepackage{mathtools}
\usepackage{amsthm}
\usepackage{ulem}
\usepackage{bbold}
\usepackage{subcaption}
\usepackage{stmaryrd}
\usepackage[table]{xcolor}

\usepackage{tikz}
\usepackage{tikzit}
\usepackage{tikz, circuitikz}
\tikzstyle{env}=[copoint,regular polygon rotate=0,minimum width=0.2cm, fill=black]

\tikzstyle{probs}=[shape=semicircle,fill=white,draw=black,shape border rotate=180,minimum width=1.2cm]

\tikzstyle{nudge}=[yshift=0.6mm]

\tikzstyle{every picture}=[baseline=-0.25em,scale=0.5]
\tikzstyle{dotpic}=[] 
\tikzstyle{diredges}=[every to/.style={diredge}]
\tikzstyle{math matrix}=[matrix of math nodes,left delimiter=(,right delimiter=),inner sep=2pt,column sep=1em,row sep=0.5em,nodes={inner sep=0pt},text height=1.5ex, text depth=0.25ex]

\tikzstyle{inline text}=[text height=1.5ex, text depth=0.25ex,yshift=0.5mm]
\tikzstyle{label}=[font=\footnotesize,text height=1.5ex, text depth=0.25ex]
\tikzstyle{left label}=[label,anchor=east,xshift=2mm]
\tikzstyle{right label}=[label,anchor=west,xshift=-2mm]

\tikzstyle{braceedge}=[decorate,decoration={brace,amplitude=2mm,raise=-1mm}]
\tikzstyle{small braceedge}=[decorate,decoration={brace,amplitude=1mm,raise=-1mm}]

\tikzstyle{doubled}=[line width=1.6pt] 
\tikzstyle{boldedge}=[doubled,shorten <=-0.17mm,shorten >=-0.17mm]
\tikzstyle{boldedgegray}=[doubled,gray,shorten <=-0.17mm,shorten >=-0.17mm]
\tikzstyle{singleedgegray}=[gray]

\tikzstyle{semidoubled}=[line width=1.4pt] 
\tikzstyle{semiboldedgegray}=[semidoubled,gray,shorten <=-0.17mm,shorten >=-0.17mm]

\tikzstyle{boxedge}=[semiboldedgegray]

\tikzstyle{boldedgedashed}=[very thick,dashed,shorten <=-0.17mm,shorten >=-0.17mm]
\tikzstyle{vboldedgedashed}=[doubled,dashed,shorten <=-0.17mm,shorten >=-0.17mm]
\tikzstyle{left hook arrow}=[left hook-latex]
\tikzstyle{right hook arrow}=[right hook-latex]
\tikzstyle{sembracket}=[line width=0.5pt,shorten <=-0.07mm,shorten >=-0.07mm]

\tikzstyle{causal edge}=[->,thick,gray]
\tikzstyle{causal nondir}=[thick,gray]
\tikzstyle{timeline}=[thick,gray, dashed]

\tikzstyle{cedge}=[<->,thick,gray!70!white]

\tikzstyle{empty diagram}=[draw=gray!40!white,dashed,shape=rectangle,minimum width=1cm,minimum height=1cm]
\tikzstyle{empty diagram small}=[draw=gray!50!white,dashed,shape=rectangle,minimum width=0.6cm,minimum height=0.5cm]

\tikzstyle{dot}=[inner sep=0mm,minimum width=2mm,minimum height=2mm,draw,shape=circle]  
\tikzstyle{Wsquare}=[white dot, shape=regular polygon, rounded corners=0.8 mm, minimum size=3.3 mm, regular polygon sides=3, outer sep=-0.2mm]
\tikzstyle{Wsquareadj}=[white dot, shape=regular polygon, rounded corners=0.8 mm, minimum size=3.3 mm, regular polygon sides=3, outer sep=-0.2mm, regular polygon rotate=180]
\tikzstyle{ddot}=[inner sep=0mm, doubled, minimum width=2.5mm,minimum height=2.5mm,draw,shape=circle]

\tikzstyle{black dot}=[dot,fill=black]
\tikzstyle{white dot}=[dot,fill=white,,text depth=-0.2mm]
\tikzstyle{white Wsquare}=[Wsquare,fill=white,,text depth=-0.2mm]
\tikzstyle{white Wsquareadj}=[Wsquareadj,fill=white,,text depth=-0.2mm]
\tikzstyle{green dot}=[white dot] 
\tikzstyle{gray dot}=[dot,fill=gray!40!white,,text depth=-0.2mm]
\tikzstyle{red dot}=[gray dot] 

\tikzstyle{black ddot}=[ddot,fill=black]
\tikzstyle{white ddot}=[ddot,fill=white]
\tikzstyle{gray ddot}=[ddot,fill=gray!40!white]

\tikzstyle{gray edge}=[gray!60!white]

\tikzstyle{small dot}=[inner sep=0.5mm,minimum width=0pt,minimum height=0pt,draw,shape=circle]

\tikzstyle{small black dot}=[small dot,fill=black]
\tikzstyle{small white dot}=[small dot,fill=white]
\tikzstyle{small gray dot}=[small dot,fill=gray!40!white]

\tikzstyle{very small dot}=[inner sep=0.3mm,minimum width=0pt,minimum height=0pt,draw,shape=circle]

\tikzstyle{very small black dot}=[very small dot,fill=black]
\tikzstyle{very small white dot}=[small dot,fill=white]
\tikzstyle{very small gray dot}=[small dot,fill=gray!40!white]

\tikzstyle{causal dot}=[inner sep=0.4mm,minimum width=0pt,minimum height=0pt,draw=white,shape=circle,fill=gray!40!white]

\tikzstyle{phase dimensions}=[minimum size=5mm,font=\footnotesize,rectangle,rounded corners=2.5mm,inner sep=0.2mm,outer sep=-2mm]
\tikzstyle{dphase dimensions}=[minimum size=5mm,font=\footnotesize,rectangle,rounded corners=2.5mm,inner sep=0.2mm,outer sep=-2mm]

\tikzstyle{white phase dot}=[dot,fill=white,phase dimensions]
\tikzstyle{white phase ddot}=[ddot,fill=white,dphase dimensions]

\tikzstyle{white rect ddot}=[draw=black,fill=white,doubled,minimum size=5mm,font=\footnotesize,rectangle,rounded corners=2.5mm,inner sep=0.2mm]
\tikzstyle{gray rect ddot}=[draw=black,fill=gray!40!white,doubled,minimum size=6mm,font=\footnotesize,rectangle,rounded corners=3mm]

\tikzstyle{gray phase dot}=[dot,fill=gray!40!white,phase dimensions]
\tikzstyle{gray phase ddot}=[ddot,fill=gray!40!white,dphase dimensions]
\tikzstyle{grey phase dot}=[gray phase dot]
\tikzstyle{grey phase ddot}=[gray phase ddot]

\tikzstyle{small phase dimensions}=[minimum size=4mm,font=\tiny,rectangle,rounded corners=2mm,inner sep=0.2mm,outer sep=-2mm]
\tikzstyle{small dphase dimensions}=[minimum size=4mm,font=\tiny,rectangle,rounded corners=2mm,inner sep=0.2mm,outer sep=-2mm]

\tikzstyle{small gray phase dot}=[dot,fill=gray!40!white,small phase dimensions]
\tikzstyle{small gray phase ddot}=[ddot,fill=gray!40!white,small dphase dimensions]

\tikzstyle{small map}=[draw,shape=rectangle,minimum height=4mm,minimum width=4mm,fill=white]

\tikzstyle{cnot}=[fill=white,shape=circle,inner sep=-1.4pt]

\tikzstyle{asym hadamard}=[fill=white,draw,shape=NEbox,inner sep=0.6mm,font=\footnotesize,minimum height=4mm]
\tikzstyle{asym hadamard conj}=[fill=white,draw,shape=NWbox,inner sep=0.6mm,font=\footnotesize,minimum height=4mm]
\tikzstyle{asym hadamard dag}=[fill=white,draw,shape=SEbox,inner sep=0.6mm,font=\footnotesize,minimum height=4mm]

\tikzstyle{hadamard}=[fill=white,draw,inner sep=0.6mm,font=\footnotesize,minimum height=4mm,minimum width=4mm]
\tikzstyle{small hadamard}=[fill=white,draw,inner sep=0.6mm,minimum height=1.5mm,minimum width=1.5mm]
\tikzstyle{small hadamard rotate}=[small hadamard,rotate=45]
\tikzstyle{dhadamard}=[hadamard,doubled]
\tikzstyle{small dhadamard}=[small hadamard,doubled]
\tikzstyle{small dhadamard rotate}=[small hadamard rotate,doubled]
\tikzstyle{antipode}=[white dot,inner sep=0.3mm,font=\footnotesize]

\tikzstyle{scalar}=[diamond,draw,inner sep=0.5pt,font=\small]
\tikzstyle{dscalar}=[diamond,doubled, draw,inner sep=0.5pt,font=\small]

\tikzstyle{small box}=[rectangle,inline text,fill=white,draw,minimum height=5mm,yshift=-0.5mm,minimum width=5mm,font=\small]
\tikzstyle{small gray box}=[small box,fill=gray!30]
\tikzstyle{medium box}=[rectangle,inline text,fill=white,draw,minimum height=5mm,yshift=-0.5mm,minimum width=8mm,font=\small]
\tikzstyle{square box}=[small box] 
\tikzstyle{medium gray box}=[small box,fill=gray!30]
\tikzstyle{semilarge box}=[rectangle,inline text,fill=white,draw,minimum height=5mm,yshift=-0.5mm,minimum width=12.5mm,font=\small]
\tikzstyle{large box}=[rectangle,inline text,fill=white,draw,minimum height=5mm,yshift=-0.5mm,minimum width=15mm,font=\small]
\tikzstyle{large gray box}=[small box,fill=gray!30]

\tikzstyle{Bayes box}=[rectangle,fill=black,draw, minimum height=3mm, minimum width=3mm]

\tikzstyle{gray square point}=[small box,fill=gray!50]

\tikzstyle{dphase box white}=[dhadamard]
\tikzstyle{dphase box gray}=[dhadamard,fill=gray!50!white]
\tikzstyle{phase box white}=[hadamard]
\tikzstyle{phase box gray}=[hadamard,fill=gray!50!white]

\tikzstyle{point}=[regular polygon,regular polygon sides=3,draw,scale=0.75,inner sep=-0.5pt,minimum width=9mm,fill=white,regular polygon rotate=180]
\tikzstyle{copoint}=[regular polygon,regular polygon sides=3,draw,scale=0.75,inner sep=-0.5pt,minimum width=9mm,fill=white]
\tikzstyle{dpoint}=[point,doubled]
\tikzstyle{dcopoint}=[copoint,doubled]

\tikzstyle{wide copoint}=[fill=white,draw,shape=isosceles triangle,shape border rotate=90,isosceles triangle stretches=true,inner sep=0pt,minimum width=1.5cm,minimum height=6.12mm]
\tikzstyle{wide point}=[fill=white,draw,shape=isosceles triangle,shape border rotate=-90,isosceles triangle stretches=true,inner sep=0pt,minimum width=1.5cm,minimum height=6.12mm,yshift=-0.0mm]
\tikzstyle{wide point plus}=[fill=white,draw,shape=isosceles triangle,shape border rotate=-90,isosceles triangle stretches=true,inner sep=0pt,minimum width=1.74cm,minimum height=7mm,yshift=-0.0mm]

\tikzstyle{wide dpoint}=[fill=white,doubled,draw,shape=isosceles triangle,shape border rotate=-90,isosceles triangle stretches=true,inner sep=0pt,minimum width=1.5cm,minimum height=6.12mm,yshift=-0.0mm]

\tikzstyle{tinypoint}=[regular polygon,regular polygon sides=3,draw,scale=0.55,inner sep=-0.15pt,minimum width=6mm,fill=white,regular polygon rotate=180] 

\tikzstyle{white point}=[point]
\tikzstyle{white dpoint}=[dpoint]
\tikzstyle{green point}=[white point] 
\tikzstyle{white copoint}=[copoint]
\tikzstyle{gray point}=[point,fill=gray!40!white]
\tikzstyle{gray dpoint}=[gray point,doubled]
\tikzstyle{red point}=[gray point] 
\tikzstyle{gray copoint}=[copoint,fill=gray!40!white]
\tikzstyle{gray dcopoint}=[gray copoint,doubled]

\tikzstyle{white point guide}=[regular polygon,regular polygon sides=3,font=\scriptsize,draw,scale=0.65,inner sep=-0.5pt,minimum width=9mm,fill=white,regular polygon rotate=180]

\tikzstyle{black point}=[point,fill=black,font=\color{white}]
\tikzstyle{black copoint}=[copoint,fill=black,font=\color{white}]

\tikzstyle{tiny gray point}=[tinypoint,fill=gray!40!white]

\tikzstyle{diredge}=[->]
\tikzstyle{ddiredge}=[<->]
\tikzstyle{rdiredge}=[<-]
\tikzstyle{thickdiredge}=[->, very thick]
\tikzstyle{pointer edge}=[->,very thick,gray]
\tikzstyle{pointer edge part}=[very thick,gray]
\tikzstyle{dashed edge}=[dashed]
\tikzstyle{thick dashed edge}=[very thick,dashed]
\tikzstyle{thick gray dashed edge}=[thick dashed edge,gray!40]
\tikzstyle{thick map edge}=[very thick,|->]

\makeatletter
\newcommand{\boxshape}[3]{%
\pgfdeclareshape{#1}{
\inheritsavedanchors[from=rectangle] 
\inheritanchorborder[from=rectangle]
\inheritanchor[from=rectangle]{center}
\inheritanchor[from=rectangle]{north}
\inheritanchor[from=rectangle]{south}
\inheritanchor[from=rectangle]{west}
\inheritanchor[from=rectangle]{east}
\backgroundpath{
\southwest \pgf@xa=\pgf@x \pgf@ya=\pgf@y
\northeast \pgf@xb=\pgf@x \pgf@yb=\pgf@y

\@tempdima=#2
\@tempdimb=#3

\pgfpathmoveto{\pgfpoint{\pgf@xa - 5pt + \@tempdima}{\pgf@ya}}
\pgfpathlineto{\pgfpoint{\pgf@xa - 5pt - \@tempdima}{\pgf@yb}}
\pgfpathlineto{\pgfpoint{\pgf@xb + 5pt + \@tempdimb}{\pgf@yb}}
\pgfpathlineto{\pgfpoint{\pgf@xb + 5pt - \@tempdimb}{\pgf@ya}}
\pgfpathlineto{\pgfpoint{\pgf@xa - 5pt + \@tempdima}{\pgf@ya}}
\pgfpathclose
}
}}

\boxshape{NEbox}{0pt}{0pt}
\boxshape{SEbox}{0pt}{-5pt}
\boxshape{NWbox}{5pt}{0pt}
\boxshape{SWbox}{-5pt}{0pt}
\boxshape{EBox}{-3pt}{3pt}
\boxshape{WBox}{3pt}{-3pt}
\makeatother

\tikzstyle{cloud}=[shape=cloud,draw,minimum width=1.5cm,minimum height=1.5cm]

\tikzstyle{map}=[draw,shape=NEbox,inner sep=2pt,minimum height=6mm,fill=white]
\tikzstyle{dashedmap}=[draw,dashed,gray,shape=NEbox,inner sep=2pt,minimum height=6mm,fill=white]
\tikzstyle{medium dashedmap}=[draw,dashed,gray,shape=NEbox,inner sep=2pt,minimum height=6mm,fill=white,minimum width=7mm]
\tikzstyle{semilarge dashedmap}=[draw,dashed,gray,shape=NEbox,inner sep=2pt,minimum height=6mm,fill=white,minimum width=9.5mm]
\tikzstyle{large dashedmap}=[draw,dashed,gray,shape=NEbox,inner sep=2pt,minimum height=6mm,fill=white,minimum width=12mm]
\tikzstyle{very large dashedmap}=[draw,dashed,gray,shape=NEbox,inner sep=2pt,minimum height=6mm,fill=white,minimum width=17mm]

\tikzstyle{dashed map}=[fill=white, draw=gray, shape=rectangle, style=map, dashed]

\tikzstyle{mapdag}=[draw,shape=SEbox,inner sep=2pt,minimum height=6mm,fill=white]
\tikzstyle{mapadj}=[draw,shape=SEbox,inner sep=2pt,minimum height=6mm,fill=white]
\tikzstyle{maptrans}=[draw,shape=SWbox,inner sep=2pt,minimum height=6mm,fill=white]
\tikzstyle{mapconj}=[draw,shape=NWbox,inner sep=2pt,minimum height=6mm,fill=white]

\tikzstyle{medium map}=[draw,shape=NEbox,inner sep=2pt,minimum height=6mm,fill=white,minimum width=7mm]
\tikzstyle{medium map dag}=[draw,shape=SEbox,inner sep=2pt,minimum height=6mm,fill=white,minimum width=7mm]
\tikzstyle{medium map adj}=[draw,shape=SEbox,inner sep=2pt,minimum height=6mm,fill=white,minimum width=7mm]
\tikzstyle{medium map trans}=[draw,shape=SWbox,inner sep=2pt,minimum height=6mm,fill=white,minimum width=7mm]
\tikzstyle{medium map conj}=[draw,shape=NWbox,inner sep=2pt,minimum height=6mm,fill=white,minimum width=7mm]
\tikzstyle{semilarge map}=[draw,shape=NEbox,inner sep=2pt,minimum height=6mm,fill=white,minimum width=9.5mm]
\tikzstyle{semilarge map trans}=[draw,shape=SWbox,inner sep=2pt,minimum height=6mm,fill=white,minimum width=9.5mm]
\tikzstyle{semilarge map adj}=[draw,shape=SEbox,inner sep=2pt,minimum height=6mm,fill=white,minimum width=9.5mm]
\tikzstyle{semilarge map dag}=[draw,shape=SEbox,inner sep=2pt,minimum height=6mm,fill=white,minimum width=9.5mm]
\tikzstyle{semilarge map conj}=[draw,shape=NWbox,inner sep=2pt,minimum height=6mm,fill=white,minimum width=9.5mm]
\tikzstyle{large map}=[draw,shape=NEbox,inner sep=2pt,minimum height=6mm,fill=white,minimum width=12mm]
\tikzstyle{large map conj}=[draw,shape=NWbox,inner sep=2pt,minimum height=6mm,fill=white,minimum width=12mm]
\tikzstyle{very large map}=[draw,shape=NEbox,inner sep=2pt,minimum height=6mm,fill=white,minimum width=17mm]
\tikzstyle{very very large map}=[draw,shape=NEbox,inner sep=2pt,minimum height=6mm,fill=white,minimum width=50mm]
\tikzstyle{large map dag}=[draw,shape=SEbox,inner sep=2pt,minimum height=6mm,fill=white,minimum width=12mm]

\tikzstyle{medium dmap}=[draw,doubled,shape=NEbox,inner sep=2pt,minimum height=6mm,fill=white,minimum width=7mm]
\tikzstyle{medium dmap dag}=[draw,doubled,shape=SEbox,inner sep=2pt,minimum height=6mm,fill=white,minimum width=7mm]
\tikzstyle{medium dmap adj}=[draw,doubled,shape=SEbox,inner sep=2pt,minimum height=6mm,fill=white,minimum width=7mm]
\tikzstyle{medium dmap trans}=[draw,doubled,shape=SWbox,inner sep=2pt,minimum height=6mm,fill=white,minimum width=7mm]
\tikzstyle{medium dmap conj}=[draw,doubled,shape=NWbox,inner sep=2pt,minimum height=6mm,fill=white,minimum width=7mm]
\tikzstyle{semilarge dmap}=[draw,doubled,shape=NEbox,inner sep=2pt,minimum height=6mm,fill=white,minimum width=9.5mm]
\tikzstyle{semilarge dmap trans}=[draw,doubled,shape=SWbox,inner sep=2pt,minimum height=6mm,fill=white,minimum width=9.5mm]
\tikzstyle{semilarge dmap adj}=[draw,doubled,shape=SEbox,inner sep=2pt,minimum height=6mm,fill=white,minimum width=9.5mm]
\tikzstyle{semilarge dmap dag}=[draw,doubled,shape=SEbox,inner sep=2pt,minimum height=6mm,fill=white,minimum width=9.5mm]
\tikzstyle{semilarge dmap conj}=[draw,doubled,shape=NWbox,inner sep=2pt,minimum height=6mm,fill=white,minimum width=9.5mm]
\tikzstyle{large dmap}=[draw,doubled,shape=NEbox,inner sep=2pt,minimum height=6mm,fill=white,minimum width=12mm]
\tikzstyle{large dmap conj}=[draw,doubled,shape=NWbox,inner sep=2pt,minimum height=6mm,fill=white,minimum width=12mm]
\tikzstyle{large dmap trans}=[draw,doubled,shape=SWbox,inner sep=2pt,minimum height=6mm,fill=white,minimum width=12mm]
\tikzstyle{large dmap adj}=[draw,doubled,shape=SEbox,inner sep=2pt,minimum height=6mm,fill=white,minimum width=12mm]
\tikzstyle{large dmap dag}=[draw,doubled,shape=SEbox,inner sep=2pt,minimum height=6mm,fill=white,minimum width=12mm]
\tikzstyle{very large dmap}=[draw,doubled,shape=NEbox,inner sep=2pt,minimum height=6mm,fill=white,minimum width=19.5mm]

\tikzstyle{muxbox}=[draw,shape=rectangle,minimum height=3mm,minimum width=3mm,fill=white]
\tikzstyle{dmuxbox}=[muxbox,doubled]

\tikzstyle{box}=[draw,shape=rectangle,inner sep=2pt,minimum height=6mm,minimum width=6mm,fill=white]
\tikzstyle{dbox}=[draw,doubled,shape=rectangle,inner sep=2pt,minimum height=6mm,minimum width=6mm,fill=white]
\tikzstyle{dmap}=[draw,doubled,shape=NEbox,inner sep=2pt,minimum height=6mm,fill=white]
\tikzstyle{dmapdag}=[draw,doubled,shape=SEbox,inner sep=2pt,minimum height=6mm,fill=white]
\tikzstyle{dmapadj}=[draw,doubled,shape=SEbox,inner sep=2pt,minimum height=6mm,fill=white]
\tikzstyle{dmaptrans}=[draw,doubled,shape=SWbox,inner sep=2pt,minimum height=6mm,fill=white]
\tikzstyle{dmapconj}=[draw,doubled,shape=NWbox,inner sep=2pt,minimum height=6mm,fill=white]

\tikzstyle{ddmap}=[draw,doubled,dashed,shape=NEbox,inner sep=2pt,minimum height=6mm,fill=white]
\tikzstyle{ddmapdag}=[draw,doubled,dashed,shape=SEbox,inner sep=2pt,minimum height=6mm,fill=white]
\tikzstyle{ddmapadj}=[draw,doubled,dashed,shape=SEbox,inner sep=2pt,minimum height=6mm,fill=white]
\tikzstyle{ddmaptrans}=[draw,doubled,dashed,shape=SWbox,inner sep=2pt,minimum height=6mm,fill=white]
\tikzstyle{ddmapconj}=[draw,doubled,dashed,shape=NWbox,inner sep=2pt,minimum height=6mm,fill=white]

\boxshape{sNEbox}{0pt}{3pt}
\boxshape{sSEbox}{0pt}{-3pt}
\boxshape{sNWbox}{3pt}{0pt}
\boxshape{sSWbox}{-3pt}{0pt}
\tikzstyle{smap}=[draw,shape=sNEbox,fill=white]
\tikzstyle{smapdag}=[draw,shape=sSEbox,fill=white]
\tikzstyle{smapadj}=[draw,shape=sSEbox,fill=white]
\tikzstyle{smaptrans}=[draw,shape=sSWbox,fill=white]
\tikzstyle{smapconj}=[draw,shape=sNWbox,fill=white]

\tikzstyle{dsmap}=[draw,dashed,shape=sNEbox,fill=white]
\tikzstyle{dsmapdag}=[draw,dashed,shape=sSEbox,fill=white]
\tikzstyle{dsmaptrans}=[draw,dashed,shape=sSWbox,fill=white]
\tikzstyle{dsmapconj}=[draw,dashed,shape=sNWbox,fill=white]

\boxshape{mNEbox}{0pt}{10pt}
\boxshape{mSEbox}{0pt}{-10pt}
\boxshape{mNWbox}{10pt}{0pt}
\boxshape{mSWbox}{-10pt}{0pt}
\tikzstyle{mmap}=[draw,shape=mNEbox]
\tikzstyle{mmapdag}=[draw,shape=mSEbox]
\tikzstyle{mmaptrans}=[draw,shape=mSWbox]
\tikzstyle{mmapconj}=[draw,shape=mNWbox]

\tikzstyle{mmapgray}=[draw,fill=gray!40!white,shape=mNEbox]
\tikzstyle{smapgray}=[draw,fill=gray!40!white,shape=sNEbox]

\makeatletter

\pgfdeclareshape{cornerpoint}{
\inheritsavedanchors[from=rectangle] 
\inheritanchorborder[from=rectangle]
\inheritanchor[from=rectangle]{center}
\inheritanchor[from=rectangle]{north}
\inheritanchor[from=rectangle]{south}
\inheritanchor[from=rectangle]{west}
\inheritanchor[from=rectangle]{east}
\backgroundpath{
\southwest \pgf@xa=\pgf@x \pgf@ya=\pgf@y
\northeast \pgf@xb=\pgf@x \pgf@yb=\pgf@y

\pgfmathsetmacro{\pgf@shorten@left}{\pgfkeysvalueof{/tikz/shorten left}}
\pgfmathsetmacro{\pgf@shorten@right}{\pgfkeysvalueof{/tikz/shorten right}}

\pgfpathmoveto{\pgfpoint{0.5 * (\pgf@xa + \pgf@xb)}{\pgf@ya - 5pt}}
\pgfpathlineto{\pgfpoint{\pgf@xa - 8pt + \pgf@shorten@left}{\pgf@yb - 1.5 * \pgf@shorten@left}}
\pgfpathlineto{\pgfpoint{\pgf@xa - 8pt + \pgf@shorten@left}{\pgf@yb}}
\pgfpathlineto{\pgfpoint{\pgf@xb + 8pt - \pgf@shorten@right}{\pgf@yb}}
\pgfpathlineto{\pgfpoint{\pgf@xb + 8pt - \pgf@shorten@right}{\pgf@yb - 1.5 * \pgf@shorten@right}}
\pgfpathclose
}
}

\pgfdeclareshape{cornercopoint}{
\inheritsavedanchors[from=rectangle] 
\inheritanchorborder[from=rectangle]
\inheritanchor[from=rectangle]{center}
\inheritanchor[from=rectangle]{north}
\inheritanchor[from=rectangle]{south}
\inheritanchor[from=rectangle]{west}
\inheritanchor[from=rectangle]{east}
\backgroundpath{
\southwest \pgf@xa=\pgf@x \pgf@ya=\pgf@y
\northeast \pgf@xb=\pgf@x \pgf@yb=\pgf@y

\pgfmathsetmacro{\pgf@shorten@left}{\pgfkeysvalueof{/tikz/shorten left}}
\pgfmathsetmacro{\pgf@shorten@right}{\pgfkeysvalueof{/tikz/shorten right}}

\pgfpathmoveto{\pgfpoint{0.5 * (\pgf@xa + \pgf@xb)}{\pgf@yb + 5pt}}
\pgfpathlineto{\pgfpoint{\pgf@xa - 8pt + \pgf@shorten@left}{\pgf@ya + 1.5 * \pgf@shorten@left}}
\pgfpathlineto{\pgfpoint{\pgf@xa - 8pt + \pgf@shorten@left}{\pgf@ya}}
\pgfpathlineto{\pgfpoint{\pgf@xb + 8pt - \pgf@shorten@right}{\pgf@ya}}
\pgfpathlineto{\pgfpoint{\pgf@xb + 8pt - \pgf@shorten@right}{\pgf@ya + 1.5 * \pgf@shorten@right}}
\pgfpathclose
}
}

\makeatother

\pgfkeyssetvalue{/tikz/shorten left}{0pt}
\pgfkeyssetvalue{/tikz/shorten right}{0pt}

\tikzstyle{kpoint common}=[draw,fill=white,inner sep=1pt,minimum height=4mm]
\tikzstyle{kpoint sc}=[shape=cornerpoint,kpoint common]
\tikzstyle{kpoint adjoint sc}=[shape=cornercopoint,kpoint common]
\tikzstyle{kpoint}=[shape=cornerpoint,shorten left=5pt,kpoint common]
\tikzstyle{kpoint adjoint}=[shape=cornercopoint,shorten left=5pt,kpoint common]
\tikzstyle{kpoint conjugate}=[shape=cornerpoint,shorten right=5pt,kpoint common]
\tikzstyle{kpoint transpose}=[shape=cornercopoint,shorten right=5pt,kpoint common]
\tikzstyle{kpoint symm}=[shape=cornerpoint,shorten left=5pt,shorten right=5pt,kpoint common]

\tikzstyle{black kpoint}=[shape=cornerpoint,shorten left=5pt,kpoint common,fill=black,font=\color{white}]
\tikzstyle{black kpoint adjoint}=[shape=cornercopoint,shorten left=5pt,kpoint common,fill=black,font=\color{white}]
\tikzstyle{black kpointadj}=[shape=cornercopoint,shorten left=5pt,kpoint common,fill=black,font=\color{white}]

\tikzstyle{black dkpoint}=[shape=cornerpoint,shorten left=5pt,kpoint common,fill=black, doubled,font=\color{white}]
\tikzstyle{black dkpoint adjoint}=[shape=cornercopoint,shorten left=5pt,kpoint common,fill=black, doubled,font=\color{white}]
\tikzstyle{black dkpointadj}=[shape=cornercopoint,shorten left=5pt,kpoint common,fill=black, doubled,font=\color{white}] 

\tikzstyle{kpointdag}=[kpoint adjoint]
\tikzstyle{kpointadj}=[kpoint adjoint]
\tikzstyle{kpointconj}=[kpoint conjugate]
\tikzstyle{kpointtrans}=[kpoint transpose]

\tikzstyle{big kpoint}=[kpoint, minimum width=1.2 cm, minimum height=8mm, inner sep=4pt, text depth=3mm]

\tikzstyle{wide kpoint}=[kpoint, minimum width=1 cm, inner sep=2pt]
\tikzstyle{wide kpointdag}=[kpointdag, minimum width=1 cm, inner sep=2pt]
\tikzstyle{wide kpointconj}=[kpointconj, minimum width=1 cm, inner sep=2pt]
\tikzstyle{wide kpointtrans}=[kpointtrans, minimum width=1 cm, inner sep=2pt]

\tikzstyle{gray kpoint}=[kpoint,fill=gray!50!white]
\tikzstyle{gray kpointdag}=[kpointdag,fill=gray!50!white]
\tikzstyle{gray kpointadj}=[kpointadj,fill=gray!50!white]
\tikzstyle{gray kpointconj}=[kpointconj,fill=gray!50!white]
\tikzstyle{gray kpointtrans}=[kpointtrans,fill=gray!50!white]

\tikzstyle{gray dkpoint}=[kpoint,fill=gray!50!white,doubled]
\tikzstyle{gray dkpointdag}=[kpointdag,fill=gray!50!white,doubled]
\tikzstyle{gray dkpointadj}=[kpointadj,fill=gray!50!white,doubled]
\tikzstyle{gray dkpointconj}=[kpointconj,fill=gray!50!white,doubled]
\tikzstyle{gray dkpointtrans}=[kpointtrans,fill=gray!50!white,doubled]

\tikzstyle{white label}=[draw,fill=white,rectangle,inner sep=0.7 mm]
\tikzstyle{gray label}=[draw,fill=gray!50!white,rectangle,inner sep=0.7 mm]
\tikzstyle{black label}=[draw,fill=black,rectangle,inner sep=0.7 mm]

\tikzstyle{dkpoint}=[kpoint,doubled]
\tikzstyle{wide dkpoint}=[wide kpoint,doubled]
\tikzstyle{dkpointdag}=[kpoint adjoint,doubled]
\tikzstyle{wide dkpointdag}=[wide kpointdag,doubled]
\tikzstyle{dkcopoint}=[kpoint adjoint,doubled]
\tikzstyle{dkpointadj}=[kpoint adjoint,doubled]
\tikzstyle{dkpointconj}=[kpoint conjugate,doubled]
\tikzstyle{dkpointtrans}=[kpoint transpose,doubled]

\tikzstyle{kscalar}=[kpoint common, shape=EBox, inner xsep=-1pt, inner ysep=3pt,font=\small]
\tikzstyle{kscalarconj}=[kpoint common, shape=WBox, inner xsep=-1pt, inner ysep=3pt,font=\small]

\tikzstyle{spekpoint}=[kpoint sc,minimum height=5mm,inner sep=3pt]
\tikzstyle{spekcopoint}=[kpoint adjoint sc,minimum height=5mm,inner sep=3pt]

\tikzstyle{dspekpoint}=[spekpoint,doubled]
\tikzstyle{dspekcopoint}=[spekcopoint,doubled]

\usetikzlibrary{circuits.ee.IEC}
 \tikzstyle{discard}=[circuit ee IEC, ground,rotate=90,scale=3,inner sep=-1mm]
 \tikzstyle{downground}=[circuit ee IEC,thick,ground,rotate=-90,scale=1.5,inner sep=-2mm]

\tikzstyle{maxmix}=[regular polygon,regular polygon sides=3,draw=black,xscale=0.4,yscale=0.3,inner sep=-0.5pt,minimum width=10mm,fill=gray,regular polygon rotate=180]

 \tikzstyle{bigground}=[regular polygon,regular polygon sides=3,draw=gray,scale=0.50,inner sep=-0.5pt,minimum width=10mm,fill=gray]

\tikzstyle{arrs}=[-latex,font=\small,auto]
\tikzstyle{arrow plain}=[arrs]
\tikzstyle{arrow dashed}=[dashed,arrs]
\tikzstyle{arrow bold}=[very thick,arrs]
\tikzstyle{arrow hide}=[draw=white!0,-]
\tikzstyle{arrow reverse}=[latex-]
\tikzstyle{cdnode}=[]

\tikzstyle{green dashed arrow}=[green, arrow dashed]
\tikzstyle{dashed blue}=[blue, dashed]
\tikzstyle{red dashed arrow}=[red, arrow dashed]
\tikzstyle{orange arrow}=[orange, arrs]
\tikzstyle{blue arrow}=[blue, arrs]
\tikzstyle{magenta arrow}=[magenta, arrs]

\tikzstyle{dotted line}=[-, style=dotted, tikzit draw=brown]
\tikzstyle{dashed line}=[-, style=dashed, tikzit draw=cyan]
\tikzstyle{green fill line}=[-, fill={green!90!black}, tikzit draw=green]
\tikzstyle{blue fill}=[-, fill=blue, tikzit fill=blue, tikzit draw={rgb,255: red,102; green,117; blue,255}]
\tikzstyle{red}=[-, draw=red, tikzit draw=red]
\tikzstyle{blue}=[-, draw=blue, tikzit draw=blue]
\tikzstyle{thick black}=[-, draw=black, tikzit draw=black, line width=1pt]
\tikzstyle{dotted red}=[-, draw=red, style=dotted, tikzit draw=red]
\tikzstyle{dotted blue}=[-, draw=blue, tikzit draw=blue, style=dotted]
\tikzstyle{dashed thick blue}=[-, draw={rgb,255: red,28; green,176; blue,255}, tikzit draw={rgb,255: red,83; green,19; blue,156}, line width=1pt, style=dashed]
\tikzstyle{dashed thick red}=[-, draw=red, tikzit draw={rgb,255: red,255; green,100; blue,10}, line width=1pt, style=dashed]
\tikzstyle{green}=[-, draw=green, tikzit draw=green]
\tikzstyle{dotted green}=[-, draw=green, tikzit draw=green, style=dotted]
\tikzstyle{arrow}=[->]
\tikzstyle{arrow green dashed}=[draw=green, ->, tikzit draw=green, style=dashed]
\tikzstyle{arrow dashed red}=[draw=red, ->, style=dashed, tikzit draw=red]
\tikzstyle{dashed green}=[-, tikzit draw=green, draw=green, style=dashed]

\usepackage{lipsum}
\newtheorem{theorem}{Theorem}
\newtheorem{definition}{Definition}

\newcommand{\s}{\texttt{SWAP}}

\newcommand{\cc}{\mathcal C}

\newcommand{\ce}{\mathcal E}
\newcommand{\cf}{\mathcal F}

\newcommand{\ch}{\mathcal H}

\newcommand{\cm}{\mathcal M}

\newcommand{\cu}{\mathcal U}

\begin{document}

\title{Which theories have a measurement problem?}

\author{Nick Ormrod} \orcid{0000-0003-2717-8709} \affiliation{Quantum Group, Department of Computer Science, University of Oxford} \email{nicholas.ormrod@cs.ox.ac.uk}

\author{V. Vilasini} \orcid{0000-0002-7035-4205} \affiliation{Institute for Theoretical Physics, ETH Zurich} \email{vilasini@phys.ethz.ch}

\author{Jonathan Barrett} \orcid{0000-0002-2222-0579} \email{jonathan.barrett@cs.ox.ac.uk} \affiliation{Quantum Group, Department of Computer Science, University of Oxford}

\maketitle

\begin{abstract}

It is shown that any theory that has certain properties has a measurement problem, in the sense that it makes predictions that are incompatible with measurement outcomes being \textit{absolute} (that is, unique and non-relational). 
These properties are \textit{Bell Nonlocality}, \textit{Information Preservation}, and \textit{Local Dynamics}. The result is extended by deriving Local Dynamics from \textit{No Superluminal Influences}, \textit{Separable Dynamics}, and \textit{Consistent Embeddings}. As well as explaining why the existing Wigner's-friend-inspired no-go theorems hold for quantum theory, these results also shed light on whether a future theory of physics might overcome the measurement problem. In particular, they suggest the possibility of a theory in which absoluteness is maintained, but without rejecting relativity theory (as in Bohm theory) or embracing objective collapses (as in GRW theory).
\end{abstract}

 \section{Introduction}

Suppose Alice performs a measurement, and observes the outcome that a light flashes red.
Is it an absolute fact that this is what she saw? Or is there some other world, context, or perspective, in which she saw it flash some other colour?

Remarkably, recent no-go results suggest that sometimes there is, at least given some well-motivated assumptions \cite{healey2018quantum, bong2020strong, haddara2022possibilistic,  leegwater2022greenberger, ormrod2022no}.\footnote{These results were inspired by a closely related no-go theorem \cite{Frauchiger2018quantum} concerning the consistency between various agents' beliefs about the outcomes they have observed.}
One assumption is that measurements can, at least in principle, be treated as unitary interactions.
In \cite{healey2018quantum, leegwater2022greenberger, ormrod2022no}, the other is that one can nevertheless apply the Born rule to obtain the statistics for a pair of simultaneous measurements in any inertial frame of reference.
Thus the two pillars of modern physics -- quantum theory and relativity -- conspire to impose some kind of relativity of observed events. 

Yet, this does not prove that observed events in \textit{nature} fail to be absolute. The claim that measurements using macroscopic apparatus are unitary interactions lies well beyond the current scope of empirical confirmation. Therefore, a legitimate response to the existing no-go results is to maintain that observed events are absolute by rejecting the version of quantum theory that suggests otherwise. One can then hope that a future, post-quantum theory of physics will avoid a similar no-go theorem, and thus overcome the infamous measurement problem of quantum theory.

Here, we investigate whether this is wishful thinking. We identify very general physical principles that inevitably lead a theory to make predictions that are inconsistent with the absoluteness of observed events. Once these principles are identified, one can judge the prospects for a future physical theory to avoid a measurement problem by dropping one of them.

To this end, we first recap the particularly simple quantum no-go theorem from \cite{ormrod2022no}, and discuss which features of quantum theory appear to make the argument go through (Section \ref{sec:simple}). We then construct a framework of more general theories that may contain an analogue of quantum theory's `Heisenberg cut', deploying the notion of a memory update from \cite{vilasini2019multi} (Section \ref{sec:framework}). We identify three properties that may be satisfied by a theory in the framework, whose conjunction leads to a breakdown in the absoluteness of observed events (Section \ref{sec:BIL}). These three properties are:
\begin{enumerate}
    \item Bell Nonlocality;
    \item Information Preservation;
    \item Local Dynamics.
\end{enumerate}

Local Dynamics is stronger than the prohibition of superluminal influences required by relativity. However, it may be derived from three further principles (Section \ref{sec:binsc}), namely
\begin{enumerate}
    \item No Superluminal Influences;
    \item Separable Dynamics;
    \item Consistent Embeddings.
\end{enumerate}
Substituting these three principles for Local Dynamics, we arrive at a deeper no-go theorem, which derives a contradiction with absolute observed events from five fundamental principles.
Our results are summarized graphically in Figure \ref{fig:implications}.

In deriving these results, we identify the precise features of quantum theory that are responsible for the existing no-go theorems, showing, for example, that unitarity per se is not essential,\footnote{This corroborates a result from \cite{vilasini2019multi}, in which a similar no-go theorem is derived in the context of PR boxes \cite{popescu1994quantum}.} and that quantum theory's peculiar combination of dynamical locality and Bell nonlocal correlations is crucial. What is particularly illuminating in this regard is that the proof of our main result is constructive, meaning that readers can see for themselves exactly how the incompatibility with the absoluteness of observed events comes about as a result of our assumptions. 

If one believes that observed events are absolute, then our first result shows that one cannot believe in a theory that fits in our framework, violates Bell inequalities, preserves information, and has local dynamics. Since the framework is rather minimal, and Bell nonlocality has already been observed, we conclude that believers in absoluteness face a dilemma between accepting the irretrievable loss of information and embracing some sort of dynamical nonlocality (Section \ref{sec:conc}). 

We note, however, that accepting dynamical nonlocality does not necessarily mean rejecting relativity theory, since one can reject Separable Dynamics instead of No Superluminal Influences. This suggests a strategy for overcoming the measurement problem: develop a theory in which the idea of nonseparability, familiar from quantum states, is extended to the dynamics, in a way that allows one to retain the absoluteness of observed events. One might then be able to avoid a measurement problem while conserving much of what is fundamental to the modern physical picture, including relativity theory and the preservation of information (Section \ref{sec:conc}).


However, with this approach one does still have to give up the idea that unitary dynamics are fundamental. For this reason, many will prefer to simply reject the absoluteness of observed events.
This raises both an ontological question, of how `relative events' might be understood and modelled mathematically, and a more practical, epistemological one, of how agents might communicate, without error, beliefs about such events.\footnote{The epistemological question was taken up in \cite{vilasini2022subjective} in the context of quantum theory; the results here might inspire a generalization of that approach to other perspectival theories.} At the end, we reflect briefly on the clues for these problems provided by our results (Section \ref{sec:relativity}).

\tableofcontents

\section{A simple quantum no-go theorem} \label{sec:simple}

\cite{ormrod2022no} derived a particularly simple version of the quantum no-go result for the absoluteness of observed events, based on the same principles as some earlier results \cite{healey2018quantum, leegwater2022greenberger}. Here, we will summarize the argument, which will be generalized in Section \ref{sec:BIL} to prove a no-go result for a broad class of theories. 

First, however, we briefly recap the original Wigner's original thought experiment \cite{wigner1995remarks}, on which the argument is an elaboration. Readers who are familiar with the experiment may wish to skip the following subsection.

\subsection{The original thought experiment}

Imagine yourself standing outside a laboratory with thick, concrete walls. Inside the lab is your friend; her only company is a qubit prepared in the state $\frac{\ket{0} + \ket{1}}{\sqrt{2}}$. She measures the qubit in the $\{\ket{0}, \ket{1}\}$ basis. She will presumably experience herself observing one of two possible outcomes. In which case, we can model the outcome -- the \textit{observed event} -- as a bit taking on the value $F=0$ or $F=1$.

But your friend is nothing special: at the end of the day, she is just a quantum system like everything else. So presumably, you should \textit{also} be able to treat her as having a quantum state. And, if all time evolution is unitarity, then presumably you should describe her interaction with the particle roughly as follows:
\begin{equation}
     \frac{\ket{0}_P + \ket{1}_P}{\sqrt{2}} \ket{{\rm ready}}_F \mapsto \frac{\ket{0}_P\ket{0}_F + \ket{1}_P\ket{1}_F}{\sqrt{2}}
\end{equation}
Intuitively, the state on the right side says that if the particle is in its $\ket{i}_P$ state, then your friend is in a state of having observed $F=i$.
She therefore ends up entangled with the particle. And this statement is by no means purely theoretical: \textit{you} can actually confirm or falsify the entanglement  (given enough trials) by performing a measurement of a basis for the lab that includes $\frac{\ket{0}_P\ket{0}_F + \ket{1}_P\ket{1}_F}{\sqrt{2}}$.

But this entanglement would appear to be in contradiction with the earlier claim that your friend observes some particular outcome, \textit{either} $F=0$ \textit{or} $F=1$. For in that case, intuition suggests that your friend and the particle should have ended up in some product state, \textit{either} $\ket{0}\ket{0}$ \textit{or} $\ket{1}\ket{1}$.



If the quantum state is a complete description of an agent-independent reality, then we have a choice. Either we accept that there is not a single, particular outcome that the friend experiences. Or, we deny that the fundamental dynamical law of quantum theory -- that time evolution is unitary -- holds inside the laboratory. 

However, this dilemma can potentially be avoided if the quantum state is epistemic. In that case, a deeper theory may reveal that the information about the friend and system summarized by an entangled quantum state is quite compatible with one or the other outcome having happened. And the dilemma can certainly be avoided if the quantum state is supplemented with additional variables, as shown by the Bohm-theoretic account of the experiment. There, all time evolution is unitary, but only one outcome is observed, depending on which branch of the wavefunction is picked out by the positions of the Bohmian particles. It follows that the argument above cannot have amounted to any no-go theorem ruling out the combination of unitarity and unique outcomes; at best it illustrates a tension between these ideas.

However, it turns out that one \textit{can} derive a no-go theorem if one (a) extends the scenario above to incorporate two spacelike separated agents, each with their own friend inside of an isolated lab, and (b) adds a locality assumption \cite{healey2018quantum, bong2020strong, haddara2022possibilistic,  leegwater2022greenberger, ormrod2022no}. This is shown by the next subsection.

\subsection{A relatively simple no-go theorem} \label{sec:simple_simple}
 
\begin{figure} 
    \centering
    \begin{tikzpicture} [scale=1.6]
	\begin{pgfonlayer}{nodelayer}
		\node [style=none] (0) at (-0.25, -2.25) {};
		\node [style=none] (1) at (0, -2.75) {$\ket{H}_{P_1P_2}$};
		\node [style=none] (2) at (-2, 0) {};
		\node [style=none] (3) at (-2, 0) {Charlie};
		\node [style=none] (4) at (2, 0) {};
		\node [style=none] (5) at (2, 0) {Daniela};
		\node [style=none] (6) at (-2, 2) {};
		\node [style=none] (7) at (-2, 2) {Alice};
		\node [style=none] (8) at (2, 2) {Bob};
		\node [style=none] (9) at (-1.75, -0.5) {};
		\node [style=none] (10) at (0.25, -2.25) {};
		\node [style=none] (11) at (1.75, -0.25) {};
		\node [style=none] (12) at (2, 0.5) {};
		\node [style=none] (13) at (2, 1.5) {};
		\node [style=none] (14) at (-2, 0.5) {};
		\node [style=none] (15) at (-2, 1.5) {};
		\node [style=none] (16) at (-2, -1) {};
		\node [style=none] (17) at (2, 3) {};
		\node [style=none] (18) at (2, -1) {};
		\node [style=none] (19) at (-2, 3) {};
	\end{pgfonlayer}
	\begin{pgfonlayer}{edgelayer}
		\draw [style=arrow plain] (0.center) to (9.center);
		\draw [style=arrow plain] (10.center) to (11.center);
		\draw [style=arrow plain] (12.center) to (13.center);
		\draw [style=arrow plain] (14.center) to (15.center);
		\draw [style=dashed] (16.center) to (17.center);
		\draw [style=dashed] (19.center) to (18.center);
	\end{pgfonlayer}
\end{tikzpicture}
    \caption{The scenario that proves the quantum no-go theorem in \cite{ormrod2022no}. Dashed lines denote the possible paths of light rays.}
\label{fig:experiment_schematic}
\end{figure}
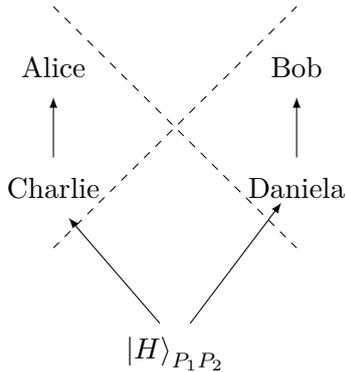

Now, let us sketch the result from \cite{ormrod2022no}.
We consider the scenario depicted schematically in Figure \ref{fig:experiment_schematic}. At the start, a `Hardy state' \cite{hardy1992quantum, hardynonlocality1993} 
\begin{equation} \label{hardy}
    \ket{H}_{P_1P_2} = \frac{1}{\sqrt{3}}(\ket{00}_{P_1P_2}+\ket{01}_{P_1P_2}+\ket{10}_{P_1P_2})
\end{equation}
of two particles is prepared. One particle is sent to Charlie, the other is sent to another, spacelike separated agent, called Daniela. Once they receive the particle, their labs are isolated from the rest of the universe. 

These agents then perform a measurement in the $\{\ket{i}\}_i$ basis. As is evident from (\ref{hardy}) that the Born probability for Charlie and Daniela both seeing the `1' outcome on the same run of the experiment is 0.
\begin{equation} \label{eq:cd}
    p(C=`1\textnormal{'}, D=`1\textnormal{'}) = 0
\end{equation}

Now, let us assume that a `superobserver' outside the labs can treat these measurements as unitary interactions. A reasonable choice of unitary in this case would be the quantum CNOT gate,\footnote{This gate either does nothing to or flips the logical value of a target qubit based on the logical value of a control qubit. Formally, we have $\texttt{CNOT} \ket{i}\ket{j}=\ket{i}\ket{j+i}$, where `+' denotes addition modulo 2.}
where the target system is a qubit representing the agent's memory (see Section \ref{sec:qpt} for a discussion of the arbitrariness of this choice).\footnote{A more sophisticated analysis of the measurement process would involve explicitly accounting for the role of many different systems involved in the measurement, including environmental systems that induce decoherence. While this additional rigor doesn't change anything important in our analysis, it does make things a lot more complicated, hence our decision to ignore it.} Once the memory of the agent is prepared in the `ready' state $\ket{0}$, these CNOTs give rise to an isometry $U$ that coherently copies the logical value of the particle onto the agent's memory. For example, for Charlie, we have 
\begin{equation} \label{coherent_copy}
    U\ket{i}_{P_1}=\ket{i}_{C} \ket{i}_{P_1}.
\end{equation} (In an abuse of notation, we are here re-using the letter $C$ to label Charlie as a quantum system, rather than to label the variable describing his outcome.) 

We consider two superobservers, Alice and Bob, who each perform a measurement on a composite system consisting of one agent's memory and the corresponding particle after the initial measurement has taken place. Given the unitary/isometric description of the initial measurements just proposed, the four-partite quantum state they will measure is given by $(U \otimes U) \ket{H}$. Alice measures the orthonormal basis $U\frac{\ket{0} \pm \ket{1}}{\sqrt{2}}$ on $\ch_C \otimes \ch_{P_1}$, and Bob performs a similar measurement on $\ch_D \otimes \ch_{P_2}$. Since $U^\dag U = I$ for an isometry, it is easy to see that the Born probabilities for this measurement are equivalent to those obtained by measuring the basis $\frac{\ket{0} \pm \ket{1}}{\sqrt{2}}$ directly on $\ket{H}$. It follows that the probability of both of the agents seeing the outcome for $\frac{\ket{0} - \ket{1}}{\sqrt{2}}$ is
\begin{equation} \label{eq:ab}
    p(A=`-\textnormal{'}, B=`-\textnormal{'}) = 1/12 
\end{equation}

So far, we have calculated joint probabilities for two pairs of measurements.
In each case, we assumed that the quantum state evolved unitarily right up to the moment of the pair of measurements in question.
Let us assume that we can do this for any pair of measurements conducted at the same time.
As Figure \ref{fig:experiment_schematic} indicates, we are also assuming that Charlie's measurement is spacelike separated from Bob's measurement, and likewise that Daniela's measurement is spacelike separated from Alice's measurement. 

Now here is the key point: if relativity theory provides the correct way of thinking about simultaneity, then \textit{it is just as legitimate to think of Alice and Daniela, or of Bob and Charlie, as measuring at the same time}. 

In that case, to calculate joint probabilities for Alice's and Daniela's outcome, we should assume Charlie's measurement can be treated unitarily, so Alice and Daniela measure the state $(U \otimes I)\ket{H}$. It follows that the probability for the agents seeing the outcomes for $\frac{\ket{0} - \ket{1}}{\sqrt{2}}$  and $\ket{0}$ respectively is
\begin{equation} \label{eq:ad}
    p(A=`-\textnormal{'}, D=`0\textnormal{'})=0
\end{equation}
and, by similar reasoning, the probability for Bob and Charlie to see these outcomes is
\begin{equation} \label{eq:bc}
    p(B=`-\textnormal{'}, C=`0\textnormal{'})=0.
\end{equation}

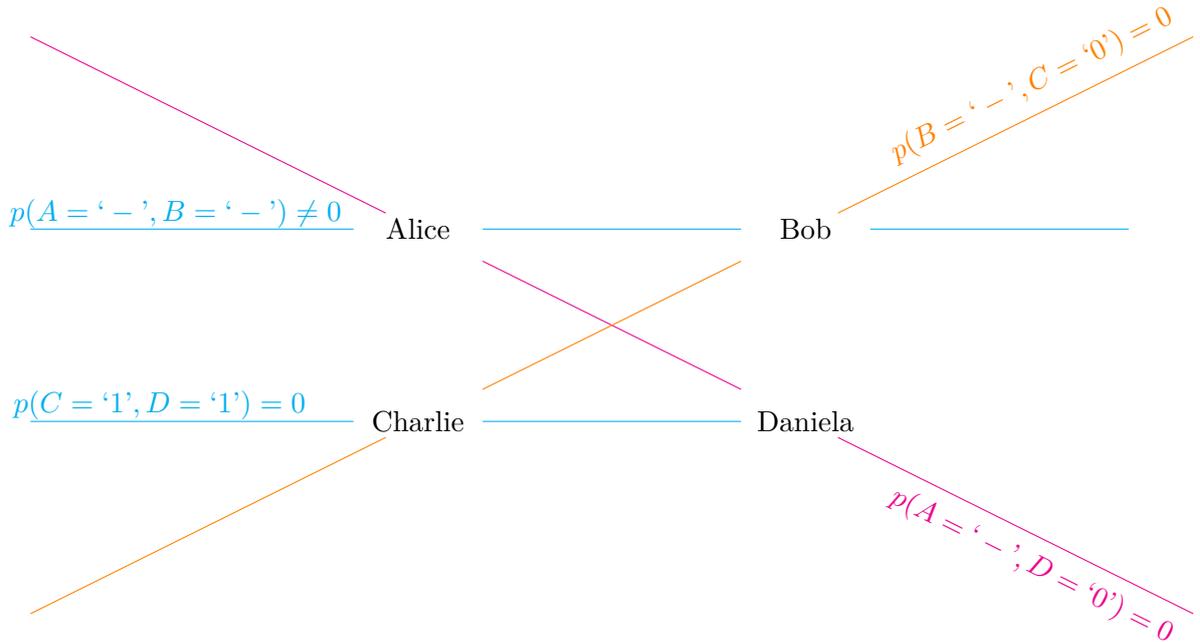
\begin{figure}
    \centering
    \begin{tikzpicture}[scale=1.7]
	\begin{pgfonlayer}{nodelayer}
		\node [style=none] (2) at (-3, -1) {};
		\node [style=none] (3) at (-3, -1) {Charlie};
		\node [style=none] (4) at (3, -1) {};
		\node [style=none] (5) at (3, -1) {Daniela};
		\node [style=none] (6) at (-3, 2) {};
		\node [style=none] (7) at (-3, 2) {Alice};
		\node [style=none] (8) at (3, 2) {Bob};
		\node [style=none] (11) at (-2, -1) {};
		\node [style=none] (18) at (2, -1) {};
		\node [style=none] (19) at (-2, 2) {};
		\node [style=none] (20) at (2, 2) {};
		\node [style=none] (21) at (-2, 1.5) {};
		\node [style=none] (22) at (2, -0.5) {};
		\node [style=none] (23) at (-3.5, 2.25) {};
		\node [style=none] (24) at (3.75, -1.25) {};
		\node [style=none] (25) at (-9, 5) {};
		\node [style=none] (26) at (9, -4) {};
		\node [style=none] (27) at (3.5, -1.25) {};
		\node [style=none] (28) at (-4, 2) {};
		\node [style=none] (29) at (-9, 2) {};
		\node [style=none] (30) at (4, 2) {};
		\node [style=none] (31) at (8, 2) {};
		\node [style=none] (32) at (-4, -1) {};
		\node [style=none] (33) at (-9, -1) {};
		\node [style=none] (34) at (4, -1) {};
		\node [style=none] (35) at (8, -1) {};
		\node [style=none] (48) at (2, 1.5) {};
		\node [style=none] (49) at (-2, -0.5) {};
		\node [style=none] (50) at (3.5, 2.25) {};
		\node [style=none] (52) at (9, 5) {};
		\node [style=none] (53) at (-9, -4) {};
		\node [style=none] (54) at (-3.5, -1.25) {};
		\node [style=none] (55) at (-6.75, 2.25) {\color{cyan} $p(A=`-\textnormal{'}, B=`-\textnormal{'}) \neq 0$};
		\node [style=none] (56) at (-7, -0.75) {$\color{cyan}  p(C=`1\textnormal{'}, D=`1\textnormal{'}) = 0$};
		\node [style=none, rotate=-26.77] (57) at (6.5, -3.25) {\color{magenta} $p(A=`-\textnormal{'}, D=`0\textnormal{'})=0$};
		\node [style=none, rotate=26.77] (58) at (6.5, 4.25) {\color{orange}$ p(B=`-\textnormal{'}, C=`0\textnormal{'})=0$};
		\node [style=none] (62) at (2, -0.5) {};
		\node [style=none] (63) at (-2, 1.5) {};
	\end{pgfonlayer}
	\begin{pgfonlayer}{edgelayer}
		\draw [style=cyan] (11.center) to (18.center);
		\draw [style=cyan] (19.center) to (20.center);
		\draw [style=magenta] (21.center) to (22.center);
		\draw [style=magenta] (25.center) to (23.center);
		\draw [style=magenta] (27.center) to (26.center);
		\draw [style=cyan] (28.center) to (29.center);
		\draw [style=cyan] (32.center) to (33.center);
		\draw [style=cyan] (30.center) to (31.center);
		\draw [style=orange] (48.center) to (49.center);
		\draw [style=orange] (52.center) to (50.center);
		\draw [style=orange] (54.center) to (53.center);
	\end{pgfonlayer}
\end{tikzpicture}
    \caption{A visual summary of the argument. The different colors represent different reference frames. Combining the quantum's predictions for measurements that are simultaneous in different frames implies that there cannot be a single global assignment of the observed outcomes on each run of the experiment.}
    \label{fig:ref_frames}
\end{figure}

But if we think of the observed outcomes $A$, $B$, $C$, and $D$ as \textit{absolute} -- that is, we assume that on any run of the experiment, there is some unique global assignment $(A=a, B=b, C=c, D=d)$ for which the quantum probabilities are predictions -- then we get a contradiction. Figure \ref{fig:ref_frames} illustrates the argument, which is summarized as follows. Suppose that on some run, Alice saw the `-' outcome. By (\ref{eq:ad}), Daniela saw `1' on the same outcome. This implies via (\ref{eq:cd}) that Charlie saw `0', which implies via (\ref{eq:bc}) that Bob saw `+'. So, whenever Alice sees `-', Bob sees `+'.  Yet (\ref{eq:ab}) implies that Alice and Bob sometimes both see `-'!\footnote{A similar no-go theorem \cite{healey2018quantum} has been criticised on the grounds that the four predictions it makes that together imply a contradiction with absoluteness can never be simultaneously observed by a single agent \cite{baumann2019comment}.
As argued in \cite{healey2019reply}, this misses the point. 
The result in \cite{healey2018quantum}, like the ones from \cite{leegwater2022greenberger, ormrod2022no}, aims to demonstrate that the absoluteness of observed events is \textit{logically inconsistent} with a particular combination of quantum theory and relativity.
The fact that the relevant predictions of that theory cannot be confirmed by a single agent poses no obstacle to this.} \footnote{In this exposition, we have prioritized conciseness above rigor and clarity. Readers who want a more detailed understanding of the no-go result are referred to \cite{ormrod2022no}, or to \cite{leegwater2022greenberger, healey2018quantum} for similar results.}

\subsection{What makes the argument go through?}

This argument relies on the following key ideas.
\begin{enumerate}
    \item \textit{A duality of perspectives on measurements.} Charlie's measurement was thought of from the `inside' as projecting onto $\{\ket{i}\}_i$, and from the `outside' as a unitary interaction.
    
    \item \textit{Nonlocal correlations.} The way that the four quantum predictions are chained together to obtain a contradiction in the argument above is precisely analogous to Hardy's proof of nonlocality \cite{hardynonlocality1993}. The argument before that just consists of tricks designed to bring that nonlocality into contradiction with the absoluteness of observed events.
    
    \item \textit{Information preservation.} Since information is preserved by isometries, Alice was able to effectively perform a measurement on $P_1$ that was incompatible with Charlie's measurement on the same particle.
    
    \item \textit{Dynamical locality.} We assumed that any pair of spacelike separated measurements could be thought of as simultaneous, allowing us to apply the Born rule to calculate joint probabilities.
\end{enumerate}

A slight modification of the result in \cite{vilasini2019multi} demonstrates that essentially the same properties in a model inspired by PR boxes \cite{popescu1994quantum} leads to a breakdown in the absoluteness of observed events. A key question of this paper is: do these properties lead to similar no-go theorems in \textit{any} physical theory? The answer is `yes', as stated in Theorem \ref{thm:BIL_AOE}. But before that theorem can be understood, we need a suitable framework of physical theories, to which we now turn.

\section{Perspectival theories} \label{sec:framework}
 
The no-go theorem above relied on a strange feature of quantum theory: that the same measurement can be thought of as either giving rise to classical outcomes, or as a unitary interaction (which does not formally involve any classical variables). The goal of this section is to formalize the idea that measurements can be viewed from two different perspectives, and to generalize it beyond quantum theory.

We will begin by introducing the highly general framework of \textit{categorical probabilistic theories} \cite{gogioso2017categorical}. 
But if the reader is not used to, or is uninterested in, category theory, there is no need to worry!
Categorical probabilistic theories are a simple idea, which we will introduce here without any heavy mathematical machinery (though a more formal introduction is found in Appendix \ref{app:CPTsformal}).

We will then supplement categorical probabilistic theories with a little bit of extra structure, allowing us uniquely associate every model of a measurement incorporating classical outcomes with another model of the same measurement that might not explicitly incorporate any classical outcomes. This gives us the framework of \textit{perspectival theories}. In some perspectival theories, such as certain classical theories or `collapse' theories, the connection between the two perspectives is very natural; they might be essentially identical. But in others, such as a unitary quantum one, the connection might feel very unnatural indeed.

\subsection{Categorical probabilistic theories}

In a nutshell, a categorical probabilistic theory is a theory that can be described by circuits, that contains a notion of normalization and the reduced states of subsystems, and that includes classical systems, so that it can directly model measurement settings and outcomes. For readers who are familiar with circuit theories, this sentence alone should permit a reasonable understanding of the next subsection; such readers should feel free to skip ahead. For the uninitiated, we introduce the main ideas here, leaving a more formal treatment for Appendix \ref{app:CPTsformal}.

\paragraph{Circuit theories.} First, we explain what we mean by a `circuit theory' (formally: a symmetric monoidal category). Such a theory includes a set of systems and a set of transformations between each pair of systems. If the theory contains a transformation $T_1: A \rightarrow B$ whose output is the system $B$, and another one $T_2 : B \rightarrow C$ that accepts $B$ as its input, then we require that the theory also contains a third transformation $T_3 := T_2 \circ T_1$, corresponding to `doing $T_1$ first and then $T_2$'. 
Diagrammatically, this means that if we draw $T_1$ and $T_2$ as boxes plugged into each other, we can think of the resulting shape as another transformation:
\begin{equation}
    \begin{tikzpicture}
	\begin{pgfonlayer}{nodelayer}
		\node [style=map] (0) at (17, 1.5) {$T_2$};
		\node [style=map] (1) at (17, -1.5) {$T_1$};
		\node [style=none] (2) at (17, 4.5) {};
		\node [style=none] (3) at (17, -4.5) {};
		\node [style=none] (4) at (15, -3.25) {};
		\node [style=none] (5) at (19, -3.25) {};
		\node [style=none] (6) at (15, 3.25) {};
		\node [style=none] (7) at (19, 3.25) {};
		\node [style=none] (8) at (19, 3.25) {};
		\node [style=none] (9) at (15.5, 2.5) {$ \color{blue} T_3$};
		\node [style=none] (10) at (17.5, -4) {$A$};
		\node [style=none] (11) at (17.5, 0) {$B$};
		\node [style=none] (12) at (17.5, 4) {$C$};
	\end{pgfonlayer}
	\begin{pgfonlayer}{edgelayer}
		\draw (3.center) to (2.center);
		\draw [style=blue] (6.center) to (8.center);
		\draw [style=blue] (8.center) to (5.center);
		\draw [style=blue] (5.center) to (4.center);
		\draw [style=blue] (4.center) to (6.center);
	\end{pgfonlayer}
\end{tikzpicture}
\end{equation}

We further require that this \textit{sequential composition} has the identity transformation $1: A \rightarrow A$ for each system $A$ as a unit, meaning that
\begin{equation}
    \begin{tikzpicture}
	\begin{pgfonlayer}{nodelayer}
		\node [style=map] (0) at (-12, 1.25) {$T$};
		\node [style=none] (1) at (-12, 2.75) {};
		\node [style=none] (2) at (-12, -2.75) {};
		\node [style=map] (3) at (-7, -1) {$T$};
		\node [style=none] (4) at (-7, 2.75) {};
		\node [style=none] (5) at (-7, -2.75) {};
		\node [style=none] (14) at (-12.5, -4.5) {};
		\node [style=none] (15) at (-5, 0) {$=$};
		\node [style=map] (16) at (-3, 0) {$T$};
		\node [style=none] (17) at (-3, 2.75) {};
		\node [style=none] (18) at (-3, -2.75) {};
		\node [style=map] (22) at (-12, -1) {$1$};
		\node [style=map] (23) at (-7, 1.25) {$1$};
		\node [style=none] (24) at (-9.5, 0) {$=$};
	\end{pgfonlayer}
	\begin{pgfonlayer}{edgelayer}
		\draw (2.center) to (1.center);
		\draw (5.center) to (4.center);
		\draw (18.center) to (17.center);
	\end{pgfonlayer}
\end{tikzpicture}
\end{equation}
for all $T$. The identity transformation can therefore be thought of as simply as a wire.
\begin{equation}
    \begin{tikzpicture}
	\begin{pgfonlayer}{nodelayer}
		\node [style=map] (0) at (-12, 0) {$1$};
		\node [style=none] (1) at (-12, 2.75) {};
		\node [style=none] (2) at (-12, -2.75) {};
		\node [style=none] (4) at (-7, 2.75) {};
		\node [style=none] (5) at (-7, -2.75) {};
		\node [style=none] (14) at (-12.5, -4.5) {};
		\node [style=none] (24) at (-9.5, 0) {$=$};
	\end{pgfonlayer}
	\begin{pgfonlayer}{edgelayer}
		\draw (2.center) to (1.center);
		\draw (5.center) to (4.center);
	\end{pgfonlayer}
\end{tikzpicture}
\end{equation}
We also require that sequential composition is associative, meaning that the picture  below is well defined; there is no need to put brackets around $T_1$ and $T_2$ or $T_2$ and $T_3$.
\begin{equation}
    \begin{tikzpicture}
	\begin{pgfonlayer}{nodelayer}
		\node [style=map] (0) at (17, 1.5) {$T_2$};
		\node [style=map] (1) at (17, -1.5) {$T_1$};
		\node [style=none] (2) at (17, 6) {};
		\node [style=none] (3) at (17, -3) {};
		\node [style=map] (13) at (17, 4.5) {$T_3$};
	\end{pgfonlayer}
	\begin{pgfonlayer}{edgelayer}
		\draw (3.center) to (2.center);
	\end{pgfonlayer}
\end{tikzpicture}

\end{equation}

We do not only want to be able to perform transformations one after the other; we also want to be able to perform any pair of transformations independently. Thus we require that we can take a \textit{tensor product} any pair of transformations $T_1: A \rightarrow B$ and $T_2: C \rightarrow D$ to form another transformation in our theory: $T_3 = T_1 \otimes T_2: A \otimes C \rightarrow B \otimes D$. To make sense of this, we also need to require that we can take a tensor product of any pair of systems to form another system in our theory. All this means that a diagram of a pair of transformations placed side-by-side always defines a new transformation:
\begin{equation}
    \begin{tikzpicture}
	\begin{pgfonlayer}{nodelayer}
		\node [style=map] (1) at (17, -1.5) {$T_1$};
		\node [style=none] (2) at (17, 1.25) {};
		\node [style=none] (3) at (17, -4.25) {};
		\node [style=map] (14) at (20, -1.5) {$T_2$};
		\node [style=none] (15) at (20, 1.25) {};
		\node [style=none] (16) at (20, -4.25) {};
		\node [style=none] (17) at (24, -1.5) {$=$};
		\node [style=map] (18) at (26, -1.5) {$T_3$};
		\node [style=none] (19) at (26, 1.25) {};
		\node [style=none] (20) at (26, -4.25) {};
		\node [style=none] (21) at (15, -3) {};
		\node [style=none] (22) at (22, -3) {};
		\node [style=none] (23) at (15, 0) {};
		\node [style=none] (24) at (22, 0) {};
		\node [style=none] (25) at (15.5, -0.5) {$\color{blue} T_3$};
		\node [style=none] (26) at (17.5, -3.75) {};
		\node [style=none] (27) at (17.5, -3.75) {$A$};
		\node [style=none] (28) at (20.5, -3.75) {$C$};
		\node [style=none] (29) at (17.5, 0.75) {$B$};
		\node [style=none] (30) at (20.5, 0.75) {$D$};
		\node [style=none] (31) at (27.5, -3.5) {$A \otimes C$};
		\node [style=none] (32) at (27.5, 0.75) {$B \otimes D$};
	\end{pgfonlayer}
	\begin{pgfonlayer}{edgelayer}
		\draw (3.center) to (2.center);
		\draw (16.center) to (15.center);
		\draw (20.center) to (19.center);
		\draw [style=blue] (23.center) to (21.center);
		\draw [style=blue] (21.center) to (22.center);
		\draw [style=blue] (22.center) to (24.center);
		\draw [style=blue] (24.center) to (23.center);
	\end{pgfonlayer}
\end{tikzpicture}
\end{equation}

We also stipulate that the tensor products on systems and on transformations are associative and unital.  The unit $1_I$ of the tensor product on transformations is the identity transformation on a special trivial unit system $I$. Associativity implies that three boxes lined up in a row have a well-defined meaning. 

A transformation $T: I \rightarrow A$ from the trivial system to some system $A$ is called a `state' on $A$. Diagramatically, the wire corresponding to the trivial system can be safely omitted, meaning that a state has the following representation.
\begin{equation}
    \begin{tikzpicture}
	\begin{pgfonlayer}{nodelayer}
		\node [style=point] (7) at (0, 1) {T};
		\node [style=none] (8) at (0, 3.5) {};
	\end{pgfonlayer}
	\begin{pgfonlayer}{edgelayer}
		\draw (7) to (8.center);
	\end{pgfonlayer}
\end{tikzpicture}
 \ \ \ \ \ \ \ \ \ \ \ \ 
\end{equation}

The last key rule is known as the interchange law. This is the requirement that the sequential composition and tensor product distribute over one another, in the sense that $(T_1 \otimes T_2) \circ (R_1 \otimes R_2) = (T_1 \circ R_1) \otimes (T_2 \circ R_2)$. Diagrammatically, this means we needn't worry about the order of boxes on different wires:
\begin{equation} \label{eq:interchange}
    \begin{tikzpicture}
	\begin{pgfonlayer}{nodelayer}
		\node [style=map] (0) at (-10, 1.25) {$T_1$};
		\node [style=none] (1) at (-10, 2.75) {};
		\node [style=none] (2) at (-10, -2.75) {};
		\node [style=map] (3) at (-7, -1) {$T_2$};
		\node [style=none] (4) at (-7, 2.75) {};
		\node [style=none] (5) at (-7, -2.75) {};
		\node [style=map] (6) at (4, 0) {$T_1$};
		\node [style=none] (7) at (4, 2.75) {};
		\node [style=none] (8) at (4, -2.75) {};
		\node [style=map] (9) at (7, 0) {$T_2$};
		\node [style=none] (10) at (7, 2.75) {};
		\node [style=none] (11) at (7, -2.75) {};
		\node [style=none] (12) at (2, 0) {$=$};
		\node [style=none] (13) at (2, 0) {};
		\node [style=none] (14) at (-12.5, -4.5) {};
		\node [style=none] (15) at (-5, 0) {$=$};
		\node [style=map] (16) at (-3, -1) {$T_1$};
		\node [style=none] (17) at (-3, 2.75) {};
		\node [style=none] (18) at (-3, -2.75) {};
		\node [style=map] (19) at (0, 1.25) {$T_2$};
		\node [style=none] (20) at (0, 2.75) {};
		\node [style=none] (21) at (0, -2.75) {};
	\end{pgfonlayer}
	\begin{pgfonlayer}{edgelayer}
		\draw (2.center) to (1.center);
		\draw (5.center) to (4.center);
		\draw (8.center) to (7.center);
		\draw (11.center) to (10.center);
		\draw (18.center) to (17.center);
		\draw (21.center) to (20.center);
	\end{pgfonlayer}
\end{tikzpicture}
\end{equation}
\textit{If} the different wires are associated with different spacelike separated regions, \textit{then} the interchange law imposes a sort of locality constraint, which corresponds to the equivalence of inertial frames. This is behind the Local Dynamics assumption of Section \ref{sec:BIL}.

If all these requirements are satisfied, plus some further ones about the existence of a `swap' transformation for each pair of systems that aren't required for the proofs in this paper, then we have a symmetric monoidal category. 
It is known that such theories always admit an adequate circuit representation \cite{joyal1991geometry, kissinger2015picturing}, so we can always think of them as `circuit theories'.

\paragraph{Categorical probabilistic theories. }A categorical probabilistic theory is a special case of a circuit theory, that also satisfies three further requirements.

\begin{enumerate}
    \item It contains a full subtheory of classical systems that also forms its own circuit theory (i.e. a symmetric monoidal category).
    \item Sums of transformations are well-defined, and include convex combinations of probability distributions in the case of classical systems. The $\circ$ and $\otimes$ operations are bilinear.
    \item The theory comes with a special `trace' transformation Tr$_A: A \rightarrow I$ for any system $A$, given by marginalization in the case of classical systems.
\end{enumerate}

Let us explain each of these in turn, starting with (1). The idea there is that there is some subset of the systems in the larger theory such that all the transformations between them define their own circuit theory. This circuit theory is classical (formally speaking, it is equivalent to $\mathbb{R}^+$-Mat, the symmetric monoidal category of positive real-valued matrices), and can therefore explicitly incorporate conditional probability distributions for measurement outcomes given settings within our theory. For example, consider the diagram below, in which dashed wires denote classical systems.
\begin{equation}
    \begin{tikzpicture}
	\begin{pgfonlayer}{nodelayer}
		\node [style=none] (0) at (-1.5, -1) {};
		\node [style=none] (1) at (1.5, -1) {};
		\node [style=none] (2) at (-1.5, -1.75) {};
		\node [style=none] (3) at (1.5, -1.75) {};
		\node [style=large map] (4) at (-1.25, 2.75) {$T_2$};
		\node [style=none] (5) at (-1, -1) {};
		\node [style=none] (6) at (0, 2.75) {};
		\node [style=none] (7) at (2, 5.5) {};
		\node [style=none] (8) at (-2.5, -3) {};
		\node [style=none] (9) at (-2.5, 2.75) {};
		\node [style=none] (10) at (0, 0.5) {};
		\node [style=none] (11) at (0, -1.5) {$\phi$};
		\node [style=none] (12) at (1, -1) {};
		\node [style=none] (13) at (-1, 2.75) {};
		\node [style=none] (14) at (-1, 5.5) {};
		\node [style=none] (15) at (1, 0.5) {};
		\node [style=semilarge map] (16) at (1, 0.5) {$T_1$};
		\node [style=none] (17) at (2, 0.5) {};
		\node [style=none] (18) at (2, 2.75) {};
		\node [style=map] (19) at (2, 2.75) {$T_3$};
	\end{pgfonlayer}
	\begin{pgfonlayer}{edgelayer}
		\draw (0.center) to (1.center);
		\draw (0.center) to (2.center);
		\draw [bend right=90, looseness=0.50] (2.center) to (3.center);
		\draw (3.center) to (1.center);
		\draw [style=dashed] (8.center) to (9.center);
		\draw [in=-90, out=90, looseness=0.75] (10.center) to (6.center);
		\draw (5.center) to (13.center);
		\draw [style=dashed] (13.center) to (14.center);
		\draw (12.center) to (15.center);
		\draw [style=dashed, in=270, out=90] (17.center) to (7.center);
	\end{pgfonlayer}
\end{tikzpicture}
\end{equation}
Out of all the boxes in the picture, only $T_3$ represents a transformation from the classical circuit subtheory, since only it acts exclusively on classical systems. Nevertheless, the combination of all the boxes into a circuit gives rise to an overall transformation that does live in the classical circuit subtheory. This overall transformation might admit an interpretation as a conditional probability distribution for a pair of measurement outcomes given a single  choice of setting.

Moving onto (2), by sums being well-defined we mean we can take a sum of any pair of transformations, $T_1, T_2: A \rightarrow B$, that act on the same systems to form another transformation $T_3 := T_1+T_2: A \rightarrow B$ that also lives in our theory.\footnote{Note that $T_3$ may not be a physical transformation. The physical transformations are picked out by the requirement that the trace is preserved, or, more generally, decreased. For example, the quantum perspectival theory discussed in Section \ref{sec:qpt} consists of completely positive maps, which give us quantum channels when we require trace-preservation.} Given this summation structure, $\circ$ and $\otimes$ are required to be bilinear operations.\footnote{
It follows from bilinearity that the possibility of taking probabilistic combinations of classical states leads to the possibility of taking probabilistic combinations of arbitrary states. For example, suppose we have a transformation $T$ from a classical system (i.e. a variable) to a nonclassical system that prepares the state $\rho_0$ when fed the classical state [0] (i.e the variable taking the `0' value), and prepares $\rho_1$ when fed the state [1]. Then the bilinearity of $\circ$ means that inputting a probabilistic combination of [0] and [1] leads to a probabilistic combination of the nonclassical outputs:
\begin{equation}
\begin{split}
    T \circ (p[0]+ (1-p)[1]) &= T(p[0]) + T((1-p)[1]) \\
    &= pT([0]) + (1-p)T([1]) \\
    &= p \rho_0 + (1-p)\rho_1  
\end{split}
\end{equation}}


Finally, the trace in (3) must satisfy ${\rm Tr}_A \otimes {\rm{Tr}}_B = {\rm Tr}_{A \otimes B}$ and the trace on the trivial system ${\rm Tr}_I$ must equal the identity transformation on that system, ${\rm Tr}_I=1_I$. The trace operation in quantum theory provides an example. This operation allows us to define a `localized' state on $B$ given some bipartite state on $A \otimes B$, to define normalized states and transformations, and to articulate causality constraints.

\subsection{Perspectival theories} \label{sec:perspectival}

From the `inside' perspective on Charlie's measurement in Section \ref{sec:simple}, outcomes are written onto classical systems with various probabilities. In a categorical probabilistic theory, this perspective is nicely represented with a \textit{probability extractor}. We define a probability extractor as any normalization-preserving transformation with a classical output, i.e.\ one satisfying:
\begin{equation}
    \begin{tikzpicture}
	\begin{pgfonlayer}{nodelayer}
		\node [style=none] (1) at (0.5, 0) {};
		\node [style=map] (2) at (0.5, 0) {$E$};
		\node [style=none] (3) at (0.5, 2) {};
		\node [style=none] (5) at (0.5, -2.5) {};
		\node [style=discard] (6) at (0.5, 2.25) {};
		\node [style=none] (7) at (3.5, -1) {};
		\node [style=none] (9) at (3.5, -2.5) {};
		\node [style=discard] (10) at (3.5, -0.75) {};
		\node [style=none] (11) at (2, 0) {$=$};
	\end{pgfonlayer}
	\begin{pgfonlayer}{edgelayer}
		\draw [style=dashed] (2) to (3.center);
		\draw (5.center) to (2);
		\draw (9.center) to (7.center);
	\end{pgfonlayer}
\end{tikzpicture}
\end{equation}
where  \begin{tikzpicture}%
\draw (0,1ex) -- (5ex,1ex);%
\draw (1ex,2ex) -- (4ex,2ex);%
\draw (2ex,3ex) -- (3ex,3ex);%
\draw (2.5ex, -2ex) -- (2.5ex, 1ex);
\end{tikzpicture} represents the trace.
Such a transformation always outputs a probability distribution when it is fed a normalized state.

Perspectival theories are categorical probabilistic theories that uniquely associate each probability extractor with a transformation of type $U:S \rightarrow M \otimes S$, called a \textit{memory update} (to borrow an idea from \cite{vilasini2019multi}).  Note that we do not require that all transformations of this type are memory updates, but only that all memory updates are of this type.\footnote{This typing assumption is mainly here to make the exposition more intuitive -- all of our theorems still hold if one drops it.} We think of $S$ as the system being measured, and $M$ as the `memory' of the agent doing the measurement. Both $S$ and $M$ can be nonclassical. 

A categorical probabilistic theory can be fleshed out into a perspectival one simply by introducing a bijective function from the probability extractors to the set of memory updates.

\begin{definition}[Perspectival theories.]
A perspectival theory is a categorical probabilistic theory equipped with a set of memory updates $\cu$; and
a bijective function $f: \mathcal{P} \rightarrow \cu$, where $\mathcal{P}$ is the set of probability extractors.\footnote{\label{note:ptgeneral} More generally, one could consider a one-to-many map from extractors to updates, with the property that no single update is associated with multiple extractors, and the proofs below would still go through. Some might prefer this version of perspectival theories since then, for example, in a quantum perspectival theory, the same POVM could be associated with multiple different update rules. We stick with bijective functions since (a) it makes the exposition easier, and (b) one can always consider a categorical probabilistic theory with multiple copies of the same extractor and map the different copies to different updates, which gives essentially the same effect as an injective map. }
\end{definition}

Let us explain how a perspectival theory accommodates a duality of perspectives on a measurement.
From the `inside' perspective, one can obtain probabilities for measurement outcomes by applying a probability extractor to a state:
\begin{equation}
    \begin{tikzpicture}
	\begin{pgfonlayer}{nodelayer}
		\node [style=point] (0) at (0, -2.25) {$\phi$};
		\node [style=none] (1) at (0.5, -1.25) {$S$};
		\node [style=none] (2) at (0, 0) {};
		\node [style=map] (4) at (0, 0) {$E$};
		\node [style=none] (5) at (0, 2) {};
		\node [style=none] (6) at (0.5, 1.25) {$X$};
	\end{pgfonlayer}
	\begin{pgfonlayer}{edgelayer}
		\draw (0) to (4);
		\draw [style=dashed] (4) to (5.center);
	\end{pgfonlayer}
\end{tikzpicture}
\end{equation}
This will be the appropriate model for the agent who actually does that measurement, who presumably takes herself to observe a unique actual outcome, among many possible outcomes that can be assigned probabilities. But in a perspectival theory, this model is uniquely associated with another model, in which it might be that no probabilities explicitly arise (perhaps because the outputs of the memory update are nonclassical):
\begin{equation}
    \begin{tikzpicture}
	\begin{pgfonlayer}{nodelayer}
		\node [style=point] (0) at (0, -2.25) {$\phi$};
		\node [style=large map] (1) at (-0.5, 0) {$f(E)$};
		\node [style=none] (2) at (0, 2.25) {};
		\node [style=none] (3) at (-1, 0) {};
		\node [style=none] (4) at (-1, 2.25) {};
		\node [style=none] (5) at (0.5, -1.25) {$S$};
		\node [style=none] (6) at (0.5, 1.25) {$S$};
		\node [style=none] (7) at (-1.5, 1.25) {$M$};
	\end{pgfonlayer}
	\begin{pgfonlayer}{edgelayer}
		\draw (2.center) to (0);
		\draw (3.center) to (4.center);
	\end{pgfonlayer}
\end{tikzpicture}
\end{equation}
This will be the appropriate model for a superobserver (that is, someone like Alice or Bob from Section \ref{sec:simple}) about to carry out a supermeasurement on $M \otimes S$. Explicitly, the probabilities for the data seen by a superobserver who carries out a measurement associated with an extractor $F$ are given by a circuit of the following form.
\begin{equation}
    \begin{tikzpicture}
	\begin{pgfonlayer}{nodelayer}
		\node [style=point] (0) at (0, -2.25) {$\phi$};
		\node [style=large map] (1) at (-0.5, 0) {$f(E)$};
		\node [style=none] (2) at (0, 2.25) {};
		\node [style=none] (3) at (-1, 0) {};
		\node [style=none] (4) at (-1, 2.25) {};
		\node [style=none] (5) at (0.5, -1.25) {$S$};
		\node [style=none] (6) at (0.5, 1.25) {$S$};
		\node [style=none] (7) at (-1.5, 1.25) {$M$};
		\node [style=large map] (9) at (-0.5, 2.5) {$F$};
		\node [style=none] (10) at (-0.5, 4.5) {};
		\node [style=none] (11) at (0, 3.75) {$Y$};
	\end{pgfonlayer}
	\begin{pgfonlayer}{edgelayer}
		\draw (2.center) to (0);
		\draw (3.center) to (4.center);
		\draw [style=dashed](9) to (10.center);
	\end{pgfonlayer}
\end{tikzpicture}
\end{equation}

On the other hand, suppose we have some circuit involving a memory update $U$:
\begin{equation} \label{eq:outside}
    \begin{tikzpicture}
	\begin{pgfonlayer}{nodelayer}
		\node [style=point] (0) at (0, -3) {$\phi$};
		\node [style=large map] (1) at (-0.5, 2) {$U$};
		\node [style=none] (2) at (0, 4.5) {};
		\node [style=none] (3) at (-1, 2) {};
		\node [style=none] (4) at (-1, 4.5) {};
		\node [style=large map] (11) at (-0.5, 4.5) {$T_2$};
		\node [style=map] (12) at (0, -0.5) {$T_1$};
		\node [style=none] (13) at (0.5, -1.75) {};
		\node [style=none] (18) at (-0.5, 6.5) {};
	\end{pgfonlayer}
	\begin{pgfonlayer}{edgelayer}
		\draw [in=90, out=-90] (2.center) to (0);
		\draw (3.center) to (4.center);
		\draw (18.center) to (11);
	\end{pgfonlayer}
\end{tikzpicture}
\end{equation}
This circuit is understood as representing a scenario where a measurement associated with $U$ is performed on a state $T_1(\phi)$, after which another transformation $T_2$ is performed on both the measured system and the memory system.

A probability distribution over the outcomes for that measurement is obtained by applying the probability extractor $f^{-1}(U)$ at the appropriate point in the circuit:
\begin{equation} \label{eq:inside}
  \begin{tikzpicture}
	\begin{pgfonlayer}{nodelayer}
		\node [style=point] (0) at (0, -3) {$\phi$};
		\node [style=large map] (1) at (-0.5, 2) {$U$};
		\node [style=none] (2) at (0, 4.5) {};
		\node [style=none] (3) at (-1, 2) {};
		\node [style=none] (4) at (-1, 4.5) {};
		\node [style=large map] (11) at (-0.5, 4.5) {$T_2$};
		\node [style=map] (12) at (0, -0.5) {$T_1$};
		\node [style=none] (13) at (0.5, -1.75) {};
		\node [style=none] (18) at (-0.5, 6.5) {};
	\end{pgfonlayer}
	\begin{pgfonlayer}{edgelayer}
		\draw [in=90, out=-90] (2.center) to (0);
		\draw (3.center) to (4.center);
		\draw (18.center) to (11);
	\end{pgfonlayer}
\end{tikzpicture} \quad \quad \quad \rightarrow \quad \quad \ \ \  \begin{tikzpicture}
	\begin{pgfonlayer}{nodelayer}
		\node [style=point] (0) at (0, -3) {$\phi$};
		\node [style=large map] (1) at (0, 2) {};
		\node [style=none] (2) at (0, 2) {$f^{-1}(U)$};
		\node [style=map] (12) at (0, -0.5) {$T_1$};
		\node [style=none] (13) at (0.5, -1.75) {};
		\node [style=none] (14) at (0, 4) {};
	\end{pgfonlayer}
	\begin{pgfonlayer}{edgelayer}
		\draw [in=90, out=-90] (2.center) to (0);
		\draw [style=dashed](14.center) to (2.center);
	\end{pgfonlayer}
\end{tikzpicture}
\end{equation}
The model on the right can be used to predict the probabilities for the outcomes of the measurement associated with $U$ (and these probabilities are associated with the time at which $U$ was implemented, so they are unaffected by $T_2$).

More generally, if a set of memory updates are applied in parallel, joint probabilities for those measurements are obtained by applying $f^{-1}$ to each one.\footnote{One might therefore wish to require that the tensor product of a set of memory updates is itself a memory update, which is associated with an extractor that is operationally equivalent to the tensor product of the extrators associated with each member of the set. We will not bother with this requirement since it is not necessary for our proofs.} For example:
\begin{equation} \label{eq:inside_pair}
    \begin{tikzpicture}
	\begin{pgfonlayer}{nodelayer}
		\node [style=point] (0) at (1.5, -3) {$\phi$};
		\node [style=none] (2) at (1.5, -0.5) {};
		\node [style=none] (3) at (-0.5, 2.25) {};
		\node [style=none] (4) at (-0.5, 4.5) {};
		\node [style=none] (7) at (0.5, -0.5) {};
		\node [style=none] (8) at (1.5, 5) {$T_2$};
		\node [style=large map] (9) at (1.5, -0.5) {$T_1$};
		\node [style=none] (11) at (2.5, 2.25) {};
		\node [style=none] (12) at (2.5, 4.5) {};
		\node [style=none] (14) at (1.5, 5.5) {};
		\node [style=none] (15) at (2.5, -0.5) {};
		\node [style=none] (16) at (0.5, 4.5) {};
		\node [style=none] (17) at (2.5, 2) {};
		\node [style=none] (18) at (4, 4.5) {};
		\node [style=none] (19) at (-1, 4.5) {};
		\node [style=none] (20) at (-1, 5.5) {};
		\node [style=none] (21) at (4, 5.5) {};
		\node [style=none] (22) at (3.5, 2.25) {};
		\node [style=none] (23) at (3.5, 4.5) {};
		\node [style=none] (24) at (1.5, 6.75) {};
		\node [style=none] (25) at (0, 2.25) {};
		\node [style=semilarge map] (26) at (-0.25, 2.25) {$U_1$};
		\node [style=semilarge map] (27) at (3.25, 2.25) {$U_2$};
	\end{pgfonlayer}
	\begin{pgfonlayer}{edgelayer}
		\draw [in=90, out=-90] (2.center) to (0);
		\draw (3.center) to (4.center);
		\draw (11.center) to (12.center);
		\draw (16.center) to (7.center);
		\draw (15.center) to (17.center);
		\draw (18.center) to (21.center);
		\draw (21.center) to (20.center);
		\draw (20.center) to (19.center);
		\draw (19.center) to (18.center);
		\draw (23.center) to (22.center);
		\draw (24.center) to (14.center);
	\end{pgfonlayer}
\end{tikzpicture}  \quad \quad \quad  \rightarrow \quad \quad \ \ \  
 \begin{tikzpicture}
	\begin{pgfonlayer}{nodelayer}
		\node [style=point] (0) at (1.5, -3) {$\phi$};
  	\node [style=none] (1) at (1.5, -0.5) {};
   		\node [style=none] (2) at (0.25, -0.5) {};
		\node [style=large map] (3) at (1.5, -0.5) {$T_1$};
		\node [style=none] (4) at (2.75, -0.5) {};
		\node [style=none] (5) at (0.25, 2.25) {};
		\node [style=none] (6) at (2.75, 2.25) {};
		\node [style=semilarge map] (7) at (0, 2.25) {};
		\node [style=semilarge map] (8) at (3, 2.25) {};
		\node [style=none] (9) at (0.25, 4.25) {};
		\node [style=none] (10) at (2.75, 4.25) {};
		\node [style=none] (11) at (0, 2.25) {$\scriptstyle f^{-1}(U_1)$};
		\node [style=none] (12) at (3, 2.25) {$\scriptstyle  f^{-1}(U_2)$};
	\end{pgfonlayer}
 \begin{pgfonlayer}{edgelayer}
		\draw [in=90, out=-90] (1.center) to (0);
		\draw (5.center) to (2.center);
		\draw (4.center) to (6.center);
		\draw [style=dashed] (9.center) to (5.center);
		\draw [style=dashed] (10.center) to (6.center);
	\end{pgfonlayer}
\end{tikzpicture}
\end{equation}
The model on the right can be used to predict the correlations between the observations of agents associated with $U_1$ and $U_2$.

This generalizes the duality of perspectives on a measurement that is familiar from quantum theory, as the next subsection makes explicit.

\subsection{A quantum perspectival theory} \label{sec:qpt}

Recalling Section \ref{sec:simple}, a natural memory update to associate with the quantum measurement of the basis $\{\ket{i}\}_i$ is $U=\sum_i \ket{ii}\bra{i}$. One can extend this idea to form a fully-fledged quantum perspectival theory, in which measurements can alternatively be viewed as POVMs or isometric interactions (which in turn arise from unitary interactions). This theory can be seen as the rigorous formalization of the `quantum theory' assumed in various no-go theorems inspired by Wigner's friend \cite{Frauchiger2018quantum, leegwater2022greenberger, healey2018quantum, bong2020strong, haddara2022possibilistic, ormrod2022no}. 

Let us briefly describe how this works. As shown in \cite{gogioso2017categorical}, one can formulate quantum theory as a categorical probabilistic theory. In this formulation, transformations between quantum systems are provided by completely positive maps, and transformations from quantum systems to classical ones include the mappings to probabilities afforded by POVMs.

To construct a quantum perspectival theory, one then only needs to associate the probability extractors with memory updates. 
One natural way of doing this relies on the Naimark dilation theorem for quantum measurements. 
This theorem says that any measurement on a quantum system can be implemented by letting that system unitarily interact with another quantum probe, initialized in some pure state, and then (projectively) measuring the probe. 
Or, diagrammatically:
\begin{equation}
    \begin{tikzpicture}
	\begin{pgfonlayer}{nodelayer}
		\node [style=none] (1) at (0.5, -1.5) {$S$};
		\node [style=none] (2) at (0, 0) {};
		\node [style=map] (3) at (0, 0) {$E$};
		\node [style=none] (4) at (0, 4.75) {};
		\node [style=none] (6) at (2, 0) {};
		\node [style=none] (8) at (0, -3) {};
		\node [style=point] (9) at (3.5, -2.5) {$\psi$};
		\node [style=none] (10) at (3.5, -0.25) {};
		\node [style=none] (11) at (3.5, 3) {};
		\node [style=none] (12) at (3.5, 3) {};
		\node [style=discard] (14) at (5.5, 4.5) {};
		\node [style=none] (15) at (6, -1.5) {$S$};
		\node [style=none] (18) at (5.5, 4.25) {};
		\node [style=none] (19) at (6, 1.5) {$S$};
		\node [style=none] (20) at (5.5, -3) {};
		\node [style=large map] (21) at (4.5, 0) {$V$};
		\node [style=none] (22) at (3, -1.5) {$P$};
		\node [style=none] (23) at (2, 0) {$=$};
		\node [style=map] (24) at (3.5, 3) {$F$};
		\node [style=none] (25) at (3.5, 4.75) {};
		\node [style=none] (26) at (3, 1.5) {$P$};
	\end{pgfonlayer}
	\begin{pgfonlayer}{edgelayer}
		\draw [style=dashed] (3) to (4.center);
		\draw (8.center) to (3);
		\draw (9) to (11.center);
		\draw (20.center) to (18.center);
		\draw [style=dashed] (11.center) to (12.center);
		\draw [style=dashed] (24) to (25.center);
	\end{pgfonlayer}
\end{tikzpicture} \ \ \ \ \ \forall E \ \exists \psi, V, F
\end{equation}
holds for extractors $E$ and $F$, a unitary $V$, and a pure state $\psi$. One could take as a measurement update for $E$ the isometry obtained by inserting $\psi$ into $V$.\footnote{This isometry is not unique, since the unitary dilation is not unique. One therefore needs to make an arbitrary choice.} 

Those who do not worship at the church of the larger Hilbert space may wish to formulate other sorts of quantum perspectival theories. For example,  objective collapse theorists will want to use non-unitary channels that induce wavefunction collapse as memory updates (such a theory is discussed in Appendix \ref{app:role_collapse}). This illustrates an important point about the framework of perspectival theories: that the two perspectives of a theory in the framework need not be in any tension with each other.

\section{A generalized no-go theorem for absolute observed events} \label{sec:BIL}

In this section, we prove that a certain class of perspectival theories are inconsistent with the absoluteness of observed events. We call them the `BIL' theories, because they violate \underline{B}ell inequalities, they preserve \underline{i}nformation, and they are dynamically \underline{l}ocal.

\subsection{BIL theories} \label{sec:BIL_bil}

Let us explain in more detail what we mean by a BIL theory, one letter at a time.

\paragraph{Local Dynamics.} 
The `L' in BIL is for a dynamical sort of locality. The basic idea is straightforward. Suppose one region contains the systems $A$ and $A'$, and another, spacelike separated,  region contains $B$ and $B'$. Consider a transformation $T: A \otimes B \rightarrow A' \otimes B'$ taking place across both regions. In any dynamically local theory, the output at $A'$ should not depend on the input at $B$, nor should the output at $B'$ depend on the input at $A$. Assuming the diagrammatic representation of $T$ is `faithful' to these independences, we should then be able to write $T$ in the following way.

\begin{equation} \label{locality_2party}
    \begin{tikzpicture}
	\begin{pgfonlayer}{nodelayer}
		\node [style=none] (0) at (-3.5, 2) {};
		\node [style=none] (1) at (-3.5, -4) {};
		\node [style=none] (2) at (-1.5, 2) {};
		\node [style=none] (3) at (-1.5, -4) {};
		\node [style=none] (4) at (-2.5, 0) {$T$};
		\node [style=none] (5) at (-4, -3.5) {$A$};
		\node [style=none] (6) at (-4, 1.5) {$A'$};
		\node [style=none] (7) at (-1, -3.5) {$B$};
		\node [style=none] (8) at (-1, 1.5) {$B'$};
		\node [style=large map] (9) at (-2.5, 0) {$T$};
		\node [style=none] (10) at (0, 0) {};
		\node [style=none] (11) at (0, 0) {};
		\node [style=none] (12) at (0, 0) {$=$};
		\node [style=none] (13) at (1.75, 2) {};
		\node [style=none] (14) at (1.75, -4) {};
		\node [style=none] (15) at (1.25, -3.5) {$A$};
		\node [style=none] (16) at (1.25, 1.5) {$A'$};
		\node [style=map] (17) at (1.75, 0) {$T_1$};
		\node [style=none] (18) at (4.25, 2) {};
		\node [style=none] (19) at (4.25, -4) {};
		\node [style=none] (20) at (4.75, -3.5) {$B$};
		\node [style=none] (21) at (4.75, 1.5) {$B'$};
		\node [style=map] (22) at (4.25, 0) {$T_2$};
		\node [style=none] (23) at (5.75, -3) {};
		\node [style=none] (24) at (7.25, -3) {};
		\node [style=none] (25) at (5.75, -3.75) {};
		\node [style=none] (26) at (7.25, -3.75) {};
		\node [style=none] (27) at (6.25, -3) {};
		\node [style=none] (28) at (6.5, -3.5) {$\psi$};
		\node [style=none] (29) at (7.25, -3) {};
		\node [style=none] (30) at (6.75, -2) {};
		\node [style=none] (31) at (6.75, -3) {};
		\node [style=none] (32) at (2, -0.5) {};
		\node [style=none] (33) at (4.75, 0) {};
	\end{pgfonlayer}
	\begin{pgfonlayer}{edgelayer}
		\draw (1.center) to (0.center);
		\draw (3.center) to (2.center);
		\draw [in=270, out=90] (14.center) to (13.center);
		\draw (19.center) to (18.center);
		\draw (23.center) to (24.center);
		\draw (23.center) to (25.center);
		\draw [bend right=90, looseness=0.50] (25.center) to (26.center);
		\draw (26.center) to (24.center);
		\draw [in=-90, out=90, looseness=0.75] (27.center) to (32.center);
		\draw [in=90, out=-90, looseness=1.25] (33.center) to (31.center);
	\end{pgfonlayer}
\end{tikzpicture}
\end{equation}
In this diagram, there is no directed path of wires from $A$ to $B'$ or from $B$ to $A'$, making the lack of dependencies clear. Generalizing to $n$ pairs of mutually spacelike separated systems, we have the following. 
\begin{equation} \tag{L} \label{local_dynamics}
    \begin{tikzpicture}
	\begin{pgfonlayer}{nodelayer}
		\node [style=none] (0) at (-3.5, 2) {};
		\node [style=none] (1) at (-3.5, -4) {};
		\node [style=none] (2) at (-1.5, 2) {};
		\node [style=none] (3) at (-1.5, -4) {};
		\node [style=none] (4) at (-2.5, 0) {$T$};
		\node [style=none] (5) at (-4, -3.5) {$A_1$};
		\node [style=none] (6) at (-4, 1.5) {$A_1'$};
		\node [style=none] (7) at (-1, -3.5) {$A_n$};
		\node [style=none] (8) at (-1, 1.5) {$A_n'$};
		\node [style=large map] (9) at (-2.5, 0) {$T$};
		\node [style=none] (10) at (0, 0) {};
		\node [style=none] (11) at (0, 0) {};
		\node [style=none] (12) at (0, 0) {$=$};
		\node [style=none] (13) at (1.75, 2) {};
		\node [style=none] (14) at (1.75, -4) {};
		\node [style=none] (15) at (1.25, -3.5) {$A_1$};
		\node [style=none] (16) at (1.25, 1.5) {$A_1'$};
		\node [style=map] (17) at (1.75, 0) {$T_1$};
		\node [style=none] (18) at (4.25, 2) {};
		\node [style=none] (19) at (4.25, -4) {};
		\node [style=none] (20) at (4.75, -3.5) {$A_n$};
		\node [style=none] (21) at (4.75, 1.5) {$A_n'$};
		\node [style=map] (22) at (4.25, 0) {$T_n$};
		\node [style=none] (23) at (5.75, -3) {};
		\node [style=none] (24) at (8.25, -3) {};
		\node [style=none] (25) at (5.75, -3.75) {};
		\node [style=none] (26) at (8.25, -3.75) {};
		\node [style=none] (27) at (6, -3) {};
		\node [style=none] (28) at (7, -3.5) {$\psi$};
		\node [style=none] (29) at (8.25, -3) {};
		\node [style=none] (30) at (6.75, -2) {};
		\node [style=none] (31) at (8, -3) {};
		\node [style=none] (32) at (2, -0.5) {};
		\node [style=none] (33) at (4.75, 0) {};
		\node [style=none] (34) at (-2.5, 1.5) {$\hdots$};
		\node [style=none] (35) at (-2.5, -3.5) {$\hdots$};
		\node [style=none] (36) at (3, 1.5) {$\hdots$};
		\node [style=none] (37) at (3, -3.5) {$\hdots$};
		\node [style=none] (38) at (6.75, -2.5) {$\hdots$};
	\end{pgfonlayer}
	\begin{pgfonlayer}{edgelayer}
		\draw (1.center) to (0.center);
		\draw (3.center) to (2.center);
		\draw [in=270, out=90] (14.center) to (13.center);
		\draw (19.center) to (18.center);
		\draw (23.center) to (24.center);
		\draw (23.center) to (25.center);
		\draw [bend right=90, looseness=0.50] (25.center) to (26.center);
		\draw (26.center) to (24.center);
		\draw [in=-90, out=90, looseness=0.75] (27.center) to (32.center);
		\draw [in=90, out=-90, looseness=1.25] (33.center) to (31.center);
	\end{pgfonlayer}
\end{tikzpicture}
\end{equation}
We consider that a theory has Local Dynamics just in case transformations on spacelike separated systems always decompose like this. 

So far this is not so formal, since we have not yet formalized the idea that systems can be spacelike separated from one another. To this end, following \cite{vilasini2022embedding}, we consider an embedding function $\ce$ that maps the input and output subsystems of a transformation to points on a Lorentzian manifold $\cm$. 

Given a transformation in our theory, we assume that an embedding function $\ce$ on its subsystems is either `valid' or `invalid'.
Local Dynamics can then be cast as a restriction on the valid embeddings of transformations.
More precisely, suppose we embedded a transformation $T: A_1 \otimes \ldots \otimes A_n \rightarrow A_1' \otimes \ldots \otimes A_n'$ in such a way that every pair $(\ce(A_i), \ce(A_i'))$ of spacetime points is spacelike separated from every other pair. Then Local Dynamics requires that this sort of embedding is valid only if $T$ decomposes as in (\ref{local_dynamics}). 

\begin{definition}[Local Dynamics.]
A perspectival theory has \textbf{Local Dynamics} if all transformations of the form $T: A_1 \otimes \ldots \otimes A_n \rightarrow A_1' \otimes \ldots \otimes A_n'$ can only be {\rm validly embedded}  into a Lorentzian manifold $\cm$ such that each pair $(\ce(A_i), \ce(A_i'))$ of spacetime points is spacelike separated from every other pair if $T$ decomposes as in (\ref{local_dynamics}). 
\end{definition}
We note that this is a rather minimal dynamical locality requirement, in the sense that it says nothing, for example, about which transformations of the form $T: A \rightarrow B$ can be embedded such that $\ce(A)$ and $\ce(B)$ are spacelike separated. This is a feature rather than a bug, since it allows us to focus on precisely what sort of dynamical locality is in tension with the absoluteness of observed events.

\paragraph{Information Preservation.} `I' is for the preservation of information.
The idea is that no information is irretrievably lost in the process of measurement, at least when we take the `outside perspective' on that measurement. This means that if two alternative measurements can be performed on a system, there is always a way to effectively perform the second measurement even after the first has been performed. Letting $U$ be the memory update for the first measurement, and $E$ be the extractor for the second, we arrive at the following formal condition for $U$ to be an \textit{information-preserving memory update} \cite{vilasini2019multi}.
\begin{equation} \label{eq:info_preservation}  \tag{IP}
    \begin{tikzpicture}
	\begin{pgfonlayer}{nodelayer}
		\node [style=none] (4) at (1, -2) {};
		\node [style=none] (5) at (1, -2) {};
		\node [style=none] (6) at (1, -0.25) {};
		\node [style=none] (7) at (1, 4) {};
		\node [style=map] (8) at (1, -0.25) {$E$};
		\node [style=none] (13) at (6, -2) {};
		\node [style=none] (14) at (6, -2) {};
		\node [style=none] (15) at (6, -0.25) {};
		\node [style=none] (16) at (6, 2) {};
		\node [style=none] (17) at (5, -0.25) {};
		\node [style=none] (18) at (5, 2) {};
		\node [style=none] (19) at (5.5, 4) {};
		\node [style=none] (20) at (3, 0) {$=$};
		\node [style=semilarge map] (21) at (5.5, -0.25) {$U$};
		\node [style=semilarge map] (22) at (5.5, 2) {$E'$};
	\end{pgfonlayer}
	\begin{pgfonlayer}{edgelayer}
		\draw (4.center) to (6.center);
		\draw [style=dashed] (6.center) to (7.center);
		\draw (13.center) to (15.center);
		\draw (15.center) to (16.center);
		\draw (17.center) to (18.center);
		\draw [style=dashed] (19.center) to (22);
	\end{pgfonlayer}
\end{tikzpicture} \ \ \ \  \forall E \  \exists E'
\end{equation}
This condition is inspired by a similar one from \cite{vilasini2019multi}. It says that for any probability extractor that can be implemented on $U$'s input system \textit{before} $U$, there exists an operationally equivalent extractor that may be implemented \textit{after} $U$.

This leads to the following definition of an information-preserving \textit{theory}.
\begin{definition}[Information-preservation.]
A perspectival theory is \textbf{Information Preserving} just in case all of its memory updates are information-preserving; that is, all memory updates satisfy (\ref{eq:info_preservation}).
\end{definition}


\paragraph{Bell nonlocality.}

B is for \textit{\underline{B}ell Nonlocal}. 
This is the requirement that our perspectival theory violates Bell inequalities in a nontrivial way. 
More precisely, it demands that the theory can model a situation in which $n$ pairwise spacelike separated\footnote{Note the importance of this requirement: even a classical and relativistic theory can violate Bell inequalities locally.} agents choose a measurement setting and record an outcome, in which the resulting data converge on a conditional probability distribution $p(A_1 \ldots A_n |X_1 \ldots X_n)$ that does not admit a local hidden variable model. Not admitting a local hidden variable model just means that it cannot be written in the form
\begin{equation} \label{eq:hvm} \tag{B1}
    p(A_1 \ldots A_n |X_1 \ldots X_n) = \sum_\lambda p(A_1|X_1 \lambda) \ldots P(A_n|X_n \lambda)p(\lambda)
\end{equation}
for any $\lambda$, $p(\lambda)$, and $p(A_i|X_i \lambda)$.

To formalize this, we consider a circuit model provided by a perspectival theory in which an $n$-partite normalized state $\phi$ is fed into a \textit{classically controlled} probability extractor, $T$:
\begin{equation}\label{eq:contextuality} \tag{B2}
    p(A_1...A_n|X_1...X_n) \ \ = \begin{tikzpicture}
	\begin{pgfonlayer}{nodelayer}
		\node [style=none] (1) at (-3.25, 3) {};
		\node [style=none] (2) at (-3.25, -2) {};
		\node [style=none] (4) at (-1.75, 3) {};
		\node [style=none] (5) at (-1.75, -2) {};
		\node [style=none] (16) at (-2.5, 3) {$T$};
		\node [style=none] (18) at (-3.75, -1.25) {$S_1$};
		\node [style=none] (19) at (-3.75, 4.5) {$A_1$};
		\node [style=none] (20) at (-1.25, -1.25) {$S_n$};
		\node [style=none] (21) at (-1.25, 4.5) {$A_n$};
		\node [style=large map] (22) at (-2.5, 3) {$T$};
		\node [style=none] (26) at (-3.75, -2) {};
		\node [style=none] (27) at (-1.25, -2) {};
		\node [style=none] (28) at (-3.75, -2.75) {};
		\node [style=none] (29) at (-1.25, -2.75) {};
		\node [style=none] (30) at (-3.25, -2) {};
		\node [style=none] (31) at (-2.5, -2.5) {$\phi$};
		\node [style=none] (32) at (-1.75, -2) {};
		\node [style=none] (33) at (-3.25, 5) {};
		\node [style=none] (34) at (-1.75, 5) {};
		\node [style=none] (35) at (-2.5, -1.25) {$\hdots$};
		\node [style=none] (36) at (-3.75, 2.25) {};
		\node [style=none] (37) at (-5.5, -2.5) {};
		\node [style=none] (38) at (-1, 3) {};
		\node [style=none] (40) at (-2.5, 4.5) {$\hdots$};
		\node [style=none] (41) at (-3.75, 3.25) {};
		\node [style=none] (42) at (-5.5, -4.25) {};
		\node [style=none] (46) at (-1.25, 2.25) {};
		\node [style=none] (47) at (0.5, -2.5) {};
		\node [style=none] (48) at (-1.25, 3.25) {};
		\node [style=none] (49) at (0.5, -4.25) {};
		\node [style=none] (50) at (-2.25, -3.25) {};
		\node [style=none] (51) at (-6, -3.75) {$X_1$};
		\node [style=none] (52) at (1, -3.75) {$X_n$};
		\node [style=none] (53) at (-2.5, -3.75) {$\hdots$};
		\node [style=none] (54) at (-4, -3.75) {$\hdots$};
		\node [style=none] (55) at (-1, -3.75) {$\hdots$};
	\end{pgfonlayer}
	\begin{pgfonlayer}{edgelayer}
		\draw (2.center) to (1.center);
		\draw (5.center) to (4.center);
		\draw (26.center) to (27.center);
		\draw (26.center) to (28.center);
		\draw [bend right=90, looseness=0.50] (28.center) to (29.center);
		\draw (29.center) to (27.center);
		\draw [style=dashed] (1.center) to (33.center);
		\draw [style=dashed] (4.center) to (34.center);
		\draw [style=dashed, in=90, out=-90, looseness=1.50] (36.center) to (37.center);
		\draw [style=dashed] (36.center) to (41.center);
		\draw [style=dashed] (42.center) to (37.center);
		\draw [style=dashed, in=90, out=-90, looseness=1.50] (46.center) to (47.center);
		\draw [style=dashed] (46.center) to (48.center);
		\draw [style=dashed] (49.center) to (47.center);
	\end{pgfonlayer}
\end{tikzpicture}
\end{equation}

By calling $T$ a classically controlled extractor, we simply mean that plugging in probability distributions on the classical inputs leads to a probability extractor on $S_1 \otimes \ldots \otimes S_n$. We require that $T$ admits a valid embedding  $\ce$ where every triplet $(\ce(S_i), \ce(X_i), \ce(A_i))$ is embedded in a region that is spacelike separated from all the others. This leads to the following definition, in which `normalized circuit model' means a circuit built exclusively out of normalized and normalization-preserving states and transformations.\footnote{We note that the predictions of a Bell Nonlocal perspectival theory do not strictly imply that Bell's local causality condition is not respected by nature, since, for \textit{that}, one also needs to assume that measurement settings can be freely chosen. However, rejecting free choice does not allow one to evade the no-go theorems from \textit{this} paper, since these are proved using scenarios that do not involve any choices of measurements.}

\begin{definition}[Bell Nonlocality.]
A perspectival theory is \textbf{Bell Nonlocal} just in case it leads to a normalized circuit model of the form (\ref{eq:contextuality}), where 
\begin{enumerate}
    \item each triplet of systems $(X_i, S_i, A_i)$ of $T$ can be validly embedded into mutually spacelike separated regions; and
    \item the resulting conditional probability distribution does not admit a local hidden variable model of the form (\ref{eq:hvm}).
\end{enumerate}
\end{definition}

To grasp the results of this paper, it will help to understand that Bell nonlocality is a kind of \textit{contextuality}.
For example, consider the classic Bell scenario, in which $n=2$ so we can write the distribution as $p(AB|XY)$. Suppose further that both settings $X$ and $Y$ are bits. 
Then it turns out that the existence of a hidden variable model of the form (\ref{eq:hvm}) is equivalent to global distribution $q(A_0A_1B_0B_1)$ that contains $p(AB|XY)$ in its marginals \cite{fine1982hidden, fine1982joint}.
Explicitly, $p(AB|XY)$ satisfies (\ref{eq:hvm}) if and only if
\begin{equation} \label{eq:proto_contextuality}
    \begin{split}
        p(A = a, B=b|X=0, Y=0) &= \sum_{A_1B_1}q(A_0=a, A_1, B_0=b, B_1) \\
        p(A = a, B=b|X=0, Y=1) &= \sum_{A_1B_0}q(A_0=a, A_1, B_0, B_1=b) \\
        p(A = a, B=b|X=1, Y=0) &= \sum_{A_0B_1}q(A_0, A_1=a, B_0=b, B_1) \\
        p(A = a, B=b|X=1, Y=1) &= \sum_{A_0B_0}q(A_0, A_1=a, B_0, B_1=b) 
    \end{split}
\end{equation}
for some $q(A_0A_1B_0B_1)$.\footnote{The proof of this theorem is not hard to grasp. The core point is that $q$ defines a hidden variable model in which each value of $\lambda$ determines each outcome precisely, and that, conversely, any such deterministic local hidden variable model defines a $q$. But \textit{any} local hidden variable model can be regarded as a probabilistic mixture of deterministic ones, and such a mixture is itself a deterministic local hidden variable model.}

This theorem generalizes to arbitrary $n$-partite Bell scenarios, meaning that the existence of a local hidden variable model $(\ref{eq:hvm})$ is equivalent to the following.
\begin{equation} \label{prob_contextuality} \tag{C1}
\begin{split}
\exists q: \forall a_1 \ldots a_n \forall x_1 \ldots x_n: \ \ \  &p(A_1=a_1\ldots A_n=a_n|X_1=x_1, \ldots X_n=x_n) \\
       &= \sum_{\overline{A_1^{x_1} \ldots A_n^{x_n}}}  q(A_1^1. \ldots  A_1^{x_1}=a_1 \ldots  A_1^{x_1^{\rm max}} \ldots A_n^1 \ldots A_n^{x_n}=a_n \ldots A_n^{x_n^{\rm max}}), \ \ 
\end{split}
\end{equation}

Thus Bell nonlocality can be understood as the impossibility of combining into a consistent whole all the data corresponding to different choices of measurements. Now, in a Bell experiment, only one choice can be made per run, meaning that this does not lead to any problems with the absoluteness of observed events. But we will soon see that, in a BIL theory, all of the different choices can effectively be made at once, leading to the impossibility of the various agents consistently combining their observations.

\paragraph{Quantum theory as a BIL theory.} First though, let us discuss how quantum theory can be understood as a BIL theory.
As discussed in Section \ref{sec:perspectival}, one can devise a quantum perspectival theory in which all memory updates are isometries. This theory is clearly Information Preserving, and will be Bell Nonlocal for any sensible specification of the valid embeddings. For the quantum perspectival theory to count as a BIL theory, it remains to establish that it has Local Dynamics. We postpone this until Section \ref{sec:binsc}, and for now ask the reader to trust us that quantum theory can be formulated as a BIL theory.

\subsection{A no-go theorem}

We aim to show that BIL theories are incompatible with the following assumption about reality.

\begin{definition}[Absoluteness of observed events (AOE).]
Every observed event is an absolute single event, not relative to anything or anyone.
\end{definition}

We take it for granted that a circuit with $N$ memory updates represents a scenario where $N$ observers have performed a measurement.\footnote{Of course, the scenario might be extremely difficult to realise -- in the quantum case, they might involve recohering large systems. But if we assume that they can be done in principle, then it follows that BIL theories in principle predict a breakdown in AOE.}
On any given run of this scenario, we take AOE to imply that there is a unique, global assignment of outcomes $\{C_i=c_i\}_{i=1}^N$ describing what each observer actually saw.

For our main result, we also assume that the relative frequencies of the global assignments converge to a probability distribution $q(\{C_i\}_i)$, and that an accurate perspectival theory would correctly predict some of the marginals of this distribution. In particular, if an accurate perspectival theory predicts a distribution for the measurements corresponding to some subset $S \subseteq \{C_i\}_{i=1}^N$, then this distribution must coincide with the corresponding marginal $q(S)$. All of this leads to the following theorem.

\begin{theorem}[BIL theories are incompatible with AOE.] \label{thm:BIL_AOE}
Any perspectival theory that is Bell Nonlocal, Information Preserving, and has Local Dynamics makes some predictions that are incompatible with AOE. 
\end{theorem}

Before proving the theorem, we make three brief comments on its implications.
Firstly, the theorem shows us that unitarity per se is not essential for a measurement problem -- what really matters about the unitary dynamics of quantum theory is that they preserve information.\footnote{A memory update being information-preserving does not logically imply that it is unitary or isometric even in the quantum case, let alone in more general theories. Consider, for example, the very trivial case of a non-isometric but information-preserving memory update that involves implementing an identity channel on the measured system and preparing the memory in a maximally mixed state. More generally, one can write down less silly information-preserving but non-isometric memory updates in which the memory system does become correlated with the measured system, but in which there is also some noise.} This contrasts both with the original Wigner's friend argument, in which unitarity seems to be at the heart of it, and the recent no-go theorems, in which unitary evolution is one of the assumptions used to derive the contradiction (but it corroborates the result from \cite{vilasini2019multi}).

Secondly, it is interesting that the theorem relies on both a locality assumption and a nonlocality assumption.
This suggests that the measurement problem of quantum theory is closely tied to the fact that it lives in a `sweet spot' in which there is enough nonlocality to violate Bell inequalities, but not enough to lead to a strong conflict with relativity theory. On one intuitive gloss \cite{shimony1984controllable}, the measurement problem is to do with the fact that quantum theory involves `passion', but not `action', at a distance.

Finally, it is interesting to note that for the nonlocality assumption, it suffices to assume that the theory violates Bell inequalities. This contrasts with some of the other no-go results, which only go through because quantum theory is nonlocal in the strictly stronger sense of violating Local Friendliness inequalities \cite{bong2020strong},  or of being \textit{possibilistically} Bell nonlocal (see Section \ref{sec:logical}) \cite{Frauchiger2018quantum, leegwater2022greenberger, haddara2022possibilistic, ormrod2022no}. A key lesson here is that nonlocality in Bell's sense is always enough to ensure a breakdown of AOE, at least when combined with some other reasonable assumptions.

\subsection{A proof: constructing a measurement problem}

This subsection will prove Theorem \ref{thm:BIL_AOE} by constructing a circuit model in an arbitrary BIL theory that is inconsistent with AOE.
We will do this explicitly for a theory that yields a nonlocal probability distribution $p(AB|XY)$ for two outcome variables and binary choices, before explaining how the argument generalizes.
Our proof will be diagrammatic, but perfectly formal -- everything we will do can be directly translated into standard linear algebra by interpreting putting boxes one after the other and next to each other as performing the $\circ$ and $\otimes$ operations respectively.

We will start by assuming that we are dealing with a BIL theory that violates a Bell inequality involving just two parties that each has two choices of measurement. For such a theory, we will construct a circuit model representing a similar situation to the one from Section \ref{sec:simple_simple}, and derive a contradiction with AOE. Then we will generalize the argument to arbitrary BIL theories.

\paragraph{Constructing the model.} Suppose some BIL theory provides a Bell nonlocal conditional distribution (i.e. one that does not satisfy (\ref{eq:hvm})) in the following way:
\begin{equation} \label{model}
    p(AB|XY)=\begin{tikzpicture}
	\begin{pgfonlayer}{nodelayer}
		\node [style=none] (0) at (-0.75, 3) {};
		\node [style=none] (1) at (-0.75, -2) {};
		\node [style=none] (2) at (0.75, 3) {};
		\node [style=none] (3) at (0.75, -2) {};
		\node [style=none] (4) at (0, 3) {$T$};
		\node [style=none] (5) at (-1.25, -1.25) {$S_1$};
		\node [style=none] (6) at (-1.25, 4.5) {$A$};
		\node [style=none] (7) at (1.25, -1.25) {$S_2$};
		\node [style=none] (8) at (1.25, 4.5) {$B$};
		\node [style=large map] (9) at (0, 3) {$T$};
		\node [style=none] (10) at (-1.25, -2) {};
		\node [style=none] (11) at (1.25, -2) {};
		\node [style=none] (12) at (-1.25, -2.75) {};
		\node [style=none] (13) at (1.25, -2.75) {};
		\node [style=none] (14) at (-0.75, -2) {};
		\node [style=none] (15) at (0, -2.5) {$\phi$};
		\node [style=none] (16) at (0.75, -2) {};
		\node [style=none] (17) at (-0.75, 5) {};
		\node [style=none] (18) at (0.75, 5) {};
		\node [style=none] (20) at (-1.25, 2.25) {};
		\node [style=none] (21) at (-3, -2.5) {};
		\node [style=none] (22) at (1.5, 3) {};
		\node [style=none] (24) at (-1.25, 3.25) {};
		\node [style=none] (25) at (-3, -4.25) {};
		\node [style=none] (26) at (1.25, 2.25) {};
		\node [style=none] (27) at (3, -2.5) {};
		\node [style=none] (28) at (1.25, 3.25) {};
		\node [style=none] (29) at (3, -4.25) {};
		\node [style=none] (31) at (-3.5, -3.75) {$X$};
		\node [style=none] (32) at (3.5, -3.75) {$Y$};
	\end{pgfonlayer}
	\begin{pgfonlayer}{edgelayer}
		\draw (1.center) to (0.center);
		\draw (3.center) to (2.center);
		\draw (10.center) to (11.center);
		\draw (10.center) to (12.center);
		\draw [bend right=90, looseness=0.50] (12.center) to (13.center);
		\draw (13.center) to (11.center);
		\draw [style=dashed] (0.center) to (17.center);
		\draw [style=dashed] (2.center) to (18.center);
		\draw [style=dashed, in=90, out=-90, looseness=1.50] (20.center) to (21.center);
		\draw [style=dashed] (20.center) to (24.center);
		\draw [style=dashed] (25.center) to (21.center);
		\draw [style=dashed, in=90, out=-90, looseness=1.50] (26.center) to (27.center);
		\draw [style=dashed] (26.center) to (28.center);
		\draw [style=dashed] (29.center) to (27.center);
	\end{pgfonlayer}
\end{tikzpicture}
\end{equation}
where $(S_1, A, X)$ and $(S_2, Y, B)$ admit a valid embedding into spacelike separated regions, and $X$ and $Y$ are binary variables.

This circuit model provides raw materials which, with the help of Information Preservation and Local Dynamics, can be used to construct a scenario much like the one in Figure \ref{fig:experiment_schematic}, in which the nonlocality of $p(AB|XY)$ is converted into the global inconsistency of four agents' results.
The first step is to use Local Dynamics to rewrite the circuit.
\begin{equation} \label{eq:contextuality_two_party}
    \begin{tikzpicture}
	\begin{pgfonlayer}{nodelayer}
		\node [style=none] (0) at (-0.75, 3) {};
		\node [style=none] (1) at (-0.75, -2) {};
		\node [style=none] (2) at (0.75, 3) {};
		\node [style=none] (3) at (0.75, -2) {};
		\node [style=none] (4) at (0, 3) {$T$};
		\node [style=none] (5) at (-1.25, -1.25) {$S_1$};
		\node [style=none] (6) at (-1.25, 4.5) {$A$};
		\node [style=none] (7) at (1.25, -1.25) {$S_2$};
		\node [style=none] (8) at (1.25, 4.5) {$B$};
		\node [style=large map] (9) at (0, 3) {$T$};
		\node [style=none] (10) at (-1.25, -2) {};
		\node [style=none] (11) at (1.25, -2) {};
		\node [style=none] (12) at (-1.25, -2.75) {};
		\node [style=none] (13) at (1.25, -2.75) {};
		\node [style=none] (14) at (-0.75, -2) {};
		\node [style=none] (15) at (0, -2.5) {$\phi$};
		\node [style=none] (16) at (0.75, -2) {};
		\node [style=none] (17) at (-0.75, 5) {};
		\node [style=none] (18) at (0.75, 5) {};
		\node [style=none] (20) at (-1.25, 2.25) {};
		\node [style=none] (21) at (-3, -2.5) {};
		\node [style=none] (22) at (1.5, 3) {};
		\node [style=none] (24) at (-1.25, 3.25) {};
		\node [style=none] (25) at (-3, -4.25) {};
		\node [style=none] (26) at (1.25, 2.25) {};
		\node [style=none] (27) at (3, -2.5) {};
		\node [style=none] (28) at (1.25, 3.25) {};
		\node [style=none] (29) at (3, -4.25) {};
		\node [style=none] (31) at (-3.5, -3.75) {$X$};
		\node [style=none] (32) at (3.5, -3.75) {$Y$};
	\end{pgfonlayer}
	\begin{pgfonlayer}{edgelayer}
		\draw (1.center) to (0.center);
		\draw (3.center) to (2.center);
		\draw (10.center) to (11.center);
		\draw (10.center) to (12.center);
		\draw [bend right=90, looseness=0.50] (12.center) to (13.center);
		\draw (13.center) to (11.center);
		\draw [style=dashed] (0.center) to (17.center);
		\draw [style=dashed] (2.center) to (18.center);
		\draw [style=dashed, in=90, out=-90, looseness=1.50] (20.center) to (21.center);
		\draw [style=dashed] (20.center) to (24.center);
		\draw [style=dashed] (25.center) to (21.center);
		\draw [style=dashed, in=90, out=-90, looseness=1.50] (26.center) to (27.center);
		\draw [style=dashed] (26.center) to (28.center);
		\draw [style=dashed] (29.center) to (27.center);
	\end{pgfonlayer}
\end{tikzpicture} = \begin{tikzpicture}
	\begin{pgfonlayer}{nodelayer}
		\node [style=none] (0) at (-0.75, 3) {};
		\node [style=none] (1) at (-0.75, -2) {};
		\node [style=none] (3) at (0.75, -2) {};
		\node [style=none] (5) at (-1.25, -1.25) {$S_1$};
		\node [style=none] (6) at (-1.25, 4.5) {$A$};
		\node [style=none] (7) at (1.25, -1.25) {$S_2$};
		\node [style=none] (8) at (4.25, 4.5) {$B$};
		\node [style=none] (10) at (-1.25, -2) {};
		\node [style=none] (11) at (1.25, -2) {};
		\node [style=none] (12) at (-1.25, -2.75) {};
		\node [style=none] (13) at (1.25, -2.75) {};
		\node [style=none] (14) at (-0.75, -2) {};
		\node [style=none] (15) at (0, -2.5) {$\phi$};
		\node [style=none] (16) at (0.75, -2) {};
		\node [style=none] (17) at (-0.75, 5) {};
		\node [style=none] (20) at (-1.25, 2.25) {};
		\node [style=none] (21) at (-3, -2.5) {};
		\node [style=none] (24) at (-1.25, 3.25) {};
		\node [style=none] (25) at (-3, -4.25) {};
		\node [style=none] (26) at (4.25, 2.25) {};
		\node [style=none] (27) at (6, -2.5) {};
		\node [style=none] (28) at (4.25, 3.25) {};
		\node [style=none] (29) at (6, -4.25) {};
		\node [style=none] (31) at (-3.5, -3.75) {$X$};
		\node [style=none] (32) at (6.5, -3.75) {$Y$};
		\node [style=none] (33) at (2.25, -2) {};
		\node [style=none] (34) at (3.75, -2) {};
		\node [style=none] (35) at (1.75, -2) {};
		\node [style=none] (36) at (4.25, -2) {};
		\node [style=none] (37) at (1.75, -2.75) {};
		\node [style=none] (38) at (4.25, -2.75) {};
		\node [style=none] (39) at (2.25, -2) {};
		\node [style=none] (40) at (3, -2.5) {$\psi$};
		\node [style=semilarge map] (43) at (-0.75, 3) {$T_1$};
		\node [style=semilarge map] (44) at (3.75, 3) {$T_2$};
		\node [style=none] (45) at (-0.25, 3) {};
		\node [style=none] (46) at (2.25, -2) {};
		\node [style=none] (47) at (2.25, -2) {};
		\node [style=none] (49) at (3.75, 3) {};
		\node [style=none] (50) at (3.75, -2) {};
		\node [style=none] (51) at (3.75, -2) {};
		\node [style=none] (57) at (0.75, -2) {};
		\node [style=none] (59) at (0.75, -2) {};
		\node [style=none] (60) at (3.25, 3) {};
		\node [style=none] (61) at (0.75, -2) {};
		\node [style=none] (62) at (0.75, -2) {};
		\node [style=none] (65) at (3.75, 5) {};
	\end{pgfonlayer}
	\begin{pgfonlayer}{edgelayer}
		\draw (1.center) to (0.center);
		\draw (10.center) to (11.center);
		\draw (10.center) to (12.center);
		\draw [bend right=90, looseness=0.50] (12.center) to (13.center);
		\draw (13.center) to (11.center);
		\draw [style=dashed] (0.center) to (17.center);
		\draw [style=dashed, in=90, out=-90, looseness=1.50] (20.center) to (21.center);
		\draw [style=dashed] (20.center) to (24.center);
		\draw [style=dashed] (25.center) to (21.center);
		\draw [style=dashed, in=90, out=-90, looseness=1.50] (26.center) to (27.center);
		\draw [style=dashed] (26.center) to (28.center);
		\draw [style=dashed] (29.center) to (27.center);
		\draw (35.center) to (36.center);
		\draw (35.center) to (37.center);
		\draw [bend right=90, looseness=0.50] (37.center) to (38.center);
		\draw (38.center) to (36.center);
		\draw [in=-90, out=90, looseness=1.50] (46.center) to (45.center);
		\draw (50.center) to (49.center);
		\draw [in=-90, out=90, looseness=1.50] (61.center) to (60.center);
		\draw [style=dashed] (49.center) to (65.center);
	\end{pgfonlayer}
\end{tikzpicture}
\end{equation}
Next, we streamline notation.
\begin{equation} \label{eq:E_def}
    \begin{tikzpicture}
	\begin{pgfonlayer}{nodelayer}
		\node [style=none] (0) at (-0.75, 3) {};
		\node [style=none] (1) at (-0.75, 0) {};
		\node [style=none] (17) at (-0.75, 5) {};
		\node [style=none] (20) at (-1.75, 2.25) {};
		\node [style=none] (21) at (-1.75, 0.75) {};
		\node [style=none] (24) at (-1.75, 3.25) {};
		\node [style=semilarge map] (43) at (-0.75, 3) {$T_1$};
		\node [style=none] (45) at (0.25, 3) {};
		\node [style=none] (46) at (0.25, 0) {};
		\node [style=point] (47) at (-1.75, 0.75) {$i$};
		\node [style=map] (48) at (3.75, 3) {$E_1^i$};
		\node [style=none] (49) at (3.25, 3) {};
		\node [style=none] (50) at (4.25, 3) {};
		\node [style=none] (51) at (3.25, 0) {};
		\node [style=none] (52) at (4.25, 0) {};
		\node [style=none] (53) at (3.75, 5) {};
		\node [style=none] (54) at (1.75, 3) {$=:$};
	\end{pgfonlayer}
	\begin{pgfonlayer}{edgelayer}
		\draw (1.center) to (0.center);
		\draw [style=dashed] (0.center) to (17.center);
		\draw [style=dashed, in=90, out=-90, looseness=1.50] (20.center) to (21.center);
		\draw [style=dashed] (20.center) to (24.center);
		\draw [style=none] (46.center) to (45.center);
		\draw [style=none] (51.center) to (49.center);
		\draw [style=none] (52.center) to (50.center);
		\draw [style=dashed] (53.center) to (48);
	\end{pgfonlayer}
\end{tikzpicture}, \quad \quad \quad \quad \quad  \begin{tikzpicture}
	\begin{pgfonlayer}{nodelayer}
		\node [style=none] (0) at (-0.75, 3) {};
		\node [style=none] (1) at (-0.75, 0) {};
		\node [style=none] (17) at (-0.75, 5) {};
		\node [style=none] (20) at (0.25, 2.25) {};
		\node [style=none] (21) at (0.25, 0.75) {};
		\node [style=none] (24) at (0.25, 3.25) {};
		\node [style=semilarge map] (43) at (-0.75, 3) {$T_2$};
		\node [style=none] (45) at (-1.75, 3) {};
		\node [style=none] (46) at (-1.75, 0) {};
		\node [style=point] (47) at (0.25, 0.75) {$i$};
		\node [style=map] (48) at (3.75, 3) {$E_2^i$};
		\node [style=none] (49) at (4.25, 3) {};
		\node [style=none] (50) at (3.25, 3) {};
		\node [style=none] (51) at (4.25, 0) {};
		\node [style=none] (52) at (3.25, 0) {};
		\node [style=none] (53) at (3.75, 5) {};
		\node [style=none] (54) at (1.75, 3) {$=:$};
	\end{pgfonlayer}
	\begin{pgfonlayer}{edgelayer}
		\draw (1.center) to (0.center);
		\draw [style=dashed] (0.center) to (17.center);
		\draw [style=dashed, in=90, out=-90, looseness=1.50] (20.center) to (21.center);
		\draw [style=dashed] (20.center) to (24.center);
		\draw [style=none] (46.center) to (45.center);
		\draw [style=none] (51.center) to (49.center);
		\draw [style=none] (52.center) to (50.center);
		\draw [style=dashed] (53.center) to (48);
	\end{pgfonlayer}
\end{tikzpicture}
\end{equation}
where $i \in {1, 2}$ labels the two possible measurement settings for each party. Note that each $E_j^i$ is a probability extractor, and therefore comes with an associated memory update. 
Defining $U_j^1 := f(E_j^1)$ as the memory update for $E_j^1$, we consider the following circuit.
\begin{equation} \label{two_mus}
    \begin{tikzpicture}
	\begin{pgfonlayer}{nodelayer}
		\node [style=none] (0) at (-2.25, 3.25) {};
		\node [style=none] (1) at (-2.25, -1.75) {};
		\node [style=none] (2) at (-0.75, -1.75) {};
		\node [style=none] (7) at (-2.75, -1.75) {};
		\node [style=none] (8) at (-0.25, -1.75) {};
		\node [style=none] (9) at (-2.75, -2.5) {};
		\node [style=none] (10) at (-0.25, -2.5) {};
		\node [style=none] (11) at (-2.25, -1.75) {};
		\node [style=none] (12) at (-1.5, -2.25) {$\phi$};
		\node [style=none] (13) at (-0.75, -1.75) {};
		\node [style=none] (14) at (-2.25, 5.25) {};
		\node [style=none] (25) at (0.75, -1.75) {};
		\node [style=none] (26) at (2.25, -1.75) {};
		\node [style=none] (27) at (0.25, -1.75) {};
		\node [style=none] (28) at (2.75, -1.75) {};
		\node [style=none] (29) at (0.25, -2.5) {};
		\node [style=none] (30) at (2.75, -2.5) {};
		\node [style=none] (31) at (0.75, -1.75) {};
		\node [style=none] (32) at (1.5, -2.25) {$\psi$};
		\node [style=none] (35) at (-1.75, 3.25) {};
		\node [style=none] (36) at (0.75, -1.75) {};
		\node [style=none] (37) at (0.75, -1.75) {};
		\node [style=none] (39) at (2.25, -1.75) {};
		\node [style=none] (40) at (2.25, -1.75) {};
		\node [style=none] (41) at (-0.75, -1.75) {};
		\node [style=none] (42) at (-0.75, -1.75) {};
		\node [style=none] (43) at (1.75, 3.25) {};
		\node [style=none] (44) at (-0.75, -1.75) {};
		\node [style=none] (45) at (-0.75, -1.75) {};
		\node [style=none] (46) at (2.25, 5.25) {};
		\node [style=none] (47) at (-3, 3.25) {};
		\node [style=large map] (49) at (-3, 3.25) {$U_1^1$};
		\node [style=none] (50) at (-1.75, 5.25) {};
		\node [style=none] (51) at (-4.25, 3.25) {};
		\node [style=none] (52) at (-4.25, 5.25) {};
		\node [style=none] (53) at (-4.75, 4.75) {};
		\node [style=none] (54) at (-5, 4.75) {$M_1^1$};
		\node [style=large map] (55) at (3, 3.25) {$U_2^1$};
		\node [style=none] (56) at (2.25, 3.25) {};
		\node [style=none] (57) at (4.25, 3.25) {};
		\node [style=none] (58) at (4.25, 5.25) {};
		\node [style=none] (59) at (5, 4.75) {$M_2^1$};
		\node [style=none] (60) at (1.75, 5.25) {};
	\end{pgfonlayer}
	\begin{pgfonlayer}{edgelayer}
		\draw (1.center) to (0.center);
		\draw (7.center) to (8.center);
		\draw (7.center) to (9.center);
		\draw [bend right=90, looseness=0.50] (9.center) to (10.center);
		\draw (10.center) to (8.center);
		\draw (27.center) to (28.center);
		\draw (27.center) to (29.center);
		\draw [bend right=90, looseness=0.50] (29.center) to (30.center);
		\draw (30.center) to (28.center);
		\draw [in=-90, out=90, looseness=1.50] (36.center) to (35.center);
		\draw [in=-90, out=90, looseness=1.50] (44.center) to (43.center);
		\draw (0.center) to (14.center);
		\draw (35.center) to (50.center);
		\draw (52.center) to (51.center);
		\draw (40.center) to (46.center);
		\draw (57.center) to (58.center);
		\draw (43.center) to (60.center);
	\end{pgfonlayer}
\end{tikzpicture}
\end{equation}

This circuit describes a pair of measurements corresponding to  $X=1$ and $Y=1$ from an outsider's perspective.
We now want to find a way of effectively performing a pair of measurements corresponding to  $X=2$ and $Y=2$ on the original systems even after these first two measurements have taken place.

To do this, we need to find appropriate supermeasurements on the outputs of each memory update $U_j^i$.
Information Preservation guarantees that such supermeasurements exist. For example, the following circuits are equivalent, for some probability extractor $\tilde{E}_1^2$.
\begin{equation} \label{supermeasurement_one}
   \begin{tikzpicture}
	\begin{pgfonlayer}{nodelayer}
		\node [style=none] (0) at (3.5, 3.25) {};
		\node [style=none] (1) at (3.5, -1.75) {};
		\node [style=none] (2) at (5, -1.75) {};
		\node [style=none] (7) at (3, -1.75) {};
		\node [style=none] (8) at (5.5, -1.75) {};
		\node [style=none] (9) at (3, -2.5) {};
		\node [style=none] (10) at (5.5, -2.5) {};
		\node [style=none] (11) at (3.5, -1.75) {};
		\node [style=none] (12) at (4.25, -2.25) {$\phi$};
		\node [style=none] (13) at (5, -1.75) {};
		\node [style=none] (14) at (3.5, 5.25) {};
		\node [style=none] (25) at (6.5, -1.75) {};
		\node [style=none] (26) at (8, -1.75) {};
		\node [style=none] (27) at (6, -1.75) {};
		\node [style=none] (28) at (8.5, -1.75) {};
		\node [style=none] (29) at (6, -2.5) {};
		\node [style=none] (30) at (8.5, -2.5) {};
		\node [style=none] (31) at (6.5, -1.75) {};
		\node [style=none] (32) at (7.25, -2.25) {$\psi$};
		\node [style=none] (35) at (4, 3.25) {};
		\node [style=none] (36) at (6.5, -1.75) {};
		\node [style=none] (37) at (6.5, -1.75) {};
		\node [style=none] (39) at (8, -1.75) {};
		\node [style=none] (40) at (8, -1.75) {};
		\node [style=none] (41) at (5, -1.75) {};
		\node [style=none] (42) at (5, -1.75) {};
		\node [style=none] (43) at (7.5, 3.25) {};
		\node [style=none] (44) at (5, -1.75) {};
		\node [style=none] (45) at (5, -1.75) {};
		\node [style=none] (46) at (8, 7) {};
		\node [style=none] (47) at (2.75, 3.25) {};
		\node [style=large map] (49) at (2.75, 3.25) {$U_1^1$};
		\node [style=none] (50) at (4, 5.25) {};
		\node [style=none] (51) at (1.5, 3.25) {};
		\node [style=none] (52) at (1.5, 5.25) {};
		\node [style=none] (53) at (1, 4.75) {};
		\node [style=none] (56) at (8, 3.25) {};
		\node [style=none] (60) at (7.5, 7) {};
		\node [style=large map] (61) at (2.75, 5.25) {$\tilde{E}_1^2$};
		\node [style=none] (62) at (2.75, 5.25) {};
		\node [style=none] (63) at (2.75, 7) {};
		\node [style=none] (64) at (2.75, 5.25) {};
		\node [style=map] (65) at (-6.5, 3.25) {$E_1^2$};
		\node [style=none] (66) at (-6.75, 3.25) {};
		\node [style=none] (67) at (-6.75, -1.75) {};
		\node [style=none] (68) at (-5.25, -1.75) {};
		\node [style=none] (69) at (-7.25, -1.75) {};
		\node [style=none] (70) at (-4.75, -1.75) {};
		\node [style=none] (71) at (-7.25, -2.5) {};
		\node [style=none] (72) at (-4.75, -2.5) {};
		\node [style=none] (73) at (-6.75, -1.75) {};
		\node [style=none] (74) at (-6, -2.25) {$\phi$};
		\node [style=none] (75) at (-5.25, -1.75) {};
		\node [style=none] (77) at (-3.75, -1.75) {};
		\node [style=none] (78) at (-2.25, -1.75) {};
		\node [style=none] (79) at (-4.25, -1.75) {};
		\node [style=none] (80) at (-1.75, -1.75) {};
		\node [style=none] (81) at (-4.25, -2.5) {};
		\node [style=none] (82) at (-1.75, -2.5) {};
		\node [style=none] (83) at (-3.75, -1.75) {};
		\node [style=none] (84) at (-3, -2.25) {$\psi$};
		\node [style=none] (85) at (-6.25, 3.25) {};
		\node [style=none] (86) at (-3.75, -1.75) {};
		\node [style=none] (87) at (-3.75, -1.75) {};
		\node [style=none] (88) at (-2.25, -1.75) {};
		\node [style=none] (89) at (-2.25, -1.75) {};
		\node [style=none] (90) at (-5.25, -1.75) {};
		\node [style=none] (91) at (-5.25, -1.75) {};
		\node [style=none] (92) at (-2.75, 3.25) {};
		\node [style=none] (93) at (-5.25, -1.75) {};
		\node [style=none] (94) at (-5.25, -1.75) {};
		\node [style=none] (95) at (-2.25, 7) {};
		\node [style=none] (96) at (-7.5, 3.25) {};
		\node [style=none] (99) at (-6.5, 4) {};
		\node [style=none] (102) at (-2.25, 3.25) {};
		\node [style=none] (103) at (-2.75, 7) {};
		\node [style=none] (106) at (-6.5, 7) {};
		\node [style=none] (107) at (0, 3) {$=$};
	\end{pgfonlayer}
	\begin{pgfonlayer}{edgelayer}
		\draw (1.center) to (0.center);
		\draw (7.center) to (8.center);
		\draw (7.center) to (9.center);
		\draw [bend right=90, looseness=0.50] (9.center) to (10.center);
		\draw (10.center) to (8.center);
		\draw (27.center) to (28.center);
		\draw (27.center) to (29.center);
		\draw [bend right=90, looseness=0.50] (29.center) to (30.center);
		\draw (30.center) to (28.center);
		\draw [in=-90, out=90, looseness=1.50] (36.center) to (35.center);
		\draw [in=-90, out=90, looseness=1.50] (44.center) to (43.center);
		\draw (0.center) to (14.center);
		\draw (35.center) to (50.center);
		\draw (52.center) to (51.center);
		\draw (40.center) to (46.center);
		\draw (43.center) to (60.center);
		\draw [style=dashed] (64.center) to (63.center);
		\draw (67.center) to (66.center);
		\draw (69.center) to (70.center);
		\draw (69.center) to (71.center);
		\draw [bend right=90, looseness=0.50] (71.center) to (72.center);
		\draw (72.center) to (70.center);
		\draw (79.center) to (80.center);
		\draw (79.center) to (81.center);
		\draw [bend right=90, looseness=0.50] (81.center) to (82.center);
		\draw (82.center) to (80.center);
		\draw [in=-90, out=90, looseness=1.50] (86.center) to (85.center);
		\draw [in=-90, out=90, looseness=1.50] (93.center) to (92.center);
		\draw (89.center) to (95.center);
		\draw (92.center) to (103.center);
		\draw [style=dashed] (65) to (106.center);
	\end{pgfonlayer}
\end{tikzpicture}
\end{equation}
We define $U_1^2:= f(\tilde{E}_1^2)$. Setting $U_2^2:= f(\tilde{E}_2^2)$ for a similarly defined $\tilde{E}_2^2$, we construct the following circuit.
\begin{equation} \label{eq:four_mus}
    \begin{tikzpicture}
	\begin{pgfonlayer}{nodelayer}
		\node [style=none] (0) at (-2.25, 3) {};
		\node [style=none] (1) at (-2.25, -2) {};
		\node [style=none] (2) at (-0.75, -2) {};
		\node [style=none] (3) at (-2.75, -2) {};
		\node [style=none] (4) at (-0.25, -2) {};
		\node [style=none] (5) at (-2.75, -2.75) {};
		\node [style=none] (6) at (-0.25, -2.75) {};
		\node [style=none] (7) at (-2.25, -2) {};
		\node [style=none] (8) at (-1.5, -2.5) {$\phi$};
		\node [style=none] (9) at (-0.75, -2) {};
		\node [style=none] (10) at (-2.25, 8.5) {};
		\node [style=none] (11) at (0.75, -2) {};
		\node [style=none] (12) at (2.25, -2) {};
		\node [style=none] (13) at (0.25, -2) {};
		\node [style=none] (14) at (2.75, -2) {};
		\node [style=none] (15) at (0.25, -2.75) {};
		\node [style=none] (16) at (2.75, -2.75) {};
		\node [style=none] (17) at (0.75, -2) {};
		\node [style=none] (18) at (1.5, -2.5) {$\psi$};
		\node [style=none] (19) at (-1.75, 3) {};
		\node [style=none] (20) at (0.75, -2) {};
		\node [style=none] (21) at (0.75, -2) {};
		\node [style=none] (22) at (2.25, -2) {};
		\node [style=none] (23) at (2.25, -2) {};
		\node [style=none] (24) at (-0.75, -2) {};
		\node [style=none] (25) at (-0.75, -2) {};
		\node [style=none] (26) at (1.75, 3) {};
		\node [style=none] (27) at (-0.75, -2) {};
		\node [style=none] (28) at (-0.75, -2) {};
		\node [style=none] (29) at (2.25, 8.5) {};
		\node [style=none] (30) at (-3, 3) {};
		\node [style=large map] (31) at (-3, 3) {$U_1^1$};
		\node [style=none] (32) at (-1.75, 8.5) {};
		\node [style=none] (33) at (-4.25, 3) {};
		\node [style=none] (34) at (-4.25, 8.5) {};
		\node [style=large map] (37) at (3, 3) {$U_2^1$};
		\node [style=none] (38) at (2.25, 3) {};
		\node [style=none] (39) at (4.25, 3) {};
		\node [style=none] (40) at (4.25, 8.5) {};
		\node [style=none] (42) at (1.75, 8.5) {};
		\node [style=large map] (43) at (-3, 6.5) {$U_1^2$};
		\node [style=none] (44) at (-2.75, 6.75) {};
		\node [style=none] (45) at (-2.75, 8.5) {};
		\node [style=large map] (46) at (3, 6.5) {$U_2^2$};
		\node [style=none] (47) at (2.75, 6.75) {};
		\node [style=none] (48) at (2.75, 8.5) {};
	\end{pgfonlayer}
	\begin{pgfonlayer}{edgelayer}
		\draw (1.center) to (0.center);
		\draw (3.center) to (4.center);
		\draw (3.center) to (5.center);
		\draw [bend right=90, looseness=0.50] (5.center) to (6.center);
		\draw (6.center) to (4.center);
		\draw (13.center) to (14.center);
		\draw (13.center) to (15.center);
		\draw [bend right=90, looseness=0.50] (15.center) to (16.center);
		\draw (16.center) to (14.center);
		\draw [in=-90, out=90, looseness=1.50] (20.center) to (19.center);
		\draw [in=-90, out=90, looseness=1.50] (27.center) to (26.center);
		\draw (0.center) to (10.center);
		\draw (19.center) to (32.center);
		\draw (34.center) to (33.center);
		\draw (23.center) to (29.center);
		\draw (39.center) to (40.center);
		\draw (26.center) to (42.center);
		\draw (44.center) to (45.center);
		\draw (47.center) to (48.center);
	\end{pgfonlayer}
\end{tikzpicture}
\end{equation}

\paragraph{Derving a contradiction with AOE.} There are four memory updates in our circuit, hence four measurements in the scenario it represents. By assumption, there exists a probability distribution $q(C_1^1, C_1^2, C_2^1, C_2^2)$ over the four outcomes.

Our theory does not directly predict any such joint distribution. But, as we shall see, it does make predictions for each of the four marginals $\sum_{C_1^i C_2^j}q(C_1^1, C_1^2, C_2^1, C_2^2)$. We will show that these predictions are inconsistent: there exists no $q$ that gives rise to each of the four predicted marginals.

First of all, we can apply (\ref{eq:inside_pair}) to find our theory's predictions for the marginal distribution $q(C_1^1, C_2^1) := \sum_{C_1^2, C_2^2} q(C_1^1, C_1^2, C_2^1, C_2^2)$.
This involves applying the $f^{-1}$ to $U_1^1$ and $U_2^1$.
It is easily seen that this prediction is given by our original conditional distribution, for the special case where $X=Y=1$. We therefore infer:
\begin{equation} \label{con1}
    p(A=a, B=b|X=1, Y=1) = \sum_{C_1^2, C_2^2} q(C_1^1=a, C_1^2, C_2^1=b, C_2^2)
\end{equation} 

We can also calculate joint probabilities for the measurements associated with $U_1^2$ and $U_2^1.$ By the interchange law (\ref{eq:interchange}), (\ref{eq:four_mus}) is equivalent to the following.
\begin{equation} \label{commute}
    \begin{tikzpicture}
	\begin{pgfonlayer}{nodelayer}
		\node [style=none] (0) at (3.5, 3) {};
		\node [style=none] (1) at (3.5, -2) {};
		\node [style=none] (2) at (5, -2) {};
		\node [style=none] (3) at (3, -2) {};
		\node [style=none] (4) at (5.5, -2) {};
		\node [style=none] (5) at (3, -2.75) {};
		\node [style=none] (6) at (5.5, -2.75) {};
		\node [style=none] (7) at (3.5, -2) {};
		\node [style=none] (8) at (4.25, -2.5) {$\phi$};
		\node [style=none] (9) at (5, -2) {};
		\node [style=none] (10) at (3.5, 12.25) {};
		\node [style=none] (11) at (6.5, -2) {};
		\node [style=none] (12) at (8, -2) {};
		\node [style=none] (13) at (6, -2) {};
		\node [style=none] (14) at (8.5, -2) {};
		\node [style=none] (15) at (6, -2.75) {};
		\node [style=none] (16) at (8.5, -2.75) {};
		\node [style=none] (17) at (6.5, -2) {};
		\node [style=none] (18) at (7.25, -2.5) {$\psi$};
		\node [style=none] (19) at (4, 3) {};
		\node [style=none] (20) at (6.5, -2) {};
		\node [style=none] (21) at (6.5, -2) {};
		\node [style=none] (22) at (8, -2) {};
		\node [style=none] (23) at (8, -2) {};
		\node [style=none] (24) at (5, -2) {};
		\node [style=none] (25) at (5, -2) {};
		\node [style=none] (26) at (7.5, 3) {};
		\node [style=none] (27) at (5, -2) {};
		\node [style=none] (28) at (5, -2) {};
		\node [style=none] (29) at (8, 12.25) {};
		\node [style=none] (30) at (2.75, 3) {};
		\node [style=large map] (31) at (2.75, 3) {$U_1^1$};
		\node [style=none] (32) at (4, 12.25) {};
		\node [style=none] (33) at (1.5, 3) {};
		\node [style=none] (34) at (1.5, 12.25) {};
		\node [style=large map] (37) at (8.75, 6.5) {$U_2^1$};
		\node [style=none] (38) at (8, 6.5) {};
		\node [style=none] (39) at (10, 6.5) {};
		\node [style=none] (40) at (10, 12.25) {};
		\node [style=none] (42) at (7.5, 12.25) {};
		\node [style=large map] (43) at (2.75, 6.5) {$U_1^2$};
		\node [style=none] (44) at (3, 6.75) {};
		\node [style=none] (45) at (3, 12.25) {};
		\node [style=large map] (46) at (8.75, 10.25) {$U_2^2$};
		\node [style=none] (47) at (8.5, 10.5) {};
		\node [style=none] (48) at (8.5, 12.25) {};
		\node [style=none] (49) at (-8.5, 3) {};
		\node [style=none] (50) at (-8.5, -2) {};
		\node [style=none] (51) at (-7, -2) {};
		\node [style=none] (52) at (-9, -2) {};
		\node [style=none] (53) at (-6.5, -2) {};
		\node [style=none] (54) at (-9, -2.75) {};
		\node [style=none] (55) at (-6.5, -2.75) {};
		\node [style=none] (56) at (-8.5, -2) {};
		\node [style=none] (57) at (-7.75, -2.5) {$\phi$};
		\node [style=none] (58) at (-7, -2) {};
		\node [style=none] (59) at (-8.5, 12.25) {};
		\node [style=none] (60) at (-5.5, -2) {};
		\node [style=none] (61) at (-4, -2) {};
		\node [style=none] (62) at (-6, -2) {};
		\node [style=none] (63) at (-3.5, -2) {};
		\node [style=none] (64) at (-6, -2.75) {};
		\node [style=none] (65) at (-3.5, -2.75) {};
		\node [style=none] (66) at (-5.5, -2) {};
		\node [style=none] (67) at (-4.75, -2.5) {$\psi$};
		\node [style=none] (68) at (-8, 3) {};
		\node [style=none] (69) at (-5.5, -2) {};
		\node [style=none] (70) at (-5.5, -2) {};
		\node [style=none] (71) at (-4, -2) {};
		\node [style=none] (72) at (-4, -2) {};
		\node [style=none] (73) at (-7, -2) {};
		\node [style=none] (74) at (-7, -2) {};
		\node [style=none] (75) at (-4.5, 3) {};
		\node [style=none] (76) at (-7, -2) {};
		\node [style=none] (77) at (-7, -2) {};
		\node [style=none] (78) at (-4, 12.25) {};
		\node [style=none] (79) at (-9.25, 3) {};
		\node [style=large map] (80) at (-9.25, 3) {$U_1^1$};
		\node [style=none] (81) at (-8, 12.25) {};
		\node [style=none] (82) at (-10.5, 3) {};
		\node [style=none] (83) at (-10.5, 12.25) {};
		\node [style=large map] (84) at (-3.25, 3) {$U_2^1$};
		\node [style=none] (85) at (-4, 3) {};
		\node [style=none] (86) at (-2, 3) {};
		\node [style=none] (87) at (-2, 12.25) {};
		\node [style=none] (88) at (-4.5, 12.25) {};
		\node [style=large map] (89) at (-9.25, 6.5) {$U_1^2$};
		\node [style=none] (90) at (-9, 6.75) {};
		\node [style=none] (91) at (-9, 12.25) {};
		\node [style=large map] (92) at (-3.25, 6.5) {$U_2^2$};
		\node [style=none] (93) at (-3.5, 6.75) {};
		\node [style=none] (94) at (-3.5, 12.25) {};
		\node [style=none] (95) at (-0.25, 3) {$=$};
		\node [style=none] (96) at (-12.75, 10.5) {};
	\end{pgfonlayer}
	\begin{pgfonlayer}{edgelayer}
		\draw (1.center) to (0.center);
		\draw (3.center) to (4.center);
		\draw (3.center) to (5.center);
		\draw [bend right=90, looseness=0.50] (5.center) to (6.center);
		\draw (6.center) to (4.center);
		\draw (13.center) to (14.center);
		\draw (13.center) to (15.center);
		\draw [bend right=90, looseness=0.50] (15.center) to (16.center);
		\draw (16.center) to (14.center);
		\draw [in=-90, out=90, looseness=1.50] (20.center) to (19.center);
		\draw [in=-90, out=90, looseness=1.50] (27.center) to (26.center);
		\draw (0.center) to (10.center);
		\draw (19.center) to (32.center);
		\draw (34.center) to (33.center);
		\draw (23.center) to (29.center);
		\draw (39.center) to (40.center);
		\draw (26.center) to (42.center);
		\draw (44.center) to (45.center);
		\draw (47.center) to (48.center);
		\draw (50.center) to (49.center);
		\draw (52.center) to (53.center);
		\draw (52.center) to (54.center);
		\draw [bend right=90, looseness=0.50] (54.center) to (55.center);
		\draw (55.center) to (53.center);
		\draw (62.center) to (63.center);
		\draw (62.center) to (64.center);
		\draw [bend right=90, looseness=0.50] (64.center) to (65.center);
		\draw (65.center) to (63.center);
		\draw [in=-90, out=90, looseness=1.50] (69.center) to (68.center);
		\draw [in=-90, out=90, looseness=1.50] (76.center) to (75.center);
		\draw (49.center) to (59.center);
		\draw (68.center) to (81.center);
		\draw (83.center) to (82.center);
		\draw (72.center) to (78.center);
		\draw (86.center) to (87.center);
		\draw (75.center) to (88.center);
		\draw (90.center) to (91.center);
		\draw (93.center) to (94.center);
	\end{pgfonlayer}
\end{tikzpicture}
\end{equation}
Thus we can apply (\ref{eq:inside_pair}) to calculate our theory's predictions for $q(C_1^2, C_2^1) := \sum_{C_1^1, C_2^2} q(C_1^1, C_1^2=a, C_2^1=b, C_2^2)$.
\begin{equation}
   \sum_{C_1^1, C_2^2} q(C_1^1, C_1^2=a, C_2^1=b, C_2^2) \ \ \ \  = \ \ \ \ \ \ \begin{tikzpicture}
	\begin{pgfonlayer}{nodelayer}
		\node [style=none] (0) at (-2.25, 0) {};
		\node [style=none] (1) at (-2.25, -5) {};
		\node [style=none] (2) at (-0.75, -5) {};
		\node [style=none] (3) at (-2.75, -5) {};
		\node [style=none] (4) at (-0.25, -5) {};
		\node [style=none] (5) at (-2.75, -5.75) {};
		\node [style=none] (6) at (-0.25, -5.75) {};
		\node [style=none] (7) at (-2.25, -5) {};
		\node [style=none] (8) at (-1.5, -5.5) {$\phi$};
		\node [style=none] (9) at (-0.75, -5) {};
		\node [style=none] (10) at (-2.25, 3.5) {};
		\node [style=none] (11) at (0.75, -5) {};
		\node [style=none] (12) at (2.25, -5) {};
		\node [style=none] (13) at (0.25, -5) {};
		\node [style=none] (14) at (2.75, -5) {};
		\node [style=none] (15) at (0.25, -5.75) {};
		\node [style=none] (16) at (2.75, -5.75) {};
		\node [style=none] (17) at (0.75, -5) {};
		\node [style=none] (18) at (1.5, -5.5) {$\psi$};
		\node [style=none] (19) at (-1.75, 0) {};
		\node [style=none] (20) at (0.75, -5) {};
		\node [style=none] (21) at (0.75, -5) {};
		\node [style=none] (22) at (2.25, -5) {};
		\node [style=none] (23) at (2.25, -5) {};
		\node [style=none] (24) at (-0.75, -5) {};
		\node [style=none] (25) at (-0.75, -5) {};
		\node [style=none] (26) at (1.75, 0) {};
		\node [style=none] (27) at (-0.75, -5) {};
		\node [style=none] (28) at (-0.75, -5) {};
		\node [style=none] (29) at (2.25, 3.5) {};
		\node [style=none] (30) at (-3, 0) {};
		\node [style=large map] (31) at (-3, 0) {$U_1^1$};
		\node [style=none] (32) at (-1.75, 3.5) {};
		\node [style=none] (33) at (-4.25, 0) {};
		\node [style=none] (34) at (-4.25, 3.75) {};
		\node [style=none] (35) at (2.25, 3.5) {};
		\node [style=none] (36) at (1.75, 3.5) {};
		\node [style=large map] (37) at (-3, 3.5) {$\tilde{E}_1^2$};
		\node [style=none] (38) at (-2.75, 3.75) {};
		\node [style=none] (41) at (-3, 6.25) {};
		\node [style=map] (42) at (2, 3.5) {$E_2^1$};
		\node [style=none] (43) at (2, 6.25) {};
	\end{pgfonlayer}
	\begin{pgfonlayer}{edgelayer}
		\draw (1.center) to (0.center);
		\draw (3.center) to (4.center);
		\draw (3.center) to (5.center);
		\draw [bend right=90, looseness=0.50] (5.center) to (6.center);
		\draw (6.center) to (4.center);
		\draw (13.center) to (14.center);
		\draw (13.center) to (15.center);
		\draw [bend right=90, looseness=0.50] (15.center) to (16.center);
		\draw (16.center) to (14.center);
		\draw [in=-90, out=90, looseness=1.50] (20.center) to (19.center);
		\draw [in=-90, out=90, looseness=1.50] (27.center) to (26.center);
		\draw (0.center) to (10.center);
		\draw (19.center) to (32.center);
		\draw (34.center) to (33.center);
		\draw (23.center) to (29.center);
		\draw (26.center) to (36.center);
		\draw [style=dashed] (41.center) to (37);
		\draw [style=dashed] (43.center) to (42);
	\end{pgfonlayer}
\end{tikzpicture}
\end{equation}
Applying the interchange law (\ref{eq:interchange}) to pull $E_2^1$ up, and then (\ref{supermeasurement_one}), (\ref{eq:E_def}), and (\ref{eq:contextuality_two_party}), we infer the following.
\begin{equation} \label{con2}
    p(A=a, B=b|X=2, Y=1) = \sum_{C_1^1, C_2^2} q(C_1^1, C_1^2=a, C_2^1=b, C_2^2)
\end{equation}
By similar reasoning, we also obtain:
\begin{equation} \label{con3}
    p(A=a, B=b|X=1, Y=2) = \sum_{C_1^2, C_2^1} q(C_1^1=a, C_1^2, C_2^1, C_2^2=b)
\end{equation}
\begin{equation} \label{con4}
    p(A=a, B=b|X=2, Y=2) = \sum_{C_1^1, C_2^1} q(C_1^1, C_1^2=a, C_2^1, C_2^2=b)
\end{equation}

We have thus shown that the four distributions $p(AB|XY)$ are precisely the marginals of $q$. But as discussed in the previous subsection, this would mean that $p(AB|XY)$ admits a local hidden variable model, contrary to our initial assumption. 

This gives us a logical contradiction between the absoluteness of observed events and the predictions of our BIL theory. It follows that either the observed events are not absolute, or that the BIL theory made false predictions for their marginals.

\paragraph{Generalizing the argument.} This argument can be straightforwardly generalized for any BIL theory and some Bell nonlocal distribution $p(A_1 \ldots A_n |X_1 \ldots X_n)$.
Again, one constructs a circuit model, analogous to (\ref{eq:four_mus}), for a measurement scenario, with one measurement for each value of each measurement setting $X_i$.
This circuit model will again describe the scenario from an outsider's perspective by using a memory update for each measurement.

To this end, for $n$ variables, one proceeds in a very similar way to construct the first `layer' of measurements, analogously to (\ref{two_mus}), and then the second layer, analogously to (\ref{eq:four_mus}). To construct the third layer, note that the property of information-preservation is closed under sequential composition. In other words, if $U$ and $V$ are both information-preserving in the sense of (\ref{eq:info_preservation}), and $V \circ U$ is well-defined, then $V \circ U$ is also information-preserving in the sense of (\ref{eq:info_preservation}).
This means that one can construct the third layer from the second in much the same way that one constructs the second layer from the first, and so on. 

Having constructed the circuit model, one can then use the interchange law (\ref{eq:interchange}) to think of any set of measurements for a global choice of settings $(X_1 = x_1, \ldots, X_n=x_n)$ as performed in parallel, before deploying a generalization of (\ref{eq:inside_pair}) to predict joint probabilities for subsets of the purportedly absolutely events. 
It is then easily verified that these predictions are equivalent to $p(A_1...A_n|X_1=x_1,...X_n=x_n)$, implying that $p(A_1...A_n|X_1...X_n)$ is Bell Nonlocal, violating our initial assumption. $\square$

\subsection{Closing a loophole} \label{sec:logical}

The argument above assumed the applicability of probability theory. In particular, it assumed that (1) the relative frequencies for all of the absolute events converged to a probability distribution, and that (2) the relative frequencies for subsets are given by marginalizing this distribution.
One might therefore speculate that a BIL theory might be reconciled with AOE via a nonclassical generalization of probability theory.

However, even this escape route can be sealed off for a subset of BIL theories.
These are the BIL theories with a strictly stronger form of Bell Nonlocality. 
Roughly speaking, \textit{Possibilistic Bell Nonlocality} (a special case of logical contextuality \cite{abramsky2015contextuality}) is Bell Nonlocality that arises not only at the level of probabilities, but at the level of possibilities.

Quantum theory is Bell nonlocal in the possibilistic sense, as is demonstrated by the result from \cite{ormrod2022no}, sketched in Section \ref{sec:simple}.\footnote{More fundamentally, it is demonstrated by Hardy's paradox \cite{hardynonlocality1993}, on which that no-go result is based.}
There, the quantum-theoretical probability for Alice and Bob both seeing `-'  is nonzero. Calling $(a, b)=(-, -)$ a \textit{local assignment}, we conclude that quantum theory predicts that this local assignment is \textit{possible.}
Quantum theory also predicts that the assignments 
\begin{equation}
    \begin{split}
        (a, d)&=(-, 0) \\
        (b, c)&=(1, 1) \\
        (c, d)&=(0, -)
    \end{split}
\end{equation}
are \textit{impossible}, because they are given probabilities of zero. 
But there is no \textit{global assignment} of the variables $(a, b, c, d)$ for which $(a, b)=(-, -)$ that does not also contain one of these three impossible local assignments.
This is the defining feature of possibilistic Bell nonlocality: there is a local assignment that cannot be embedded into any possible global assignment (i.e.\ a global assignment that does not contain any impossible local assignments). A general and formal definition of possibilistic Bell nonlocality can be found in Appendix \ref{app:possibilistic}.

This property is the reason why the quantum no-go result did not require us to assume the existence of a global probability distribution over the purportedly absolute observed events. Our next result generalizes this: all of the \underline{P}IL theories suffer a similar result.

\begin{theorem}[PIL theories are incompatible with AOE, without probabilities.] Any perspectival theory that is Possibilistically Bell Nonlocal, Information Preserving, and has Local Dynamics makes some predictions that are incompatible with AOE -- even when the applicability of probability theory is not assumed. \label{theorem:possibilistic}
\end{theorem}

Theorem \ref{theorem:possibilistic} is proven in Appendix \ref{app:possibilistic}.

\subsection{A trilemma?}

Bell Nonlocality is experimentally well-supported. Therefore, believers in AOE can only adopt perspectival theories that either fail to be Information Preserving or else lack Local Dynamics. 

This might suggest that one faces a trilemma between (1) rejecting AOE; (2) allowing information to be destroyed; and (3) rejecting relativity theory. But this would be too quick a judgement, since Local Dynamics is not simply the prohibition of superluminal influences suggested by relativity. The next section shows that this prohibition can indeed be used to derive Local Dynamics, but only in combination with the assumption of a sort of dynamical separability (together with a rather minimal consistency constraint). 

And, as we will discuss in Section \ref{sec:conc}, this suggests a relatively conservative way of retaining AOE.

\section{From deeper principles} \label{sec:binsc}

In this section, we will derive a contradiction with AOE from even deeper physical principles.
To this end, we will show that all `NSC' perspectival theories have Local Dynamics (Theorem \ref{thm:nsc}).
The NSC theories are those with \textit{No Superluminal Influences, Separable Dynamics}, and \textit{Consistent Embeddings}. This leads to a more fundamental no-go theorem, to the effect that all `BINSC' theories are incompatible with AOE (Theorem \ref{thm:binsc}).
The theorem suggests an interesting approach to maintaining AOE, in which one rejects Local Dynamics and yet avoids superluminal influences by rejecting Separable Dynamics.

\subsection{NSC theories have Local Dynamics} \label{sec:binsc_nsc}

To define the three NSC properties, we must first assume that there is a subset of the transformations in our perspectival theory that can be considered \textit{fundamental}.
We stipulate that any transformation $T$ in the theory should be obtained from a fundamental one $V$ and some state $\psi$ in the following way.
\begin{equation} \label{fundamental}
    \begin{tikzpicture}
	\begin{pgfonlayer}{nodelayer}
		\node [style=none] (0) at (-1, 2.5) {};
		\node [style=none] (1) at (-1, -2.5) {};
		\node [style=none] (2) at (1, 2) {};
		\node [style=none] (3) at (1, -1.75) {};
		\node [style=none] (4) at (0, 0) {$T$};
		\node [style=large map] (9) at (0, 0) {$V$};
		\node [style=none] (10) at (-2.5, 0) {$=$};
		\node [style=point] (11) at (1, -1.75) {$\psi$};
		\node [style=discard] (12) at (1, 2.25) {};
		\node [style=map] (13) at (-4, 0) {$T$};
		\node [style=none] (14) at (-4, -2.5) {};
		\node [style=none] (15) at (-4, 2.5) {};
	\end{pgfonlayer}
	\begin{pgfonlayer}{edgelayer}
		\draw (1.center) to (0.center);
		\draw (3.center) to (2.center);
		\draw (15.center) to (14.center);
	\end{pgfonlayer}
\end{tikzpicture}
\end{equation}
(Although this resembles the stinespring dilation in quantum theory, note that there is no assumption here that $\psi$ is `pure'.)

We also must assume that for each input subsystem of a fundamental transformation, there is a fact of the matter about whether it exerts a causal influence on any given output subsystem. The precise definition of causal influence is irrelevant to the derivation, though a natural candidate definition is that $A$ causally influences $D$ through a fundamental transformation $V$ if and only if the following holds, where $T$ is some trace-preserving transformation.

\begin{equation} \label{ni}
    \begin{tikzpicture}
	\begin{pgfonlayer}{nodelayer}
		\node [style=none] (0) at (-3.5, 2) {};
		\node [style=none] (1) at (-3.5, -2.5) {};
		\node [style=none] (2) at (-1.5, 2.5) {};
		\node [style=none] (3) at (-1.5, -2.5) {};
		\node [style=none] (4) at (-2.5, 0) {$T$};
		\node [style=large map] (5) at (-2.5, 0) {$V$};
		\node [style=none] (6) at (0, 0) {$=$};
		\node [style=discard] (7) at (-3.5, 2.25) {};
		\node [style=none] (8) at (1.5, -1) {};
		\node [style=none] (9) at (1.5, -2.5) {};
		\node [style=discard] (10) at (1.5, -0.75) {};
		\node [style=none] (11) at (3.5, -2.5) {};
		\node [style=none] (12) at (3.5, 2.5) {};
		\node [style=map] (13) at (3.5, 0) {$T$};
		\node [style=none] (14) at (-4, -2) {$A$};
		\node [style=none] (15) at (-1, -2) {$B$};
		\node [style=none] (16) at (-4, 1.5) {$C$};
		\node [style=none] (17) at (-1, 1.5) {$D$};
		\node [style=none] (18) at (1, -2) {$A$};
		\node [style=none] (19) at (4, -2) {$B$};
		\node [style=none] (20) at (4, 1.5) {$D$};
	\end{pgfonlayer}
	\begin{pgfonlayer}{edgelayer}
		\draw (1.center) to (0.center);
		\draw (3.center) to (2.center);
		\draw (9.center) to (8.center);
		\draw (11.center) to (12.center);
	\end{pgfonlayer}
\end{tikzpicture} \ \ \ \ \ \ \ \ \not\exists T
\end{equation}
Then No Superluminal Influences may be defined as follows.

\begin{definition}[No Superluminal Influences.]
A perspectival theory respects \textbf{No Superluminal Influences} just in case there is no valid embedding of any fundamental transformations such that an input subsystem exerts a causal influence on a spacelike separated output subsystem.
\end{definition}

Our next principle concerns a fundamental transformation of the type $V: A_1 \otimes \ldots \otimes A_n \otimes \lambda \rightarrow A_1' \otimes \ldots \otimes A_n' \otimes F$ with a particular sort of causal structure. Namely, a causal structure such that each input $A_i$ does not influence $A_j'$ for any $j\neq i$. In that case, we want to assume that $V$ \textit{separates} into a set of transformations, such that there are only directed paths of wires between systems that might influence each other:
\begin{equation} \label{separable_dynamics}
    \begin{tikzpicture}
	\begin{pgfonlayer}{nodelayer}
		\node [style=none] (0) at (-7.25, 1.25) {};
		\node [style=none] (1) at (-7.25, -3.75) {};
		\node [style=none] (2) at (-5.25, 1.25) {};
		\node [style=none] (3) at (-5.25, -3.75) {};
		\node [style=none] (5) at (-7.75, -3.25) {$A_1$};
		\node [style=none] (6) at (-7.75, 7.25) {$A_1'$};
		\node [style=none] (7) at (-4.75, -3.25) {$A_n$};
		\node [style=none] (8) at (-4.75, 7.25) {$A_n'$};
		\node [style=none] (10) at (-2, 2) {};
		\node [style=none] (11) at (-2, 2) {};
		\node [style=none] (12) at (-2, 2) {$=$};
		\node [style=none] (13) at (-0.25, 7.75) {};
		\node [style=none] (14) at (-0.25, -3.75) {};
		\node [style=none] (15) at (-0.75, -3.25) {$A_1$};
		\node [style=none] (16) at (-0.75, 7.25) {$A_1'$};
		\node [style=map] (17) at (-0.25, 2) {$T_1$};
		\node [style=none] (18) at (2.25, 7.75) {};
		\node [style=none] (19) at (2.25, -3.75) {};
		\node [style=none] (20) at (2.75, -3.25) {$A_n$};
		\node [style=none] (21) at (2.75, 7.25) {$A_n'$};
		\node [style=map] (22) at (2.25, 2) {$T_n$};
		\node [style=none] (23) at (4, -1) {};
		\node [style=none] (24) at (4.75, 0) {};
		\node [style=none] (25) at (6, -1) {};
		\node [style=none] (26) at (0, 1.5) {};
		\node [style=none] (27) at (2.75, 2) {};
		\node [style=none] (28) at (-6.25, 7.25) {$\hdots$};
		\node [style=none] (29) at (-6.25, -3.25) {$\hdots$};
		\node [style=none] (30) at (1, 7.25) {$\hdots$};
		\node [style=none] (31) at (1, -3.25) {$\hdots$};
		\node [style=none] (32) at (4.75, -0.5) {$\hdots$};
		\node [style=large map] (33) at (5, -1.5) {$T_0$};
		\node [style=none] (34) at (5, -3.75) {};
		\node [style=none] (35) at (4, 5) {};
		\node [style=none] (36) at (6, 5) {};
		\node [style=none] (37) at (0, 2.5) {};
		\node [style=none] (38) at (2.75, 2) {};
		\node [style=none] (39) at (4.75, 4.5) {$\hdots$};
		\node [style=large map] (40) at (5, 5.5) {$T_{n+1}$};
		\node [style=none] (41) at (5, 7.75) {};
		\node [style=none] (42) at (5.5, -3.25) {$\lambda$};
		\node [style=none] (43) at (5.5, 7.25) {$F$};
		\node [style=none] (44) at (-3, -3.25) {$\lambda$};
		\node [style=none] (45) at (-3, 7.25) {$F$};
		\node [style=none] (46) at (-3.5, 1.25) {};
		\node [style=none] (47) at (-3.5, -3.75) {};
		\node [style=none] (48) at (-5.5, 2) {$V$};
		\node [style=none] (49) at (-7.75, 1.25) {};
		\node [style=none] (50) at (-3, 1.25) {};
		\node [style=none] (51) at (-3, 2.75) {};
		\node [style=none] (52) at (-7.75, 2.75) {};
		\node [style=none] (53) at (-7.25, 7.75) {};
		\node [style=none] (54) at (-7.25, 2.75) {};
		\node [style=none] (55) at (-5.25, 7.75) {};
		\node [style=none] (56) at (-5.25, 2.75) {};
		\node [style=none] (59) at (-3.5, 7.75) {};
		\node [style=none] (60) at (-3.5, 2.75) {};
	\end{pgfonlayer}
	\begin{pgfonlayer}{edgelayer}
		\draw (1.center) to (0.center);
		\draw (3.center) to (2.center);
		\draw [in=270, out=90] (14.center) to (13.center);
		\draw (19.center) to (18.center);
		\draw [in=-90, out=90, looseness=0.75] (23.center) to (26.center);
		\draw [in=90, out=-90, looseness=1.25] (27.center) to (25.center);
		\draw (34.center) to (33);
		\draw [in=90, out=-90, looseness=0.75] (35.center) to (37.center);
		\draw [in=-90, out=90, looseness=1.25] (38.center) to (36.center);
		\draw (41.center) to (40);
		\draw (47.center) to (46.center);
		\draw (52.center) to (49.center);
		\draw (49.center) to (50.center);
		\draw (50.center) to (51.center);
		\draw (51.center) to (52.center);
		\draw (54.center) to (53.center);
		\draw (56.center) to (55.center);
		\draw (60.center) to (59.center);
	\end{pgfonlayer}
\end{tikzpicture}
\end{equation}
Note that we do not assume that the $T_i$ are fundamental transformations. Our second NSC principle is summarized as follows.

\begin{definition}[Separable Dynamics]
A theory has \textbf{Separable Dynamics} just in case any fundamental transformation $V: A_1 \otimes \ldots \otimes A_n \otimes \lambda \rightarrow A_1' \otimes \ldots \otimes A_n' \otimes F$ with the property that $A_i$ does not influence $A_j'$ for all $j \neq i$ decomposes as in (\ref{separable_dynamics}).
\end{definition}

The final principle we need to derive Local Dynamics is just a consistency constraint on the valid embeddings. The idea is that, for any valid embedding of a non-fundamental transformation, there is a consistent valid embedding of the fundamental one from which it can arise. For example, suppose the input of a quantum channel is a system $A$ at the location $x$, and its output is a system $B$ at the location $y$. If this channel actually arises from a more fundamental, unitary, channel, then obviously the input subsystem $A$ and the output subsystem $B$ of that unitary are located at $x$ and $y$ respectively (though its other subsystems might be elsewhere). So the fact that we can embed the original channel a certain way means we must also be able to embed some unitary in a corresponding way.

Consistent Embeddings is the obvious generalization of this idea. If the embedding of $T$ maps its input/output subsystems to a certain set of spacetime points, then one should be able to map the same subsystems to the same set of points when one considers them as inputs and outputs to some fundamental transformation $V$, from which $T$ arises via (\ref{fundamental}). More formally, given an embedding function $\cf$ for $V$, one can consider the restriction of the function $\cf_{| {\rm Sys}(T)}$ to one that acts only on the systems shared by $T$ and $V$. We then have the following definition.
\begin{definition} [Consistent Embeddings.]
A perspectival theory has \textbf{Consistent Embeddings} just in case $\ce$ is only a valid embedding of a transformation $T$ if there exists a fundamental transformation $V$ satisfying (\ref{fundamental}) that can be validly embedded with a function $\cf$ such that $\cf_{| {\rm Sys}(T)}= \ce$. Denoting the proposition that $\ce$ is a valid embedding of $T$ as $(T, \ce)$, we have
\begin{equation} 
    (T, \ce) \implies \Big( \exists V: {\rm  ((\ref{fundamental}) \ holds}) \ \land \  (\exists \cf: (V, \cf) \ \land  \ \cf_{| {\rm Sys}(T)} = \ce) \Big) 
\end{equation} 
\end{definition}

We then obtain the following theorem.
\begin{theorem}[NSC theories have Local Dynamics.]
Any perspectival theory with No Superluminal Influences, Separable Dynamics, and Consistent Embeddings also has Local Dynamics. \label{thm:nsc}
\end{theorem}

\paragraph{Proof.} If the theory has Consistent Embeddings, then any transformation $T: A_1 \otimes \ldots \otimes A_n \rightarrow A_1' \otimes \ldots \otimes A_n'$ can be embedded such that each pair $(A_i, A_i')$ is spacelike separated only if it can be obtained via (\ref{fundamental}) from a fundamental transformation $V: A_1 \otimes \ldots \otimes A_n \otimes \lambda \rightarrow A_1' \otimes \ldots \otimes A_n' \otimes F$, which can itself be embedded in such a way that each pair $(A_i, A_i')$ is spacelike separated.
In that case, given No Superluminal Influences, the antecedent condition of the Separable Dynamics condition is satisfied by $V$. Thus, given Separable Dynamics, $V$ decomposes as in (\ref{separable_dynamics}). Inserting a state on $\lambda$ and tracing out $F$ in (\ref{separable_dynamics}) to obtain $T$ via (\ref{fundamental}) leads to a circuit diagram that can easily be simplified to one of the form (\ref{local_dynamics}). $\square$

\paragraph{Quantum theory and Local Dynamics.}
Before stating another result, let us finally make good on our promise to show that quantum theory can be formulated as a BIL theory. Consider again the quantum perspectival theory with isometric memory updates from Section \ref{sec:qpt}. Let us designate the unitary channels as the fundamental transformations, and adopt the definition of causal influences in (\ref{ni}). We can then impose by fiat that the valid embeddings are restricted so that both No Superluminal Influences and Consistent Embeddings are respected. Appendix \ref{app:sep} shows that the resulting perspectival theory has Separable Dynamics, implying, by Theorem \ref{thm:nsc}, that it also has Local Dynamics.

\subsection{Recasting the no-go result}

Combining Theorems \ref{thm:BIL_AOE} and \ref{thm:nsc} leads to the following no-go result.

\begin{theorem}[BINSC theories are incompatible with AOE.] \label{thm:binsc}
Any perspectival theory that 
\begin{itemize}
    \item is Bell Nonlocal;
    \item is Information Preserving;
    \item satisfies No Superluminal Influences;
    \item has Separable Dynamics; and
    \item has Consistent Embeddings
\end{itemize}
makes some predictions that are incompatible with AOE.
\end{theorem}
Thus in order to maintain AOE one would have to either embrace a perspectival theory that lacked one of these five properties, or else adopt a theory that could not be formulated in the perspectival framework. We note that a similar theorem can be derived to the effect that `PINSC' theories are incompatible with AOE, by swapping Bell Nonlocality for Possibilistic Bell Nonlocality, and that this theorem would not rely on the validity of probability theory.

\begin{figure}
    \centering
    \includegraphics[scale=0.6]{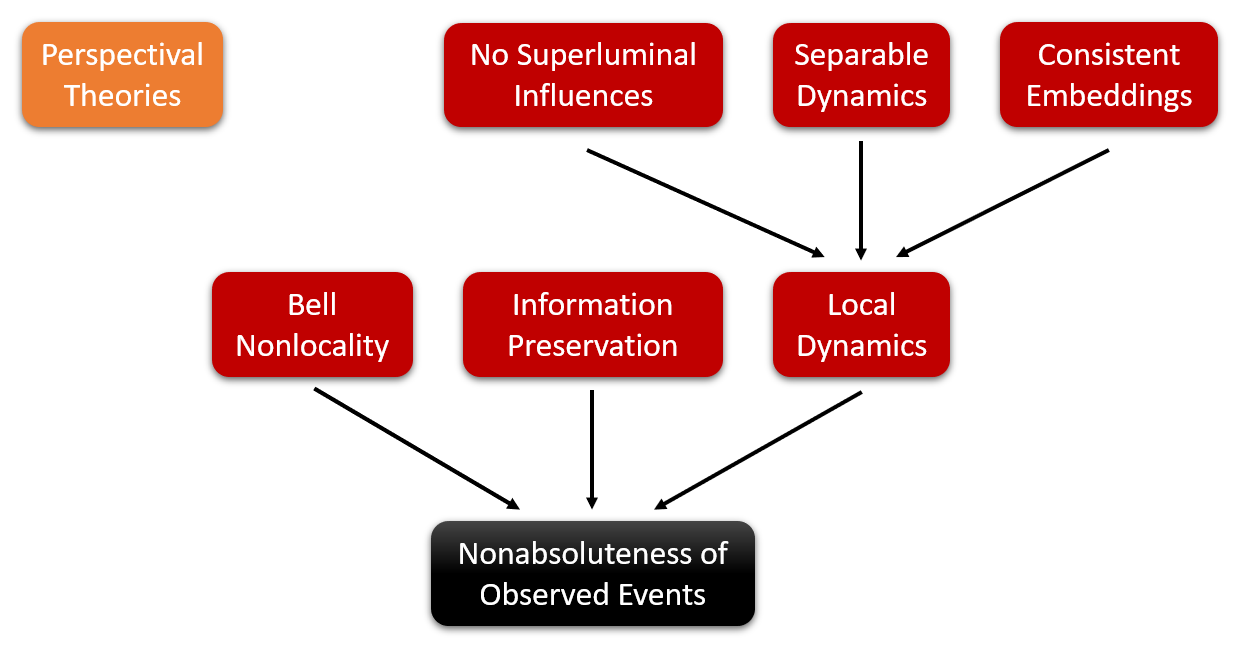}
    \caption{A graphical summary of theorems \ref{thm:BIL_AOE}, \ref{thm:nsc}, and \ref{thm:binsc}.}    \label{fig:implications}
\end{figure}

Our no-go results admit a nice graphical representation, in the style of \cite{wiseman2015causarum, Cavalcanti2021implications}. In Figure \ref{fig:implications}, each principle with parents is implied by their conjunction. The disconnected orange node denotes the background assumption that one's theory can be formulated as a perspectival theory, which is required to make sense of the other principles. Rejecting a principle in the graph that has parents requires one to reject at least one of its ancestors and/or the disconnected node. 

\subsection{Notions of dynamical nonlocality}

Since we will argue that distinguishing between different notions of dynamical (non)locality may hold the key to solving the measurement problem, it is worth elaborating on them a bit.

\begin{figure}
    \centering
    \includegraphics[scale=0.6]{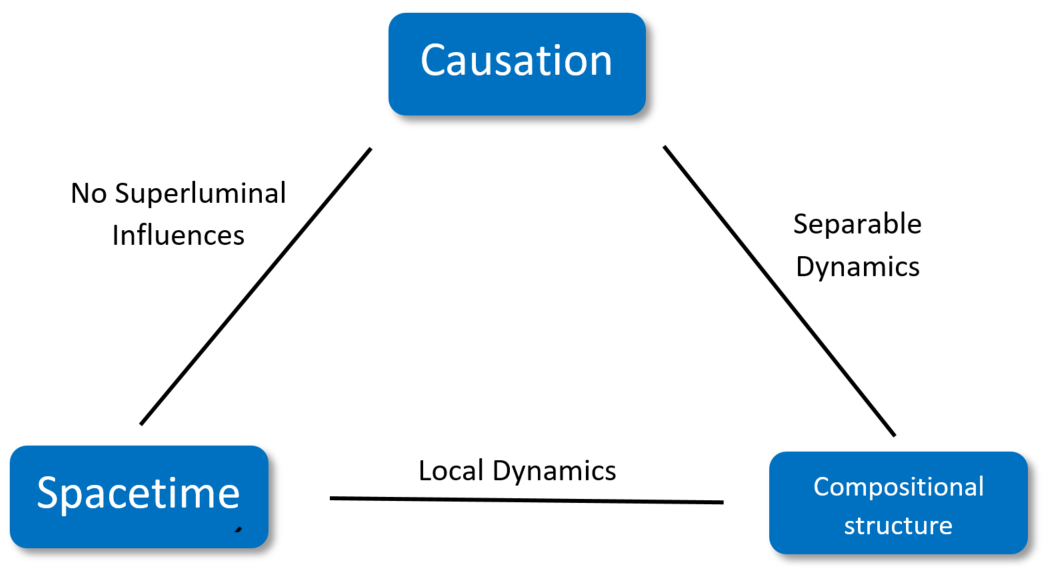}
    \caption{Different notions of dynamical locality.}    \label{fig:notions}
\end{figure}

Local Dynamics is the assumption that transformations must decompose a certain way when they take place across spacelike separated regions: it is a connection between \textit{spacetime} and \textit{compositional structure}. 
Given Consistent Embeddings, one can split up Local Dynamics into two further dynamical locality assumptions. No Superluminal Influences says that causal influences are local: it links \textit{spacetime} with \textit{causation}. On the other hand, Separable Dynamics says that transformations with certain causal structures must decompose a certain way: it links \textit{causation} with \textit{compositional structure}. Therefore, one can think of the connection between spacetime and compositional structure provided by Local Dynamics as mediated by causation, as in Figure \ref{fig:notions}.

Clearly, relativity theory strongly suggests that spacetime should constrain causation. On the other hand, it is not obvious that it suggests causation should constrain compositional structure. This observation is key to the argument we will give in Section \ref{sec:conc} that rejecting Separable Dynamics is a possible way of avoiding a measurement problem.

\subsection{Are the assumptions individually necessary?}

We have shown that certain sets of properties are \textit{jointly sufficient} for a breakdown of AOE, but not that they are \textit{individually necessary}. That is, we have not shown that there exist perspectival theories that remain consistent with AOE by rejecting only one of the BINSC properties, while clinging to the remaining four.

In fact, one can avoid a breakdown in absoluteness by only rejecting Bell Nonlocality, or by only rejecting No Superluminal Influences, as shown in Appendix \ref{app:role}. That appendix also shows that any BINSC theory that rejects Information Preservation by embracing `collapses' is consistent with AOE, even if it has all four other properties. As well as providing some reassurance that we have not assumed any unnecessary properties in our no-go theorems, the arguments there also help shed light on the roles played by the individual assumptions in the breakdown of AOE.

However, Appendix \ref{app:role} does not answer whether one can recover AOE, and thus avoid a measurement problem, by rejecting only Separable Dynamics. We take this up in the next section.

\section{Roads back to absoluteness} \label{sec:conc}

In this penultimate section, we assess the prospects for a future theory of physics to avoid a measurement problem, in the sense of being consistent with AOE. Our strategy will be to see whether our results suggest any paths to restoring AOE that keep core aspects of the modern physical worldview intact. We will be particularly concerned to retain relativity theory. 

While previous no-go results have shown that maintaining AOE and relativity theory requires one to give up the idea that all interactions are unitary, we will argue that one might nevertheless be able to cling to the deeper principle that all interactions preserve information. This new road back to absoluteness appears, to us, to be the most conservative way of avoiding a measurement problem.

It may at first appear that there are other, less painful, strategies. For instance, our theorems only hold for perspectival theories -- couldn't one evade our no-go results simply by rejecting that framework?
The problem one then faces is identifying a principle of perspectival theories that should be rejected, and, moreover, one which was essential to our derivations. This is an unenviable task. Recall that perspectival theories are just categorical probabilistic theories endowed with the resources to discuss two different perspectives on measurements. Categorical probabilistic theories are, in turn, essentially just circuit theories containing classical circuit subtheories that handle empirical predictions, and which come with a summation structure that allows one to talk about probabilistic mixtures. 

Now, rejecting circuit theories themselves would apparently amount to rejecting \textit{compositionality}, the idea that if a set of transformations are possible, then combinations of them are also possible. This move would appear ad hoc, and would require both a new mathematical framework for describing which combinations are possible, and an explanation of why some are not.\footnote{Circuit theories also assume that the combinations are associative, so one might hope to avoid the results just by rejecting associativity. But the charge of ad-hoc-ness and the need for new frameworks still apply, on top of the fact that it is not clear that associativity is necessary for the derivations in this paper.} And one clearly cannot reject the idea behind categorical probabilistic theories that a physical circuit theory should make empirical predictions. Finally, rejecting the idea from perspectival theories that theories should have diverging outside perspectives is a non-starter, since a theory without a divergent outside perspective can always be thought of as a theory with an outside perspective that is the same as its inside one. Such theories are already accommodated for by the perspectival framework.

What can one do without rejecting the framework? If we stick to the idea that we should be able to formulate a physical theory as a perspectival one, then our results tell us that any theory that is consistent with AOE must fail to violate Bell inequalities, to preserve information, or else to it must exhibit some form of dynamical nonlocality. But clearly, a theory that does not violate Bell inequalities cannot explain the experimental violations. Such a theory is empirically inadequate, hence unacceptable.

If we want to maintain AOE, that leaves us with just two options. The first option is to embrace the idea that information is sometimes destroyed. This means that when a friend inside a lab performs some particular measurement, it can somehow lead to an unrecoverable loss of the information associated with the incompatible measurements that she chose not to perform. This idea has been advocated for some time by objective collapse theorists. The second option is to embrace some sort of dynamical nonlocality, and has been long advocated (in one form) by pilot-wave theorists, especially in light of Bell's theorem.

It is easy to see how both of these camps might see their approaches as vindicated by this result. Objectives collapse theorists will be pleased to see that the destruction of information is one of the only two possible roads back to absoluteness, and can argue for their chosen path in both a positive and negative way. Positively, they might claim that the black hole paradox provides an independent reason for believing that information may be destroyed \cite{preskill1992black, oppenheim2018post}, and may also argue that working experimental quantum physicists implicitly rely on this assumption all the time. Negatively, they might argue that theirs is the only strategy that does not have a fundamental problem with relativity theory, which apparently rules out dynamical nonlocality.

On the other hand, pilot-wave theorists can argue that the preservation of information is an extremely fundamental principle, and the rejection of it appears ad-hoc. They will accept that the difficulty of reconciling their own approach with relativity is unfortunate, but may argue that it was already the message of Bell's theorem that relativity must be revised, and the measurement no-go theorems only serve to confirm it.

But both the foregoing argument for objective collapses, and the one for pilot wave theories, involves an assumption -- that relativity theory enforces the sort of dynamical locality associated with Local Dynamics. It is not at all clear that this is true. As far as we know, Local Dynamics cannot be derived from No Superluminal Influences without the help of both Separable Dynamics and Consistent Embeddings. It is very plausible that relativity theory rules out superluminal influences, and it is true that Consistent Embeddings is hard to imagine revising. But it is not obvious that relativity theory implies Separable Dynamics. Perhaps nonseparability provides a form of dynamical nonlocality that is strong enough to avoid a conflict with AOE, but weak enough to avoid a conflict with relativity theory. 

There is an irresistible comparison to be made with a common response to Bell's theorem. That theorem showcases a tension between relativity theory, which appears to require locality, and quantum theory, whose violation of Bell inequalities means it must be somehow nonlocal. But the tension can be diffused when we make a distinction between different notions of nonlocality. In particular, one can distinguish between nonlocal causal influences and nonseparable states (or correlations between systems that are not underwritten by their non-holistic properties). There are good arguments that relativity prohibits the former, but it is much harder to argue that it rules out the latter. We might say that relativity theory rules out action at a distance, while allowing passion at a distance \cite{shimony1984controllable}. This would appear to be roughly the view of most researchers in quantum physics and its foundations, who generally do not believe that Bell's theorem calls for an upheaval of relativity theory (and are therefore obligated to explain the Bell inequality violations in terms of passion, rather than action, at a distance).

Here, the idea is similar. Even if relativity theory rules out dynamics that support nonlocal causal influences, it might still be consistent with nonseparable dynamics. Perhaps the lesson of Bell is that the states of distant particles are inextricably linked, and the lesson of the new measurement no-go theorems is that their dynamics are, too. 

If the kind of theory we are imagining could be constructed, then there is a clear sense in which it would be preferable to the existing worked-out theories that avoid the measurement problem -- to objective collapse theories in that it would preserve information, and to Bohm theory in that it would avoid the need for superluminal influences and a preferred foliation. It is therefore worth describing in more detail what sort of theory we are proposing, and how it might be pursued.

There is nothing in our results that indicates that a `BINC' perspectival theory -- one with Bell Nonlocality, Information Preservation, No Superluminal Influences, and Consistent Embeddings -- must be inconsistent with AOE. So the first question to ask is whether it is in fact possible to construct a BINC perspectival theory that is consistent with AOE. Of course, such a theory would have to violate Separable Dynamics. If this sort of theory is possible, then one should be able to go further than the bare perspectival framework -- which does not assume that measurement outcomes are absolute -- and explicitly model the absolute events using the theory. In other words, one should be able to move beyond the \textit{operational }perspectival framework and towards a \textit{ontological} BINC-type  theory\footnote{Which would, however, lead to a BINC perspectival theory describing its operational predictions.} that characterises the absolute events in terms of a precisely specified, agent-independent ontology.

If that is possible, it would be very interesting from a conceptual point of view, since it would confirm that dynamical separability is an essential ingredient for deriving a measurement problem. But it wouldn't necessarily mean that dynamical nonseparability provides an attractive way of avoiding the measurement problem. For that, we would need a BINC theory that is (1) consistent with AOE and (2) recovers and explains the empirical predictions of the physics that has been verified so far. And the latter would involve both (2a) having a theory from which quantum predictions can be recovered in an appropriate limit, and (2b) being able to formulate the theory in an explicitly relativistically covariant way.

If all that can be done, then we can ask an even more interesting question: whether such a theory can make interesting, novel empirical predictions, and, in particular, whether it can do so in regimes of quantum gravity.

In pursuing these questions, one might start by trying to identify an appropriate set of fundamental transformations for the theory. One might even consider certain sets of stochastic matrices or quantum channels, but ultimately one might then want to explore less familiar sorts of transformations. The fundamental transformations should preserve information in the sense of (\ref{eq:info_preservation}), but should also sometimes fail to separate in the sense of (\ref{separable_dynamics}). When an appropriate set is identified, one can then investigate whether there is some natural way by which those transformations might be associated with absolute observed events. If there is such a way, then one can explore whether the theory can be fleshed out into a BINC theory.

While exciting, this suggestion is no silver bullet.  Embracing dynamical separability amounts to rejecting unitary dynamics, which are separable in the relevant sense. For many, rejecting unitarity will feel equally or even more difficult than rejecting relativity theory. One might therefore be inclined to simply accept the failure of AOE, with all of its problematic consequences.

\section{Roads to relativity} \label{sec:relativity}

Even if one declares that observed events are not absolute, the problem does not immediately
vanish. For one thing, it is not at all clear how to make sense of the failure of AOE.
For another, it is unclear that scientific theories can be empirically confirmed except by
the observation of absolute events that they accurately predicted.



Let us elaborate. If observed events are not absolute, then it seems they must be
somehow relative. But relative to what? Suggestions diverge considerably; observed events
might be relative to emergent quasi-classical ‘worlds’; to consistent sets of histories; or
else to certain other events when the corresponding systems are interacting. But neither
Everett, nor consistent histories, nor Rovelli’s relationalism has been able to command
consensus, and it is even controversial whether any one of these approaches constitutes a
precise physical theory that can recover the predictions of Copenhagen interpretation (see,
for instance, \cite{greaves2007probability, dowker1996consistent}).

The basic problem is: if the ontology of my theory provides many inconsistent stories about what happens,
depending on some choice of reference, then I must choose a reference before I get
unambiguous predictions. But if the theory claims that all the references are equally valid,
it is not clear this can be done in a principled way. Hence theories that do not make predictions
for absolute observed events have difficulties with making unequivocal predictions
for one’s experiences, meaning that it can be hard to see how any set of experiences could confirm the theory. In a nutshell, the worry is that predictions made by theories with a relational ontology are too ambiguous to provide ways of empirically confirming it. Therefore, the question of how one should make sense of the claim
that observed events are not absolute remains wide open.

We note that there is a closely related problem here, of a more purely epistemological variety. Namely: in a world where observed events are not absolute, how can inter-subjective agreement be achieved? This question was already tackled from a quantum perspective in \cite{vilasini2022subjective}, which resolves paradoxes arising from multiple agents combining their beliefs about non-absolute events. To address the problem of inter-subjectivity in a much more general context, one could extend the quantum circuits framework for the subjective perspectives of agents from \cite{vilasini2022subjective} to arbitrary BINSC theories.
\footnote{That framework might also provide hints about which sorts of relational ontology can permit inter-subjective agreement.}

Our results provide also provide a clue for addressing the more ontological problem, of how to think of and model `relative events'. In a BIL perspectival theory, it is inevitable that various
inconsistent outside perspectives arise. But, if AOE is not assumed, then the inconsistent
perspectives might be allowed to peacefully co-exist. One might attempt to define an
event not as a single classical variable taking a value, but as a whole collection of variables,
corresponding to different possible perspectives. This brings us to something like the
consistent histories perspective, in which different consistent sets of families of projections
are on an equal ontological footing. Then the challenge is to reign in the inconsistency to
the extent that the theory makes clear and unambiguous predictions. Forthcoming work
by two of us takes this on, also guided by clues from the study of quantum causal structures
\cite{ormrod2022, lorenz2020causal}.

\section*{Acknowledgements}

We are pleased to thank
Eric Cavalcanti,
James Hefford,
Richard Howl,
Hl\'er Kristj\'ansson,
Tein van der Lugt,
Nicola Pinzani, and 
Augustin Vanrietvelde
for helpful discussions. 
Particular thanks go to James Hefford for elegant explanations of the category theory working behind the scenes.

N.O. acknowledges funding from the UK Engineering and Physical Sciences Research Council (EPSRC). 
V.V. is supported by an ETH Postdoctoral Fellowship and acknowledges financial support from the Swiss National Science Foundation (SNSF) Grant Number 200021\_188541.
This publication was made possible through the support of the grant 61466 ‘The Quantum Information Structure of Spacetime (QISS)’ (qiss.fr) from the John Templeton Foundation. The opinions expressed in this
publication are those of the authors and do not necessarily reflect the views of the John Templeton Foundation.

\bibliography{references}

\appendix

\section{Categorical probabilistic theories done more formally} \label{app:CPTsformal}

Here, we provide a rigorous definition of categorical probabilistic theories that does not assume any prior knowledge of category theory. We do so in four steps. First, we define a \textit{symmetric monoidal category} (SMC), which is the categorical term for what we have called a circuit theory. Then, we introduce the concept of traces. We then offer a particular example of an SMC, known as `matrices over the positive reals', or $\mathbb{R}^+$-Mat. This will put us in a good position to define categorical probabilistic theories, which are SMCs that include (a category that is equivalent to) $\mathbb{R}^+$-Mat as a special \textit{sub}-SMC.

\subsection{(Strict) symmetric monoidal categories}

Giving a general formal definition of SMCs is quite involved, but so-called \textit{strict} SMCs are much simpler.
And, as luck would have it, any SMC is equivalent\footnote{Equivalent categories contain more or less the same systems (more precisely: for every system in one category, there is an isomorphic system in that category that is associated with a system in the other category), and their transformations exhibit the same compositional structure.} to some strict SMC. 
Since equivalent categories are, practically speaking, the same, for our purposes it will suffice to explicitly define only strict SMCs.

Any SMC $\cc$ comes with a collection of systems Sys($\cc$). At the heart of the SMC lies its transformations, which come in a set $\cc(A, B)$ mapping $A$ to $B$ for every $A, B \in \ $Sys($\cc$). These are subject to a function $\circ: \cc(B, C) \times \cc(A, B) \rightarrow \cc(A, C)$ that defines a third transformation corresponding to doing a first one followed by a second one, so long as output system of the first transformation is the input system of the second. It is required that $\circ$ is associative and unital
\begin{equation}
    \begin{split}
        (V \circ U) \circ T = V \circ (U \circ T) \\
        1_B \circ T = T \circ 1_A,
    \end{split}
\end{equation}
for which we must require that there is a special identity transformation $1_A \in \cc(A, A)$ for any system $A \in $ Sys($\cc$). (Above we have assumed that $T$ is of the type $T: A \rightarrow B$.)

Having all this structure ensures we have a \textit{category}. To make it monoidal, we need a notion of performing transformations independently, rather than one after the other. To this end, we introduce a new function on systems $\otimes: {\rm Sys}(\cc) \times {\rm Sys}(\cc) \rightarrow {\rm Sys}(\cc)$ and a closely related one on transformations $\otimes: \cc(A, B) \times \cc(C, D) \rightarrow \cc(A \otimes C, B \otimes D)$ that defines a third transformation corresponding to doing any pair of transformations independently. For a strict SMC, we require both of these operations to be associative and unital
\begin{equation} \label{eq:monoidal}
    \begin{split}
        (A \otimes B) \otimes C =  A \otimes (B \otimes C) \\
        A \otimes I = A = I \otimes A \\
        (T \otimes U) \otimes V =  T \otimes (U \otimes V) \\
        T \otimes 1_I = T = 1_I \otimes T
    \end{split}
\end{equation}
for which we must require that there is a special unit or `trivial' object $I \in {\rm Sys}(\cc)$. Further imposing the interchange law
\begin{equation}
    (T_1 \otimes T_2) \circ (R_1 \otimes R_2) = (T_1 \circ R_1) \otimes (T_2 \circ R_2)
\end{equation}
gives us a \textit{monoidal category}.

To make this a strict \textit{symmetric} monoidal category, we simply add a transformation $\s_{A, B}: A \otimes B \rightarrow B \otimes A$ for every $A, B \in {\rm Sys}(\cc)$ that `swaps around' transformations in the sense that
\begin{equation}
     \s_{A, B}(T \otimes U) = (U \otimes T) \s_{A, B},
\end{equation}
and also satisfies the following.
\begin{equation}
\begin{split}
    \s_{B, A} \circ \s_{A, B} = 1_A \otimes 1_B \\
     (1_B \otimes \s_{A, C})(\s_{A, B} \otimes 1_C) = \s_{A, B \otimes C} \\
\end{split}
\end{equation}

The reader might be surprised that we have defined a sort of theory that is supposed to be appropriate for physics without so much as mentioning the word `state'. In fact, states emerge naturally from what we have said above. Defining a state as a transformation $T:I \rightarrow A$ whose input is the unit system, we see from the final line of (\ref{eq:monoidal}) that its input can be effectively ignored, and we see from the second line that the $\otimes$-product of any pair of states is another state. This is at the heart of the key conceptual shift brought on by a categorical approach to physics: transformations are no longer defined by states but are \textit{primitive}, and states are just special cases of transformations.

\subsection{The trace}

If we want to ensure that states are normalized and transformations are normalization-preserving, or if we want to ignore a part of a system and focus on a smaller subsystem, then it is very useful to introduce a \textit{trace} transformation, Tr: $A \rightarrow I$, for each system $A \in {\rm Sys}(\cc)$. 
We require that tracing out two subsystems individually is equivalent to tracing out the composite system 
\begin{equation}
    { \rm Tr}_A \otimes { \rm Tr}_B = { \rm Tr}_{A \otimes B}
\end{equation}
and that tracing out the trivial system is equivalent to doing nothing at all
\begin{equation}
    { \rm Tr}_I = 1_I.
\end{equation}

Let us call a transformation $T: A \rightarrow B$ \textit{causal} (i.e.\ normalization-preserving) just in case the trace pulls through
\begin{equation}
    { \rm Tr}_B \circ T =  { \rm Tr}_A.
\end{equation}
It is not hard to see that the causal transformations in an SMC form a \textit{sub-SMC}. This means that they form an SMC in their own right, using the identity and swap transformations from the larger SMC. One example of an SMC and its sub-SMC of causal transformations is the matrices over positive numbers and stochastic matrices, to which we now turn.

\subsection{$\mathbb{R}^+$-Mat} \label{sec:rmat}

For our purposes, a particularly important SMC is $\mathbb{R}^+$-Mat.
In this SMC, the systems are natural numbers and the transformations in $\cc(A, B)$ are all of the $\mathbb{R}^+$-valued matrices with $A$ columns and $B$ rows. The $\circ$ and $\otimes$ operations are given by matrix multiplication and the Kroenecker product respectively. The trivial system is the number `1', transformations on which are one dimensional $\mathbb{R}^+$-valued matrices with one entry, giving us a notion of positive numbers. The identity transformations are identity matrices, and \s  \ is defined in an obvious way. It is straightforward to verify that this theory satisfies all the axioms of an SMC given above. 

The traces in this theory are row matrices with a `1' for each entry. Multiplying a matrix with the trace from the left results in a new row matrix where each entry is the sum of all of the elements in the corresponding column of the original matrix. This ensures that the causal transformations are left stochastic matrices and that the causal states are probability distributions. Hence $\mathbb{R}^+$-Mat, and especially its sub-SMC of causal transformations, provides useful tools for modelling operational procedures, in which measurement settings may be chosen and outcomes obtained with various probabilities.

\subsection{Categorical probabilistic theories}

We can now finally define categorical probabilistic theories. These are SMCs with three specific features. Firstly, they contain $\mathbb{R}^+$-Mat, or an equivalent category, as a full sub-SMC.\footnote{A sub-SMC is an SMC comprised of some of the systems from a larger SMC, and some of the transformations between them. The sub-SMC inherits the relevant compositional structure from the larger SMC, and must share its trivial system, identity transformations, and swap transformations.
A sub-SMC is \textit{full} just in case it contains \textit{all} of the transformations from the larger SMC that go between any pair of systems that they share.} We will call the smaller theory the `classical sub-SMC'.

Secondly, a categorical probabilistic theory comes with traces for all its systems, and the traces in the classical sub-SMC are just (equivalent to) the usual traces from $\mathbb{R}^+$-Mat. 

Finally, a categorical probabilistic theory must come with a notion of summing a pair of transformations $T, U \in  \cc(A, B)$ to form another transformation $T+U \in \cc(A, B)$, satisfying three requirements. Firstly, there is a unit of summation $0 \in \cc(A, B)$ for each $A$ and $B$. Secondly, the sums on the classical sub-SMC are given by the usual notion of adding matrices (when we restrict to the causal states, this means taking convex combinations of probability distributions). And thirdly,  the $\circ$ and $\otimes$ operations are bilinear with respect to the sums, in the sense that the following equations are respected.

\begin{multicols}{2}
\begin{equation} \nonumber
\begin{split}
(T+U) \circ V = T \circ V + U \circ V \\
T \circ (U+V) = T \circ U + T \circ V  \\
T \circ 0 = 0 \\
0 \circ T = 0 \\
\end{split}
\end{equation}

   \begin{equation}
\begin{split}
(T+U) \otimes V = T \otimes V + U \otimes V \\
T \otimes (U+V) = T \otimes U + T \otimes V  \\
T \otimes 0 = 0 \\
0 \otimes T = 0 \\
\end{split}
\end{equation}

\end{multicols}

Finally, let us state the definition more compactly.

\begin{definition}
A categorical probabilistic theory is an SMC with a full classical sub-SMC, with traces (which are the usual traces for $\mathbb{R}^+$-Mat on the sub-SMC) and sums (which are unital, are the addition of matrices in the classical sub-SMC, and with respect to which $\otimes$ and $\circ$ are bilinear). 
\end{definition}

\section{PIL theories and absoluteness} \label{app:possibilistic}

This appendix elaborates on Section \ref{sec:logical}. We will explain the idea of possibilistic Bell nonlocality more thoroughly, before proving Theorem \ref{app:possibilistic}.

The following is a formal definition of a conditional probability distribution being possibilistically Bell nonlocal.

\begin{equation} \label{eq:logical_contextuality}
    \begin{split}
        &\exists x_1 \ldots x_n a_1 \ldots a_n:  \Big(p(A_1=a_1, \ldots A_n=a_n|X_1 =x_1, \ldots X_n =x_n) \neq 0 \Big) \\  & \land  \not\exists q: \Big( [ \sum_{\overline{A_1^{x_1} \ldots, A_n^{x_n}}} q(A_1^1 \ldots,  A_1^{x_1}=a_1 \ldots,  A_1^{x_1^{\rm max}} \ldots A_n^1, \ldots, A_n^{x_n}=a_n, \ldots, A_n^{x_n^{\rm max}}) \neq 0 ] \\ &  \ \ \ \ \ \ \ \ \ \ \ \ \ \    \land  [p(A_1=a_1', \ldots, A_n=a_n'|X_1 =x_1', \ldots, X_n =x_n') =0 \\
        & \ \ \ \ \ \ \ \ \ \ \ \ \ \ \implies  \sum_{\overline{A_1^{x_1'} \ldots A_n^{x_n'}}} q(A_1^1, \ldots,  A_1^{x_1}=a_1', \ldots,  A_1^{x_1'^{\rm max}}, \ldots, A_n^1, \ldots, A_n^{x_n'}=a_n', \ldots, A_n^{x_n^{\rm max}}) =0 ]\Big)
    \end{split}
\end{equation}

That is, there exists some measurement context $x_1 \ldots x_n$ and some value-assignment $a_1 \ldots a_n$ that is possible according to $p(A_1 \ldots A_n|X_1 \ldots X_n)$, but which cannot be embedded into a distribution $q$ over a global assignment that makes impossible all the assignments that should be impossible according to $p(A_1 \ldots A_n|X_1 \ldots X_n)$. 

Note that this definition need not be stated in terms of a full-blown conditional probability distribution. It could also be given in terms of a conditional \textit{possibility} distribution, obtainable from the former by replacing each nonzero entry with a `1'. (In this case, the sum should be thought of as Boolean.)

We can now define the possibilistic Bell nonlocality of a perspectival theory in much the same way that we defined standard Bell Nonlocality. 

\begin{definition}[Possibilistic Bell Nonlocality.]
A perspectival theory exhibits \textbf{Possibilistic Bell Nonlocality} just in case it provides a normalized circuit model of the form (\ref{eq:contextuality}), where 
\begin{enumerate}
    \item each triplet of systems $(X_i, S_i, A_i)$ of $T$ can be validly embedded into mutually spacelike separated regions; and
    \item the resulting conditional probability distribution does not satisfy (\ref{eq:logical_contextuality}).
\end{enumerate}
\end{definition}

We now sketch a proof of Theorem \ref{theorem:possibilistic}. A PIL theory must give a possibilistically Bell nonlocal distribution $p(A_1 \ldots A_n|X_1 \ldots X_n)$ via a circuit model of the form \ref{eq:contextuality}.
We proceed, via the same argument from Section \ref{sec:BIL}, to construct a circuit model with a memory update for every value of every setting, where any set of measurements that can be thought of as performed in parallel give the same probabilities as $p(A_1 \ldots A_n|X_1 \ldots X_n)$ for some choice of settings $X_1=x_1, \ldots X_n=x_n$. 

If AOE is true, then on any particular run of the experiment, there is a unique global assignment describing the events that were observed:
\begin{equation}
    (C_1^1=c_1^1, \ldots C_1^{x_1^{\rm max}}=c_1^{x_1^{\rm max}}, \ldots, C_n^1=c_n^1, \ldots, C_n^{x_n^{\rm max}}=c_n^{x_n^{\rm max}})
\end{equation}
Consider again the definition (\ref{eq:logical_contextuality}) of the possibilistic Bell nonlocality of a conditional distribution.
It implies that the only way the global assignment above could contain the possible assignment $a_1 \ldots a_n$ for the context $x_1 \ldots x_n$ is if it contains at least one of the local assignments $a_1' \ldots a_n'$ for a context $x_1' \ldots x_n'$ for which $p(A_1=a_1', \ldots A_n=a_n'|X_1 =x_1', \ldots X_n =x_n') =0$. That is, 
\begin{equation}
\begin{split}
    &(C_1^{x_1}=a_1, \ldots, C_n^{x_n}=a_n) \\
    & \implies \exists x_1' \ldots x_n' a_1' \ldots a_n': p(A_1=a_1', \ldots A_n=a_n'|X_1 =x_1', \ldots X_n =x_n') =0 \\
    & \ \ \ \ \ \ \ \ \ \ \ \ \ \ \ 
    \quad \quad \quad \quad \quad \quad \quad 
    \land (C_1^{x_1'}=a_1', \ldots, C_n^{x_n'}=a_n')
\end{split}
\end{equation}
In other words, if the theory correctly predicts that sometimes the events $(C_1^{x_1}=a_1, \ldots, C_n^{x_n}=a_n)$ are observed,  then it wrongly predicts that some other set of events $(C_1^{x_1'}=a_1', \ldots, C_n^{x_n'}=a_n')$ are never observed.
So if AOE is true, then the PIL theory makes false predictions. $\square$

For example, consider the argument in \cite{ormrod2022no}, sketched in Section \ref{sec:simple}, in which we attempted to use quantum theory to make predictions for observed events that we assumed were absolute. Quantum theory predicts that sometimes Alice and Bob both see `-'. AOE then implies that it must either be the case that Charlie and Daniela both see `1', or else that at least one of them sees `zero'. A fortiori, one of the following statements must then be true:
\begin{itemize}
    \item Charlie and Daniela both see `1';
    \item Alice sees `-' and Daniela sees `0'; or
    \item Bob sees `-' and Charlie sees `0'.
\end{itemize}
But quantum predictions are that none of these statements are ever true. Because this is a scenario in which its possibilistic Bell nonlocality is on display, quantum theory's accurate prediction about what Alice and Bob see leads it to rule out a set of statements whose disjunction is a logical consequence of that very same prediction when one also assumes AOE.

\section{Quantum theory has Separable Dynamics} \label{app:sep}

Here, we show that unitary transformations of the form
\begin{equation}
    V: A_1 \otimes \ldots \otimes A_n \otimes \lambda \rightarrow A_1' \otimes \ldots \otimes A_n' \otimes F,
\end{equation}
with the property that each $A_i$ does not influence $A_j'$ for any $j\neq i$, are separable, in the sense of satisfying (\ref{separable_dynamics}). The notion of influence used here is given by (\ref{ni}), which is equivalent to the possibility of signalling through the unitary, as well as a number of other natural conceptions of causal influence (see Theorem III.1 of \cite{ormrod2022}).  An immediate corollary is that any perspectival theory that has unitaries as fundamental transformations, and the same definition of influence, has 
 Separable Dynamics. This includes the quantum perspectival theory with isometric memory updates introduced Section \ref{sec:qpt} and developed in sections \ref{sec:BIL_bil} and \ref{sec:binsc_nsc}.

It is a consequence of Theorem 3 of \cite{lorenz2020causal} that any unitary of the type $V: A \otimes B \otimes \lambda \rightarrow A' \otimes B' \otimes F$ for which $A$ does not influence $B'$ (again, given the definition (\ref{ni})) and $B$ does not influence $A'$ has the form
\begin{equation} \label{sdproof}
    \begin{tikzpicture}
	\begin{pgfonlayer}{nodelayer}
		\node [style=none] (0) at (1, -0.75) {};
		\node [style=none] (1) at (1, -5.75) {};
		\node [style=none] (2) at (3, -0.75) {};
		\node [style=none] (3) at (3, -5.75) {};
		\node [style=none] (4) at (0.5, -5.25) {$A$};
		\node [style=none] (5) at (0.5, 5.25) {$A'$};
		\node [style=none] (6) at (3.5, -5.25) {$B$};
		\node [style=none] (7) at (3.5, 5.25) {$B'$};
		\node [style=none] (8) at (6.25, 0) {};
		\node [style=none] (9) at (6.25, 0) {};
		\node [style=none] (10) at (6.25, 0) {$=$};
		\node [style=none] (11) at (8, 5.75) {};
		\node [style=none] (12) at (8, -5.75) {};
		\node [style=none] (13) at (7.5, -5.25) {$A$};
		\node [style=none] (14) at (7.5, 5.25) {$A'$};
		\node [style=map] (15) at (8, 0) {$W$};
		\node [style=none] (16) at (10.5, 5.75) {};
		\node [style=none] (17) at (10.5, -5.75) {};
		\node [style=none] (18) at (11, -5.25) {$B$};
		\node [style=none] (19) at (11, 5.25) {$B'$};
		\node [style=map] (20) at (10.5, 0) {$S$};
		\node [style=none] (21) at (12.25, -3) {};
		\node [style=none] (22) at (12.75, -2.5) {$X$};
		\node [style=none] (23) at (14.25, -3) {};
		\node [style=none] (24) at (8.25, -0.5) {};
		\node [style=none] (25) at (11, 0) {};
		\node [style=large map] (26) at (13.25, -3.5) {$T_1$};
		\node [style=none] (27) at (13.25, -5.75) {};
		\node [style=none] (28) at (12.25, 3) {};
		\node [style=none] (29) at (14.25, 3) {};
		\node [style=none] (30) at (8.25, 0.5) {};
		\node [style=none] (31) at (11, 0) {};
		\node [style=large map] (32) at (13.25, 3.5) {$T_2$};
		\node [style=none] (33) at (13.25, 5.75) {};
		\node [style=none] (34) at (13.75, -5.25) {$\lambda$};
		\node [style=none] (35) at (13.75, 5.25) {$F$};
		\node [style=none] (36) at (1, -0.75) {};
		\node [style=none] (37) at (3, -0.75) {};
		\node [style=none] (38) at (4.75, -0.75) {};
		\node [style=none] (39) at (2.75, 0) {$V$};
		\node [style=none] (40) at (0.5, -0.75) {};
		\node [style=none] (41) at (5.25, -0.75) {};
		\node [style=none] (42) at (5.25, 0.75) {};
		\node [style=none] (43) at (0.5, 0.75) {};
		\node [style=none] (44) at (1, 0.75) {};
		\node [style=none] (45) at (3, 0.75) {};
		\node [style=none] (46) at (4.75, 0.75) {};
		\node [style=none] (47) at (5.25, -5.25) {$\lambda$};
		\node [style=none] (48) at (5.25, 5.25) {$F$};
		\node [style=none] (49) at (4.75, -0.75) {};
		\node [style=none] (50) at (4.75, -5.75) {};
		\node [style=none] (51) at (4.75, 5.75) {};
		\node [style=none] (52) at (4.75, 0.75) {};
		\node [style=none] (53) at (1, 5.75) {};
		\node [style=none] (54) at (1, 0.75) {};
		\node [style=none] (55) at (3, 5.75) {};
		\node [style=none] (56) at (3, 0.75) {};
		\node [style=none] (57) at (1, 5.75) {};
		\node [style=none] (58) at (3, 5.75) {};
		\node [style=none] (59) at (14.75, -2.5) {$Y$};
		\node [style=none] (60) at (12.75, 2.5) {$X$};
		\node [style=none] (61) at (14.75, 2.5) {$Y$};
	\end{pgfonlayer}
	\begin{pgfonlayer}{edgelayer}
		\draw (1.center) to (0.center);
		\draw (3.center) to (2.center);
		\draw [in=270, out=90] (12.center) to (11.center);
		\draw (17.center) to (16.center);
		\draw [in=-90, out=90, looseness=0.75] (21.center) to (24.center);
		\draw [in=90, out=-90, looseness=1.25] (25.center) to (23.center);
		\draw (27.center) to (26);
		\draw [in=90, out=-90, looseness=0.75] (28.center) to (30.center);
		\draw [in=-90, out=90, looseness=1.25] (31.center) to (29.center);
		\draw (33.center) to (32);
		\draw (43.center) to (40.center);
		\draw (40.center) to (41.center);
		\draw (41.center) to (42.center);
		\draw (42.center) to (43.center);
		\draw (50.center) to (49.center);
		\draw (52.center) to (51.center);
		\draw (54.center) to (53.center);
		\draw (56.center) to (55.center);
	\end{pgfonlayer}
\end{tikzpicture}
\end{equation}
where $T_1$ is an isometry, $T_2$ is a channel, and $W$ and $S$ are unitaries. Furthermore, these transformations have some useful properties. In particular, the Hilbert space of the systems $X$ and $Y$ can be decomposed as
\begin{equation}
    \begin{split}
        \ch_X &= \bigoplus_{i=1}^n \ch_X^i \\
        \ch_Y &= \bigoplus_{j=1}^n \ch_Y^j
    \end{split}
\end{equation}
such that $T_1$ can produce any pure state that lives a subspace $\ch_X^i \otimes \ch_Y^i$ of $\ch_X \otimes \ch_Y$ where the indices match, i.e.\
\begin{equation} \label{produce}
        \forall i \ \forall \ket{\psi} \in \ch_X^i \otimes \ch_Y^i \ \ \exists \ket{\phi} \in \ch_{\lambda}: \ \ \ \ket{\psi}\bra{\psi} = T_1(\ket{\phi}\bra{\phi}),  
\end{equation}
and such that $W$ and $S$ decompose into a direct sum of unitaries that act on the different $\ch_X^i$ and $\ch_Y^i$ respectively, i.e.
\begin{equation}
\begin{split}
    W &= \bigoplus_{i=1}^n W_i \\
    S &= \bigoplus_{j=1}^n S_j.
\end{split}
\end{equation}

Now, suppose that $A$ and $A'$ each factor into a pair of systems
\begin{equation}
    \begin{split}
        A &= C \otimes D \\
        A' &= C' \otimes D'
    \end{split}
\end{equation}
such that $C$ does not influence $D'$ or $B'$, and $D$ does not influence $C'$ or $B'$, through $V$. It follows that the same no-influence relations hold through each $W_i$.\footnote{This is because 
\begin{enumerate}
    \item if it were possible to signal from, say, $C$ to $D'$ through some $W_i$ then it would be possible to signal through some $W_i$, then it would be possible to signal in the special case where $X$ is prepared in some pure state $\ket{\psi} \in \ch_X^i$ (as can be proved using (\ref{ni}) and the fact that pure density operators span the space of linear operators on a Hilbert space), but
    \item  if \textit{that} were possible, then it would be possible to signal from $C$ to $D'$ through $V$ (as follows from (\ref{sdproof}) and (\ref{produce})).
\end{enumerate}} Thus one can apply Theorem 3 of \cite{lorenz2020causal} to (\ref{sdproof}) again to derive the following
\begin{equation} \label{sdproof2}
    \begin{tikzpicture}
	\begin{pgfonlayer}{nodelayer}
		\node [style=none] (10) at (-0.5, 0) {};
		\node [style=none] (11) at (-0.5, 0) {};
		\node [style=none] (12) at (-0.5, 0) {$=$};
		\node [style=none] (13) at (1.25, 5.75) {};
		\node [style=none] (14) at (1.25, -5.75) {};
		\node [style=none] (15) at (0.75, -5.25) {$C$};
		\node [style=none] (16) at (0.75, 5.25) {$C'$};
		\node [style=map] (17) at (1.25, 0) {$Q$};
		\node [style=none] (18) at (6.25, 5.75) {};
		\node [style=none] (19) at (6.25, -5.75) {};
		\node [style=none] (20) at (6.75, -5.25) {$B$};
		\node [style=none] (21) at (6.75, 5.25) {$B'$};
		\node [style=map] (22) at (6.25, 0) {$S$};
		\node [style=none] (23) at (8.75, -3) {};
		\node [style=none] (24) at (9.5, -2) {};
		\node [style=none] (25) at (10.75, -3) {};
		\node [style=none] (26) at (1.5, -0.5) {};
		\node [style=none] (27) at (6.75, 0) {};
		\node [style=large map] (28) at (9.75, -3.5) {$T_3$};
		\node [style=none] (29) at (9.75, -5.75) {};
		\node [style=none] (30) at (8.75, 3) {};
		\node [style=none] (31) at (10.75, 3) {};
		\node [style=none] (32) at (1.5, 0.5) {};
		\node [style=none] (33) at (6.75, 0) {};
		\node [style=large map] (34) at (9.75, 3.5) {$T_4$};
		\node [style=none] (35) at (9.75, 5.75) {};
		\node [style=none] (36) at (10.25, -5.25) {$\lambda$};
		\node [style=none] (37) at (10.25, 5.25) {$F$};
		\node [style=none] (38) at (3.75, 5.75) {};
		\node [style=none] (39) at (3.75, -5.75) {};
		\node [style=none] (40) at (4.25, -5.25) {$D$};
		\node [style=none] (41) at (4.25, 5.25) {$D'$};
		\node [style=map] (42) at (3.75, 0) {$R$};
		\node [style=none] (43) at (4, -0.5) {};
		\node [style=none] (44) at (4, 0.5) {};
		\node [style=none] (45) at (9.75, -3) {};
		\node [style=none] (46) at (4, 0) {};
		\node [style=none] (47) at (9.75, 3) {};
		\node [style=none] (48) at (4, 0) {};
		\node [style=none] (49) at (-5.75, -0.75) {};
		\node [style=none] (50) at (-5.75, -5.75) {};
		\node [style=none] (51) at (-3.75, -0.75) {};
		\node [style=none] (52) at (-3.75, -5.75) {};
		\node [style=none] (53) at (-6.25, -5.25) {$A$};
		\node [style=none] (54) at (-6.25, 5.25) {$A'$};
		\node [style=none] (55) at (-3.25, -5.25) {$B$};
		\node [style=none] (56) at (-3.25, 5.25) {$B'$};
		\node [style=none] (57) at (-5.75, -0.75) {};
		\node [style=none] (58) at (-3.75, -0.75) {};
		\node [style=none] (59) at (-2, -0.75) {};
		\node [style=none] (60) at (-4, 0) {$V$};
		\node [style=none] (61) at (-6.25, -0.75) {};
		\node [style=none] (62) at (-1.5, -0.75) {};
		\node [style=none] (63) at (-1.5, 0.75) {};
		\node [style=none] (64) at (-6.25, 0.75) {};
		\node [style=none] (65) at (-5.75, 0.75) {};
		\node [style=none] (66) at (-3.75, 0.75) {};
		\node [style=none] (67) at (-2, 0.75) {};
		\node [style=none] (68) at (-1.5, -5.25) {$\lambda$};
		\node [style=none] (69) at (-1.5, 5.25) {$F$};
		\node [style=none] (70) at (-2, -0.75) {};
		\node [style=none] (71) at (-2, -5.75) {};
		\node [style=none] (72) at (-2, 5.75) {};
		\node [style=none] (73) at (-2, 0.75) {};
		\node [style=none] (74) at (-5.75, 5.75) {};
		\node [style=none] (75) at (-5.75, 0.75) {};
		\node [style=none] (76) at (-3.75, 5.75) {};
		\node [style=none] (77) at (-3.75, 0.75) {};
		\node [style=none] (78) at (-5.75, 5.75) {};
		\node [style=none] (79) at (-3.75, 5.75) {};
	\end{pgfonlayer}
	\begin{pgfonlayer}{edgelayer}
		\draw [in=270, out=90] (14.center) to (13.center);
		\draw (19.center) to (18.center);
		\draw [in=-90, out=90, looseness=0.75] (23.center) to (26.center);
		\draw [in=90, out=-90, looseness=1.25] (27.center) to (25.center);
		\draw (29.center) to (28);
		\draw [in=90, out=-90, looseness=0.75] (30.center) to (32.center);
		\draw [in=-90, out=90, looseness=1.25] (33.center) to (31.center);
		\draw (35.center) to (34);
		\draw (39.center) to (38.center);
		\draw [in=90, out=-90] (46.center) to (45.center);
		\draw [in=-90, out=90] (48.center) to (47.center);
		\draw (50.center) to (49.center);
		\draw (52.center) to (51.center);
		\draw (64.center) to (61.center);
		\draw (61.center) to (62.center);
		\draw (62.center) to (63.center);
		\draw (63.center) to (64.center);
		\draw (71.center) to (70.center);
		\draw (73.center) to (72.center);
		\draw (75.center) to (74.center);
		\draw (77.center) to (76.center);
	\end{pgfonlayer}
\end{tikzpicture}
\end{equation}
where $T_3$ is an isometry, $T_4$ is a channel, and $Q$ and $R$ are unitaries. Furthermore, we can always set things up so that we use some systems $X'$, $Y'$ and $Z'$ as outputs of $T_3$ and inputs of $T_4$ that admit decompositions
\begin{equation}
    \begin{split}
        \ch_{X'} &= \bigoplus_{i=1}^{n'} \ch_X^i \\
        \ch_{Y'} &= \bigoplus_{j=1}^{n'} \ch_Y^j \\
        \ch_{Z'} &= \bigoplus_{k=1}^{n'} \ch_Z^k
    \end{split}
\end{equation}
such that any pure state that lives in some $\ch_{X'}^i \otimes \ch_{Y'}^i \otimes \ch_{Z'}^i$ can be produced by $T_3$, and with respect to which $Q$, $R$, and $S$ decompose into direct sums of unitaries.

One can then argue that if $C'$ further factorizes into two systems that do not influence each other through $V$, then $Q$ should decompose as well. Proceeding iteratively in this way, one can derive the full Separable Dynamics condition.\footnote{One can further prove that $V$ decomposes in a similar way but where all the transformations are unitary, as long as one uses the extended quantum circuits introduced in \cite{lorenz2020causal} and developed in \cite{vanrietvelde2020routed, vanrietvelde2021universal}. One simply generalizes the iterative procedure in the proof of Theorem 11 of \cite{lorenz2020causal}.}

\section{On the individual necessity of the assumptions} \label{app:role}

This appendix studies how just one of the five BINSC properties might be dropped to avoid a conflict with AOE.
In doing so, it sheds light on the precise roles the individual properties play in the breakdown of the absoluteness of observed events.

We will explicitly construct one perspectival theory that retains consistency with AOE just by dropping Bell Nonlocality, and another that does so just by dropping Local Dynamics.
While we don't have an example of a theory that does so by only giving up on Information Preservation while clinging to the other four properties, we do show that any theory with `collapses' is consistent with AOE. A corollary of this is that \textit{if} one can construct a collapse theory that has all the BINSC properties except Information Preservation, \textit{then} that theory will be consistent with AOE. 

We will conclude by constructing a quantum collapse theory, and explaining why it does not retain all the other BINSC properties. We leave open whether there exists a perspectival theory that has all BINSC properties besides Information Preservation, and that is consistent with AOE.

We do not discuss the possibility of rejecting Consistent Embeddings, since we doubt that anyone will want to. This leaves open the possibility, discussed in Section \ref{sec:conc}, of dropping just Separable Dynamics in order to recover AOE.

\subsection{Bell Nonlocality}

Bell Nonlocality was in the punchline of the proof of Theorem \ref{thm:BIL_AOE}.
It established that the predictions of any BIL perspectival theory for the marginals of the assumed global distribution were inconsistent with one another.
If the perspectival theory were instead Bell \textit{Local}, as well as being Information Preserving and having Local Dynamics, then we would have proven quite the opposite: that the predictions \textit{were} consistent.

Let us see an example. The ideas that information is preserved, dynamics are local, and Bell inequalities cannot be violated are all at the heart of the classical physical worldview. It is therefore not surprising that one can easily write down a classical perspectival theory of this description that is manifestly compatible with AOE. 

The perspectival theory we have in mind is based on the categorical probabilistic theory of matrices over reals, $\mathbb{R}^+$-Mat, defined in Appendix \ref{sec:rmat}.
The probability extractors are stochastic matrices. The memory updates simply copy the measured system and then implement the extractor on the copy. Explicitly, the memory update for some extractor $S: A \rightarrow B$ is given by a matrix $U_S: A \rightarrow A \otimes A$ of the form
\begin{equation}
    U_S = (S \otimes I) \circ \texttt{COPY}
\end{equation}
where \texttt{COPY}$: A \rightarrow A \otimes A$ has the matrix elements $\delta_i^{j}\delta_i^k$.  It is immediately clear that this theory is Information Preserving since from the outside perspective the measurements do not disturb the system. 

Now, let us designate the matrices that represent reversible functions as the fundamental transformations, and define causal relations via (\ref{ni}) as functional dependences. Then the theory has Separable Dynamics.\footnote{This follows from the fact that (a) any function with many outputs can be implemented by copying the inputs and feeding them into a function for each output, and (b) when an output does not depend on an input in the original function, it is not necessary for its own particular function in the decomposition to receive that input.}
By fiat, we restrict the embeddings so that No Superluminal Influences and Consistent Embeddings are also respected.\footnote{Doing this is easy -- we simply start by restricting the embeddings on the reversible functions so that No Superluminal Influences is respected, and then we rule out all embeddings on the other transformations that would then lead to a conflict with Consistent Embeddings.} It follows from Theorem \ref{thm:nsc} that the theory has Local Dynamics. Recalling that the transformations are classical in our theory, Local Dynamics is equivalent to the failure of Bell Nonlocality. Thus the theory is not Bell Nonlocal.

As mentioned above, since this theory is Information Preserving and has Local Dynamics but is not Bell Nonlocal, one can show that it \textit{is} consistent with AOE in the sort of scenario we used to prove Theorem \ref{thm:BIL_AOE}. This is not surprising given how natural the connection is between the inside and the outside perspectives.

\subsection{No Superluminal Influences}

One only has to modify the theory above slightly so that it becomes Bell Nonlocal but contains superluminal influences. One simply stipulates that now \textit{all} embeddings are valid. Consistent Embeddings is still satisfied, though now in a rather trivial way.
But No Superluminal Influences is not -- since we can now embed reversible functions however we like, we can certainly embed them such that certain outputs depend on inputs outside their past light cone. 

By a similar argument, it follows that this new theory is Bell Nonlocal. (And since the theory has classical transformations, it immediately follows that it lacks Local Dynamics.) Since the memory updates are unchanged, the theory is still Information Preserving. And since the fundamental transformations and the definition of influence have not changed, the theory retains Separable Dynamics. It thus has every one of the BINSC properties except for No Superluminal Influences. And it is not hard to convince oneself that the theory remains consistent with AOE, at least in the sort of scenario considered in the proof of Theorem \ref{thm:BIL_AOE}.

More realistically, a perspectival formulation of Bohm theory would plausibly have all five BINSC properties except No Superluminal Influences.

\subsection{Information Preservation}
\label{sec:collapses}

Information Preservation ensures that any set of measurements can effectively be performed on a system in a single run of the experiment, even if they are `incompatible'.
In the context of a BIL or BINSC theory, it thereby enables us to bring into the observed part of reality all of the probability distributions for fixed measurement settings which, by Bell Nonlocality (and, in particular, (\ref{eq:contextuality})), cannot be unified into a consistent whole.

But if we do not assume Information Preservation, then whatever kind of nonabsoluteness our perspectival theory requires might be confined to the unobserved domain.
In fact, if we assume that we are dealing with a theory that, far from being measurement-preserving, involves \textit{collapses}, then we can prove that such a theory is compatible with AOE, even if it has Local Dynamics and Bell Locality. 

By a theory `having collapses' we mean that it has the property that whenever applying an extractor to a state gives a probability distribution of the form $E(\phi)=p(A)$, applying associated measurement update gives a state $M(\phi)=\sum_a  p(A=a)\psi_a$ that can be written as a probabilistic mixture of normalized states $\psi_a$ with the same coefficients. That is, 
\begin{equation} \label{eq:collapse}
    E(\phi)=p(A) \implies (f(E))(\phi)=\sum_a  p(A=a)\psi_a
\end{equation}

Let us explain why theories with collapses are consistent with AOE, starting with a very simple example. Consider the following circuit model, which describes the original `Wigner's Friend' scenario \cite{wigner1995remarks}.
\begin{equation} \label{eq:wigners_friend}
    \begin{tikzpicture}
	\begin{pgfonlayer}{nodelayer}
		\node [style=point] (0) at (0.75, -2) {$\phi$};
		\node [style=large map] (1) at (0, 0.25) {$U_1$};
		\node [style=none] (2) at (0.75, 5) {};
		\node [style=none] (3) at (-0.75, 0.25) {};
		\node [style=none] (4) at (-0.75, 5) {};
		\node [style=none] (9) at (0, 4.75) {};
		\node [style=large map] (11) at (0, 2.75) {$U_2$};
		\node [style=none] (12) at (0.25, 3) {};
		\node [style=none] (13) at (0.25, 5) {};
		\node [style=none] (14) at (0.25, 3) {};
	\end{pgfonlayer}
	\begin{pgfonlayer}{edgelayer}
		\draw (2.center) to (0);
		\draw (3.center) to (4.center);
		\draw (13.center) to (14.center);
	\end{pgfonlayer}
\end{tikzpicture}
\end{equation}
It follows from the convex linearity of transformations in a perspectival theory that, in a theory with collapses, predictions for probability distribution over the outcomes $C_2$ of the supermeasurement have the form $\sum_{C_1}p(C_2|C_1)p(C_1)$, where $p(C_1)$ is the distribution over the outcomes of the first measurement, and $p(C_2|C_1=c)$ is the probability for the supermeasurement outcome given that the $c$th outcome was observed. But this means that both distributions are marginals of the global distribution $q(C_1C_2):=p(C_2|C_1)p(C_1)$.

Slightly more generally, consider the circuit model for a scenario with four (super)measurements from (\ref{eq:four_mus}). 
It is easily derived that in a theory with collapses the predictions for each pair of measurements are the marginals of a distribution of the form
\begin{equation}
\begin{split}
    q(C_1^1, C_2^1, C_1^2, C_2^2) &= p(C_1^2, C_2^2|C_1^1, C_2^1)p(C_1^1, C_2^1) \\
   &= p(C_1^2|C_1^1)p(C_2^2|C_2^1)p(C_1^1, C_2^1)
\end{split}
\end{equation} 
where $p(C_1^1, C_2^1)$ are the joint probabilities for the pair of initial measurements, and $p(C_1^2, C_2^2|C_1^1=c, C_2^1=d)$ are the probabilities for the pair of supermeasurements given some fixed outcomes for the initial measurements. A similar result can be derived for arbitrary normalized circuit models with measurement updates. Thus any theory with collapses is consistent with AOE, at least in scenarios like the one from the proof of Theorem \ref{thm:BIL_AOE}.

\subsection{A quantum collapse theory} \label{app:role_collapse}

Consider again a categorical probabilistic formulation of quantum theory, in which the transformations are completely positive maps. The probability extractors in this theory are provided by POVMs. A POVM -- or \textit{positive operator-valued measurement} -- is a measurement in which an outcome $i$ corresponds to one of a set of positive operators $\{\sigma^i_A\}_i$ that sums to the identity operator, and has a probability given by tracing that operator together with a state operator $\rho$, so that ${\rm Prob}(i)={\rm Tr}(\sigma_A^i \rho)$.

Instead of choosing isometries as a memory updates, to formulate a quantum collapse perspectival theory we can choose some non-isometric channels. We will associate the POVM $\{\sigma^i_A\}_i$ with the following channel
\begin{equation}
    \cc(\cdot) = \sum_i {\rm Tr}( \sigma^i_A(\cdot)) \frac{\sigma^i_A}{{\rm Tr}(\sigma^i_A)} \otimes \ket{i}\bra{i}_M
\end{equation}

This channel leaves the memory system in the basis state $\ket{i}$ with a probability equal to that of obtaining the $i$ outcome, and it re-prepares the measured system in the state corresponding to that outcome.\footnote{More generally, one might want to be able to associate a single POVM with many different memory updates, which re-prepare the state in different ways. One can accomplish this by either (a) considering a theory with multiple copies of the same POVM, or (b) using an injective mapping, rather than a bijective function, from extractors to updates (c.f.\ footnote \ref{note:ptgeneral}).} One can see that the function is bijective, since one can recover the set that defines the original POVM as follows
\begin{equation}
    \{ \bra{i}_M \cc(I) \ket{i}_M \}_i
\end{equation}

 It is clear that $\cc$ respects (\ref{eq:collapse}). We therefore have a perspectival theory with collapses, and which is therefore not Information Preserving. Of course, the theory is also Bell Nonlocal. This leaves open whether we can find a set of valid embeddings, a set of fundamental transformations, and a definition of causal relations such that the remaining three BINSC properties are satisfied.

It is not obvious that we can do so in any reasonable way. The unitaries would not be a natural choice of fundamental transformations for this theory,\footnote{It is possible that some advocates of `subjective collapse' theories, such as certain variants of the Copenhagen interpretation, will want to defend the idea that unitaries should be considered fundamental even a quantum theory with collapses. But in our framework the point of designating a set of transformations as fundamental is that one defines causal relations in terms of them, and we think it more likely that such people would want to define causal relations in terms of arbitrary quantum operations, or even just at the level of experimental data (as in \cite{gogioso2022topology}).} but if we take arbitrary quantum channels as fundamental, then we do not have the separability property. It is possible that there exists some natural closed set of fundamental transformations for the theory that do have the separability property (\ref{separable_dynamics}) for a suitable notion of a causal relation, but we have no positive evidence for this. 

Thus, while the theory can certainly be fleshed out, in a natural enough way, to ensure that No Superluminal Influences and Consistent Embeddings are respected, it is not clear that it can be naturally fleshed out to ensure that it also has Separable Dynamics (and/or Local Dynamics). We leave the construction of such a collapse theory as an open problem.

\end{document}